\newif\iffigs\figstrue
\documentclass[a4paper,11pt]{article}
\usepackage{latexsym,amssymb,amsmath,lscape,graphics, graphics, setspace}
\usepackage{graphicx}        % pacchetto grafico standard LaTeX
\usepackage{longtable}
\usepackage{youngtab}
\usepackage{multirow}
\usepackage{color}
\usepackage{slashed,epsfig}
\usepackage{amsfonts}

\newcommand{\e}{\textrm{e}}

\newtheorem{definizione}{Definition}[section]

\newcommand{\bd}{\begin{definizione}}
\newcommand{\ed}{\end{definizione}}
\newcommand{\mathsym}[1]{{}}
\newcommand{\unicode}[1]{{}}

\def\Gslat{\relax{\slash\kern-.58em G}}
\def\Dslat{\relax{\slash\kern-.68em D}}
\def\Fslat{\relax{\slash\kern-.68em F}}
\def\Phislat{\relax{\slash\kern-.65em \Phi}}
\def\IC{\relax\,\hbox{$\inbar\kern-.3em{\rm C}$}}
\def\IG{\relax\,\hbox{$\inbar\kern-.3em{\rm G}$}}
\def\IB{\relax{\rm I\kern-.18em B}}
\def\ID{\relax{\rm I\kern-.18em D}}
\def\IL{\relax{\rm I\kern-.18em L}}
\def\IF{\relax{\rm I\kern-.18em F}}
\def\IH{\relax{\rm I\kern-.18em H}}
\def\II{\relax{\rm I\kern-.17em I}}
\def\IN{\relax{\rm I\kern-.18em N}}
\def\IP{\relax{\rm I\kern-.18em P}}
\def\IQ{\relax\,\hbox{$\inbar\kern-.3em{\rm Q}$}}
\def\bfzero{\relax\,\hbox{$\inbar\kern-.3em{\rm 0}$}}
\def\IK{\relax{\rm I\kern-.18em K}}
\def\IG{\relax\,\hbox{$\inbar\kern-.3em{\rm G}$}}
 \font\cmss=cmss10 \font\cmsss=cmss10 at 7pt
\def\IR{\relax{\rm I\kern-.18em R}}
\def\ZZ{\relax\ifmmode\mathchoice
{\hbox{\cmss Z\kern-.4em Z}}{\hbox{\cmss Z\kern-.4em Z}}
{\lower.9pt\hbox{\cmsss Z\kern-.4em Z}} {\lower1.2pt\hbox{\cmsss Z\kern-.4em
Z}}\else{\cmss Z\kern-.4em Z}\fi}
\def\bfone{\relax{\rm 1\kern-.35em 1}}

\def\inbar{\vrule height1.5ex width.4pt depth0pt}
\def\bfzero{\relax{\rm I\kern-.18em 0}}
\def\bfone{\relax{\rm 1\kern-.35em 1}}

\DeclareFontFamily{U}{rsf}{} \DeclareFontShape{U}{rsf}{m}{n}{
  <5> <6> rsfs5 <7> <8> <9> rsfs7 <10-> rsfs10}{}
\DeclareMathAlphabet\Scr{U}{rsf}{m}{n}

\def\e{\epsilon}

\newcommand{\ft}[2]{{\textstyle\frac{#1}{#2}}}
\def\tilde{\widetilde}

\def\1bar{1\hskip -.275cm -}
\def\2bar{2\hskip -.275cm -}
\def\3bar{3\hskip -.275cm -}

\newsavebox{\uuunit}
\sbox{\uuunit}
                 {\setlength{\unitlength}{0.825em}
                      \begin{picture}(0.6,0.7)
                                      \thinlines
                                      \put(0,0){\line(1,0){0.5}}
                                      \put(0.15,0){\line(0,1){0.7}}
                                      \put(0.35,0){\line(0,1){0.8}}
                                     \multiput(0.3,0.8)(-0.04,-0.02){10}{\rule{0.5pt}{0.5pt}}
                      \end {picture}}

\makeatletter \@addtoreset{equation}{section} \makeatother
\def\bfone{\relax{\rm 1\kern-.35em 1}}

\def\bfone{\relax{\rm 1\kern-.35em 1}}
\font\cmss=cmss10 \font\cmsss=cmss10 at 7pt
\begin{document}
\begin{titlepage}
\begin{center}
%\begin{flushright}
%DFTT07/23\\
%JINR-E2-2007-145
%\end{flushright}
\vskip 0.2cm
%%%%%%%%%%%%%%%%%%%%%%%%%%%%%%%%%%%%%%%%%%%%%%%%%%%%%%%%%%%%%%%%%%%%
\vskip 0.2cm
{\Large\sc Supersymmetric  M2-branes with Englert Fluxes,   \\
\vskip 0.2cm  and \vskip 0.2cm
 the simple  group $\mathrm{PSL(2,7)}$}  \\[1cm]
%%%%%%%%%%%%%%%%%%%%%%%%%%%%%%%%%%%%%%%%%%%%%%%%%%%%%%%%%%%%%%%%%%%%
%%%%%%%%%%%%%%%%%%%%%%%%%%%%%%%%%%%%%%%%%%%%%%%%%%%%%%%%%%%%%%%%%%%%
{\sc Pietro~Fr\'e${}^{\; a,b,e}$}\\[10pt]
 \vspace{6pt}
{${}^{a}$\sl\small Dipartimento di Fisica\footnote{Prof. Fr\'e is presently
fulfilling the duties of Scientific Counselor of the Italian Embassy in the
Russian Federation, Denezhnij pereulok, 5, 121002 Moscow, Russia.
\emph{e-mail:} \quad {\small {\tt pietro.fre@esteri.it}}},
Universit\`a di Torino\\${}^{b}$INFN -- Sezione di Torino \\
via P. Giuria 1, \ 10125 Torino \ Italy\\}
\emph{e-mail:} \quad {\small {\tt fre@to.infn.it}}\\
\vspace{5pt}
{{\em $^{c}$\sl\small  National Research Nuclear University MEPhI\\ (Moscow Engineering Physics Institute),}}\\
{\em Kashirskoye shosse 31, 115409 Moscow, Russia}~\quad\\
%%%%%%%%%%%%%%%%%%%%%%%%%%%%%%%%%%%%%%%%%%%%%%%
\vspace{15pt}
\begin{abstract}
A new class is introduced of M2-branes solutions of d=11 supergravity that include internal fluxes obeying
Englert equation in 7-dimensions. A simple criterion  for the existence of Killing spinors in such
backgrounds is established. Englert equation is viewed as the generalization to d=7 of Beltrami equation
defined in d=3 and it is treated accordingly. All 2-brane solutions of minimal d=7 supergracity can be
uplifted to d=11 and have $\mathcal{N} \ge 4$ supersymmetry. It is shown that the simple group
$\mathrm{PSL(2,7)}$ is crystallographic in d=7 having an integral action on the A7 root lattice. By means of
this point-group and of the $\mathrm{T^7}$ torus obtained quotiening $\mathbb{R}^7$ with the A7 root lattice
we were able to construct new M2 branes with Englert fluxes and $\mathcal{N} \le 4$. In particular we exhibit
here an $\mathcal{N}=1$ solution depending on 4-parameters and admitting a large non abelian discrete
symmetry, namely $\mathrm{G_{21}}\equiv \mathbb{Z}_3\ltimes \mathbb{Z}_7 \subset \mathrm{PSL(2,7)}$. The dual
$d=3$ field theories have the same symmetries and have complicated non linear interactions.
\end{abstract}
\end{center}
\end{titlepage}
\tableofcontents \noindent {}
\newpage
{\small
\begin{flushright}
{\it This paper is dedicated by the author to his very good and distinguished friend \\
the Nobel Laureate {Francois} Englert, remembering the good time when he first\\
announced in the Trieste Spring School 
his newly found solution of M-theory.\\
The author hopes that {Francois} will like this new, unexpected
 Englert solution\\  with $\mathcal{N}=1$ supersymmetry.}
\end{flushright}
}
\section{Introduction}
\label{introibo}
The finite group:
\begin{equation}\label{L168}
  \mathrm{L_{168}}\, \equiv \, \mathrm{PSL(2,\mathbb{Z}_7)}
\end{equation}
is the second smallest simple group after the alternating group $A_5$ which has
60 elements and coincides with the symmetry group of the regular icosahedron or
dodecahedron. As anticipated by its given name, $\mathrm{L_{168}}$  has 168
elements: they can be identified with all the possible $2\times2$ matrices with
determinant one whose entries belong to the finite field $\mathbb{Z}_7$,
counting them up to an overall sign. In projective geometry, $\mathrm{L_{168}}$
is classified as a \textit{Hurwitz group} since it is the automorphism group of
a Hurwitz Riemann surface, namely a surface of genus $g$ with the maximal
number $84\,(g-1)$ of conformal automorphisms\footnote{Hurwitz's automorphisms
theorem proved in 1893 \cite{hurvizzo} states that the order $|\mathcal{G}|$ of
the group $\mathcal{G}$ of orientation-preserving conformal automorphisms, of a
compact Riemann surface of genus $g > 1$ admits the following upper bound
$|\mathcal{G}| \le 84(g - 1)$}. The Hurwitz surface pertaining to the Hurwitz
group $\mathrm{L_{168}}$ is the Klein quartic \cite{kelinquart}, namely the
locus $\mathcal{K}_4$ in $\mathbb{P}_2(\mathbb{C})$ cut out by the following
quartic polynomial constraint on the homogeneous coordinates $\{x,y,z\}$:
\begin{equation}\label{kleinusquart}
   x^3 \,y + y^3 \, z \, + z^3 \, x \, = \, 0
\end{equation}
Indeed $\mathcal{K}_4$ is a genus $g=3$ compact Riemann surface and it can be
realized as the quotient of the hyperbolic Poincar\'e plane $\mathbb{H}_2$ by a
certain  group $\Gamma$ that acts freely on $\mathbb{H}_2$ by isometries.
\par
The $\mathrm{L_{168}}$ group, which is also isomorphic to
$\mathrm{GL(3,\mathbb{Z}_2)}$, has received a lot of attention in Mathematics
and it has  important applications in algebra, geometry, and number theory: for
instance, besides being associated with the Klein quartic, $\mathrm{L_{168}}$
is the automorphism group of the Fano plane. The reason why we consider
$\mathrm{L_{168}}$ in this paper is associated with another property of this
finite simple group which was proved fifteen years ago in \cite{kingus},
namely:
\begin{equation}\label{caralino}
    \mathrm{L_{168}} \, \subset \, \mathrm{G_{2(-14)}}
\end{equation}
This means that $\mathrm{L_{168}}$ is a finite subgroup of the compact form of
the exceptional Lie group $\mathrm{G_2}$ and the $7$-dimensional fundamental
representation of the latter is irreducible upon restriction to
$\mathrm{L_{168}}$.
\par
This fact is quite inspiring in the context of $\mathrm{M}$-theory since it
suggests possible connections with manifolds of $\mathrm{G_{2}}$-holonomy and
alludes to scenarios where $\mathrm{L_{168}}$ plays some key role in
compactifications  $11\to4 $ or in $M2$-brane solutions.
\par
These suggestions become much more circumstantial if we focus on the following
linear equation:
\begin{equation}\label{englertus}
    \star \mathrm{d}\mathbf{Y}^{[3]}\, = \, - \, \frac{\mu}{4} \,\mathbf{Y}^{[3]}
\end{equation}
that implies
\begin{equation}\label{cassionevole}
    \mathrm{d}\star\mathbf{Y}^{[3]} \, = \, 0
\end{equation}
and which we  name Englert equation for the reasons we presently explain. In
1982, after the Freund Rubin compactification of d=11 supergravity on the round
$7$-sphere had been found \cite{freundrubin} and other similar solutions on
different $7$-manifolds had also been proposed \cite{su3su2u1},\cite{Awada:1982pk} Francois
Englert introduced a new solution on the round $7$-sphere \cite{Englert:1982vs} where the $4$-form
$\mathcal{F}_{\hat{\mu}\hat{\nu}\hat{\rho}\hat{\sigma}}$ has two distinct
non-vanishing parts, the first $e \,\epsilon_{\mu\nu\rho\sigma}$ living on the
reduced space-time $\mathcal{M}_4$ and proportional to the epsilon symbol
through a constant parameter (the Freund Rubin parameter $e$), the second
$\mathcal{F}_{IJKL}$ living on the internal $\mathcal{M}_7$ manifold and giving
rise to a $3$-form $\Phi$ and a dual $4$-form $\null^\star\Phi$   according to:
\begin{eqnarray}\label{Phiquesto}
    \null^\star\Phi & = & \mathcal{F}_{IJKL} \, e^I\wedge e^J \wedge e^K \wedge e^L \nonumber\\
    \Phi & \equiv & \ft {1}{24} \epsilon_{ABCIJKL} \, \mathcal{F}_{IJKL} \, e^A \, \wedge \,
  e^B\, \wedge \, \e^C
\end{eqnarray}
where $e^I$ denote the vielbein $1$-forms on $\mathcal{M}_7$. The conditions
that $\Phi$ has to satisfy in order to fit into an exact solution of $d=11$
supergravity are the following ones:
\begin{eqnarray}
  \null^\star\mathrm{d} \Phi & = & 12\, e \, \Phi \quad ; \quad \mathrm{d} \null^\star\Phi \, = \, 0
  \label{ollaolla}
\end{eqnarray}
As one sees these conditions just coincide with what we named Englert equation:
it suffices to set $\mathbf{Y}^{[3]}\, = \, \Phi$ and  identify $\frac{\mu}{4}
\, = \, -12 e$. In \cite{Fre':2006ut}, together with M. Trigiante I showed that the existence of a form $\Phi$ satisfying eq.s (\ref{ollaolla}) is equivalent to the definition,
introduced in \cite{weakG2}, of manifolds $\mathcal{M}_7$ of weak
$\mathrm{G_2}$-holonomy. This is the new phrasing, in modern parlance, of the
notion originally introduced at the beginning of the eighties by the authors of
\cite{riccapietroEngMan} under the name of Englert manifolds. On Englert
manifolds $\mathcal{M}_7$ we have non trivial solutions $\eta$ of the following
weak Killing spinor equation:
\begin{equation}
   \mathcal{D} \, \eta \, = \, - \, \ft 32 \, e \, \tau_I \, \eta \, e^I
\label{weakkispinor}
\end{equation}
where $\tau_I$ are a set of real gamma matrices:
\begin{equation}\label{rumanzo}
    \left\{\tau_I \, , \, \tau_J \right \} \, = \, - \, 2 \, \delta_{IJ}
\end{equation}
It suffice to set:
\begin{equation}\label{ferdinandus}
    \Phi \, \equiv \, \ft {1}{24} \epsilon_{ABCIJKL} \, \left(\eta^T \, \tau_{IJKL} \,\eta \right) e^A \, \wedge \,
  e^B\, \wedge \, \e^C
\end{equation}
and in force of eq.(\ref{weakkispinor}), Englert equation (\ref{Phiquesto}) is
satisfied.
\par
In the recent paper \cite{Fre':2015lds} it was observed by some of us that
Englert equation (\ref{englertus}) is the natural generalization to
$7$-dimensions of Beltrami equation:
\begin{equation}\label{Beltramone}
   \star \mathrm{d}\mathbf{Y}^{[1]}\, = \,\mu \,\mathbf{Y}^{[1]}
\end{equation}
which, introduced in 1889 \cite{beltramus} by the famous italian mathematician
Eugenio Beltrami, has a distinguished  and rich history in Mathematics, in
particular in connection with hydrodynamics\cite{Henon,contactgeometria} and
with an important theorem on chaotic streamlines that was proved in the early
seventies by Vladimir Arnold \cite{arnoldus,ArnoldBook}. For almost three
decades a particular very simple solution of eq.(\ref{Beltramone}), derived on
a cubic $3$-torus $\mathrm{T^3}$ and named  the ABC-flow, has been extensively
investigated in the physical-mathematical literature \cite{ABCFLOW} providing a
working ground for both numerical and analytical studies. At the beginning of
this year together with A.Sorin I presented a systematic algorithm \cite{arnolderie} 
for the solution of Beltrami equation on metric torii of the form:
\begin{equation}\label{torellini}
 \mathrm{T^3} \, \simeq \, \frac{\mathbb{R}^3}{\Lambda_{crys}}
\end{equation}
where $\Lambda_{crys}$ denotes a crystallographic lattice. The catch of such
algorithm is the use of the Point Group $\mathrm{\mathbb{G}_{Point}}$ of the
lattice and of its orbits $\mathfrak{S}_k$ in the dual momentum lattice
$\null^\star\Lambda_{crys}$. Summing on periodic functions associated with each
of the momenta in a given orbit one obtains a solution of Beltrami equation
that depends on a number of parameters which is uniquely determined by the
length of the orbit. This solution can be later decomposed into irreducible
representation either of the Point Group or of its Space-Group extensions that
include also appropriate discrete translations. A complete classification of
solutions of (\ref{Beltramone}) was obtained in \cite{arnolderie} for the case
where the crystallographic lattice is the cubic one $\Lambda_{cubic}$ and,
consequentely, the Point Group is the order 24 octahedral group
$\mathrm{O_{24}}$.
\par
In a series of further papers
\cite{arnoldtwobranes,pietroantoniosorin,Fre':2015lds} appeared this year, it
was shown that Arnold-Beltrami Flows, namely one-forms $\mathbf{Y}^{[1]}$
satisfying Beltrami equation (\ref{Beltramone})  can be used to construct
fluxes in the transverse space to $2$-brane solutions of $d=7$ minimal
supergravity \cite{PvNT,bershoffo1,SalamSezgin} whose residual supersymmetries
can also be counted \cite{Fre':2015lds}.
 \par
 As it was briefly sketched in \cite{Fre':2015lds}, all these $2$-branes solutions of
 $d=7$ supergravity can be uplifted to $M2$-branes solutions of d=11 supergravity and
 constitute particular cases of a more general class of $M2$-branes that is the purpose
 of the present paper to describe.
 \par
 The key issue is Englert equation (\ref{englertus}) and
 the question whether we are able to solve it on $7$-dimensional metric torii of the form:
 \begin{equation}\label{rosticceria}
  \mathrm{ T^7}\,\simeq  \, \frac{\mathbb{R}^7}{\Lambda}
 \end{equation}
 where $\Lambda$ is some approrpiate lattice in $d=7$.
 Unfortunately little is known in Mathematics on the classification of finite
 subgroups of higher dimensional rotation groups and on the classification of crystallographic
 lattices in $d > 5$, yet some intuition and educated guesses can take us a long way ahead.
 Knowing for sure that eq.(\ref{caralino}) holds true, it is tempting to guess
 that $\mathrm{L_{168}}$ is crystallographic in $d=7$.  This means that there should be a basis of vectors:
 \begin{equation}\label{basilea}
  {\mathbf{v}}_1  , \,  {\mathbf{v}}_2 \, , \dots ,  \, {\mathbf{v}}_7 \,
 \end{equation}
 in which the $7 \times 7$ matrices $\mathrm{D[\gamma]}$ representing
 the elements of $\mathrm{L_{168}}$ have \textit{integer valued} entries,
 have determinant one and are orthogonal with respect to the flat metric:
 \begin{equation}\label{flattaetta}
   \eta_{ij} \,  \equiv \,{\mathbf{v}}_i \cdot {\mathbf{v}}_j
 \end{equation}
 namely:
 \begin{equation}\label{ortogonallus}
   \forall \gamma \in \mathrm{L_{168}} \quad : \quad \mathrm{D[\gamma]}^T \, \eta \,  \mathrm{D[\gamma]} \, = \, \eta
 \end{equation}
 As we show in this paper such a guess is true. The
 appropriate crystallographic lattice $\Lambda$ is the
 \textit{root lattice} $\Lambda_{root}$ of the $A_7$ simple
 Lie algebra and the invariant metric $\eta$ is the
 corresponding Cartan matrix $\mathcal{C}$. Thanks to this
 we were able to extend the algorithms of \cite{arnolderie}
 to Englert equation (\ref{englertus}) and construct rich
 families of its solutions associated with
 $\mathrm{L_{168}}$ orbits in the $A_7$ weight lattice. On
 the other hand each solution of (\ref{englertus}) gives
 rise to an exact $M2$-brane solution of M-theory.
 The new vast bestiary of branes has now to be investigated for properties and applications.
 \par
 One important point to stress is that the choice of the lattice $\Lambda \subset \mathbb{R}^7$
 is equivalent to the choice of a flat metric  $\eta_{ij}$ as it is mentioned in eq. ({\ref{flattaetta}).
 The space of flat metrics on $\mathrm{T^7}$ is a well known coset:
 \begin{equation}\label{t7modulli}
  \mathcal{ M}_{moduli}\left[\mathrm{T^7}\right] \, = \, \frac{\mathrm{SL(7,\mathbb{R})}}{\mathrm{O(7)}}
 \end{equation}
 hence when we choose a crystallographic group $\mathbb{G}_{point}$ and a lattice $\Lambda$ we just choose a particular point in the moduli space (\ref{t7modulli}) of $\mathrm{T^7}$ and the metric $\eta_{ij}$ should be regarded as an item in the M2-brane solution that we construct.
 \par
 Obviously, just as in $d=3$, also in $d=7$  there is not a single crystallographic lattice and the A7-root lattice with the point group $\mathrm{L_{168}}$ is not the only choice. One can find  solutions of Englert equation also on torii identified by different lattices.  For instance the uplifting to $d=11$ of the solutions with Arnold-Beltrami fluxes that we found in $d=7$  corresponds to the choice of a 7-dimensional lattice of the following form:
 \begin{equation}\label{caronte}
   \Lambda^{[7]} \, = \, \Lambda_{cubic}^{[3]} \otimes \Lambda^{[4]}_{cubic}
 \end{equation}
 which can be easily proved to admit no crustallographic embedding in the A7-root lattice.
 Hence the $2$-branes solutions of $d=7$ minimal supergravity with Arnold-Beltrami fluxes are uplifted to $d=11$ in different points of the $T^7$ moduli space (\ref{t7modulli}).
 \par
 The choice of the A7-lattice and of the crystallographic group $\mathrm{L_{168}}$ has one distinguished advantage. The presence of a $\mathbb{Z}_7$ subgroup and the immersion in $\mathrm{G_{2(-14)}}$ appears to be the basis for the existence of $\mathcal{N}=1$ supersymmetric M2-brane solutions with Englert fluxes.
 \par
 In this paper we discuss the general criterion for the existence of Killing spinors in M2-brane solutions with Englert fluxes. We show that the uplifting of Arnold-Beltrami solutions of $d=7$ supergravity to $d=11$ has always a large residual supersymmetry, namely $\mathcal{N} \,\ge\, 4$ in $d=3$ for a total of $\#$ of \textit{supercharges} $\ge\, 8$.
 \par
 On the other hand using the A7-lattice, the number of Killing spinors is bounded from below by zero: $\mathcal{N} \, \ge \, 0$.
 \par
 In this paper we explicitly construct a solution with $\mathcal{N}=1$ which has a discrete symmetry group of order 21.
 \par
 We do not feel it necessary to insert the traditional illustration of the content of the various sections since it is evident from the table of contents at the beginning of this paper.
\section{M2-branes with Englert fluxes}
\label{genM2Engle} In this section we consider the general form of $M2$-brane
solutions with fluxes and their relation with Englert equation. To this effect
let us consider the effective low energy lagrangian of $M$-theory, namely
$d=11$ supergravity for which we utilize the geometric rheonomic formulation of
\cite{D'Auria:1982nx,FDAgauge}.
\subsection{Summary of d=11 supergravity in the rheonomy framework}
\label{summario} The complete set of curvatures defining the relevant Free
Differential Algebra is given below (\cite{D'Auria:1982nx,FDAgauge}):
\begin{eqnarray}
\mathfrak{T}^{a} & = & \mathcal{D}V^a - {\rm i} \ft 12 \, \overline{\psi} \, \wedge \, \Gamma^a \, \psi \nonumber\\
\mathfrak{R}^{ab} & = & d\omega^{ab} - \omega^{ac} \, \wedge \, \omega^{cb}
\nonumber\\
\rho & = & \mathcal{D}\psi \equiv d \psi - \ft 14 \, \omega^{ab} \, \wedge \, \Gamma_{ab} \, \psi\nonumber\\
\mathbf{F^{[4]}} & = & d\mathbf{A^{[3]}} - \ft 12\, \overline{\psi} \, \wedge
\, \Gamma_{ab} \, \psi \,
\wedge \, V^a \wedge V^b \nonumber\\
\mathbf{F^{[7]}} & = & d\mathbf{A^{[6]}} -15 \, \mathbf{F^{[4]}} \, \wedge \,
\mathbf{A^{[3]}} - \ft {15}{2} \, \, V^{a}\wedge V^{b} \, \wedge \, {\bar \psi}
\wedge \, \Gamma_{ab} \, \psi
\, \wedge \, \mathbf{A^{[3]}} \nonumber\\
\null & \null & - {\rm i}\, \ft {1}{2} \, \overline{\psi} \, \wedge \,
\Gamma_{a_1 \dots a_5} \, \psi \, \wedge \, V^{a_1} \wedge \dots \wedge V^{a_5}
\label{FDAcompleta}
\end{eqnarray}
From their very definition, by taking a further exterior derivative one obtains
the Bianchi identities:
\begin{eqnarray}
  \mathcal{D} \mathfrak{R}^{ab} &=& 0 \label{LorentzBianchiM}\\
  \mathcal{D} \mathfrak{T}^{a} \, +\, \mathfrak{R}^{ab}\, \wedge\, V_{b} \, - \, {\rm i}
  \, \bar{\psi} \,\wedge \, \Gamma^a \, \rho  &=& 0 \label{TorsionBianchiM}\\
  \mathcal{D} \rho \, + \, \ft 14 \, \Gamma^{ab} \, \psi \, \wedge\,
  \mathfrak{R}^{ab}  &=& 0 \label{gravitinoBianchiM}\\
  d\mathbf{F^{[4]}} \, - \, \bar{\psi }\, \wedge \, \Gamma_{ab} \,\rho \, \wedge V^a \, \wedge\, V^b \, +
 \, \bar{\psi}\, \Gamma_{ab} \, \psi \, \wedge \, \mathfrak{T}^a \, \wedge \, V^b &=& 0
 \label{F4BianchiM}\\
 d\mathbf{F^{[7]}} \, - \,{\rm i} \bar{\psi }\, \wedge \, \Gamma_{a_1\cdots a_5} \,\rho \,
 \wedge \, V^{a_1} \, \wedge\,\ldots\wedge V^{a_5} \, \nonumber\\
 \, - \, \ft 53 {\rm i} \, \bar{\psi} \,\wedge \, \Gamma_{a_1\cdots a_5} \,\psi \, \wedge \,
 \mathfrak{T}^{a_1} \,\wedge \, V^{a_2} \, \wedge\,\ldots\wedge V^{a_5} \nonumber\\
 \, - \, 15 \, \, \bar{\psi }\, \wedge \, \Gamma_{ab} \,\rho \, \wedge V^a \, \wedge\, V^b \,
 \wedge \, \mathbf{F^{[4]}}\, - \, 15 \,
  \mathbf{F^{[4]}}\, \wedge \, \mathbf{F^{[4]}}\,&=& 0 \label{F7BianchiM}
\end{eqnarray}
There is a unique rheonomic parametrization of the curvatures
(\ref{FDAcompleta}) which solves the Bianchi identities and it  is the
following one:
\begin{eqnarray}
\mathfrak{T}^a & = & 0 \nonumber\\
\mathbf{F^{[4]}} & = & F_{a_1\dots a_4} \, V^{a_1} \, \wedge \dots \wedge \, V^{a_4} \nonumber\\
\mathbf{F^{[7]}} & = & \ft {1}{84} F^{a_1\dots a_4} \, V^{b_1} \, \wedge \dots
\wedge \,
V^{b_7} \, \epsilon_{a_1 \dots a_4 b_1 \dots b_7} \nonumber\\
\rho & = & \rho_{a_1a_2} \,V^{a_1} \, \wedge \, V^{a_2} - {\rm i} \ft 13 \,
\left(\Gamma^{a_1a_2 a_3} \psi \, \wedge \, V^{a_4} + \ft 1 8 \Gamma^{a_1\dots
a_4 m}\, \psi \, \wedge \, V^m
\right) \, F^{a_1 \dots a_4} \nonumber\\
\mathfrak{R}^{ab} & = & R^{ab}_{\phantom{ab}cd} \, V^c \, \wedge \, V^d + {\rm
i} \, \rho_{mn} \, \left( \ft 12 \Gamma^{abmn} - \ft 2 9 \Gamma^{mn[a}\,
\delta^{b]c} + 2 \,
\Gamma^{ab[m} \, \delta^{n]c}\right) \, \psi \wedge V^c\nonumber\\
 & &+\overline{\psi} \wedge \, \Gamma^{mn} \, \psi \, F^{mnab} + \ft 1{24} \overline{\psi} \wedge \,
 \Gamma^{abc_1 \dots c_4} \, \psi \, F^{c_1 \dots c_4}
\label{rheoFDA}
\end{eqnarray}
The expressions (\ref{rheoFDA}) satisfy the Bianchi.s provided the space--time
components of the curvatures satisfy the following constraints
\begin{eqnarray}
0 & = & \mathcal{D}_m F^{mc_1 c_2 c_3} \, + \, \ft 1{96} \, \epsilon^{c_1c_2c_3
a_1 a_8} \, F_{a_1 \dots a_4}
\, F_{a_5 \dots a_8}  \label{maxwell}\\
0 & = & \Gamma^{abc} \, \rho_{bc} \label{gravitino}\\
R^{am}_{\phantom{bm}cm} & = & 6 \, F^{ac_1c_2c_3} \,F^{bc_1c_2c_3} - \, \ft 12
\, \delta^a_b \, F^{c_1 \dots c_4} \,F^{c_1 \dots c_4}\label{fieldeque}
\end{eqnarray}
which are the  space--time field equations.
\subsection{M2-brane solutions with $\mathbb{R}_+\times \mathrm{T^7}$ in the transverse dimensions }
\label{EngleM2T7} Among the many possible solutions of the field equations
(\ref{maxwell}-\ref{fieldeque}) we are interested in those that describe
$M2$-branes of the following sort.
\par
Inspired by our previous results in $d=7$
\cite{pietroantoniosorin,Fre':2015lds}, we give the $11$-dimensional manifold
the following topology:
\begin{equation}\label{rabolatus}
    \mathcal{M}_{11} \, = \, \mathrm{Mink}_{1,2} \times \mathbb{R}_+\times \mathrm{T^7}
\end{equation}
where $\mathrm{Mink}_{1,2}$ is Minkowski space in $1+2$ dimensions and
represents the world-volume of the $M2$-brane, while $T^7$ is a flat compact
seven-torus. $\mathbb{R}_+\times \mathrm{T^7}$ is the eight-dimensional space
transverse to the brane.
\par
Then, according to the general rules of brane-chemistry (see for instance
\cite{pietrobook}, page 288 and following ones), we introduce the following
$d=11$ metric:
\begin{equation}\label{colabrodo}
    ds_{11}^2 \, = \, H(y)^{-\ft{4 \tilde{d}}{9\Delta}}\left(d\xi^\mu \otimes d\xi^\nu\,
    \eta_{\mu\nu}\right)\, - \,H(y)^{\ft{4 {d}}{9\Delta}}\left(dy^I \otimes dy^J\,
    \delta_{IJ}\right)
\end{equation}
where:
\begin{equation}\label{worldindex}
    \xi^\mu \quad ; \quad \mu \, = \, \underline{0},\underline{1},\underline{2}
\end{equation}
are the coordinates on $\mathrm{Mink}_{1,2}$, while:
\begin{equation}\label{trasversale}
    y^I \quad ; \quad I \, = \, 0,1,2,\,\dots ,7
\end{equation}
are the coordinates of the 8-dimensional transverse space. Since in $d=11$
there is no dilaton  we have
\begin{equation}\label{bamboccione}
    \Delta \, = \, 2 \frac{\tilde{d} \, d}{9} \, = \, 2 \, \frac{6 \times 3}{9} \, = \, 4 \quad ; \quad d=3 \, ;
    \quad \tilde{d} =6
\end{equation}
and the appropriate $M2$ ansatz for the metric becomes:
\begin{equation}\label{colabrodone}
    ds_{11}^2 \, = \, H(y)^{-\ft{2}{3}}\left(d\xi^\mu \otimes d\xi^\nu\,
    \eta_{\mu\nu}\right)\, - \,H(y)^{\ft{1}{3}}\left(dy^I \otimes dy^J\,
    \delta_{IJ}\right)
\end{equation}
Because of the chosen topology of the transverse space it is convenient to set:
\begin{equation}\label{faccialei}
    y^0 \, = \, U \in \mathbb{R}_+ \quad ; \quad y^i \, = \, x^i \in \mathrm{T^7} \quad (i=1,\dots,7)
\end{equation}
The next point is to choose an appropriate ansatz for the three-form
$\mathbf{A^{[3]}}$. We set:
\begin{equation}\label{coffilasio}
    \mathbf{A^{[3]}}\, = \, \frac{2}{H(y)} \, \Omega^{[3]} \, + \, e^{-\mu \,U} \mathbf{Y}^{[3]}
\end{equation}
where:
\begin{eqnarray}
% \nonumber to remove numbering (before each equation)
  \Omega^{[3]} &=& \ft{1}{6} \epsilon_{\mu\nu\rho} \, d\xi^\mu \wedge d\xi^\nu \wedge d\xi^\rho \label{wvol}\\
 \mathbf{Y}^{[3]} &=& Y_{ijk}(x) dx^i \wedge dx^j \wedge dx^k
\end{eqnarray}
The essential point in the above formula is that the antisymmetric tri-tensor
$Y_{ijk}(x)$ depends only on the coordinates $x$ of the seven-torus
$\mathrm{T^7}$.
\subsubsection{Analysis of Maxwell equation}
\label{maxello} With the above data we are in a position to work out the
explicit form of the $4$-form field strength, its intrinsic anholonomic
components $F_{abcd}$ and insert them into the field equation (\ref{maxwell}).
An essential ingredient in this calculation is provided by the transcription of
the metric (\ref{colabrodone}) and by the calculation of the spin connection.
\par
We set:
\begin{eqnarray}
  V^{\underline{a}} &=& H(y)^{-\ft 13}\, d\xi^{\underline{a}} \\
V^{I}&=& H(y)^{\ft 16}\, dy^{I}
\end{eqnarray}
and we obtain:
\begin{eqnarray}
% \nonumber to remove numbering (before each equation)
  \omega^{\underline{ab}} &=& 0 \\
  \omega^{\underline{a} I} &=& - \frac{1}{3}\, H^{-\ft 7 6}\, V^{\underline{a}}\, \partial^I H(y) \\
  \omega^{IJ} &=& \frac{1}{6} \,H^{-\ft 7 6}\, \left ( \partial^IH(y)\, V^J \, - \, \partial^JH(y) V^I \right) \\
\end{eqnarray}
With the ansatz (\ref{coffilasio}) the non vanishing components of the $4$-form
$\mathbf{F}^{[4]}$ are the following ones:
\begin{eqnarray}
% \nonumber to remove numbering (before each equation)
  F_{\underline{abc}I} &=& \frac{1}{12 }\,H(y)^{-\ft 76}\,\partial_I H(y) \\
 F_{0ijk} &=& - \frac{\mu}{4} e^{-\mu \, U}\, H(y)^{-\ft 23}\, Y_{ijk}  \\
F_{ijk\ell} &=& H(y)^{-\ft 23} \, e^{-\mu \,U} \partial_i Y_{jk\ell}
\end{eqnarray}
Then we can easily verify that the Maxwell field equation (\ref{maxwell}) is
satisfied provided the following two differential constraints hold true:
\begin{eqnarray}\label{harmoniusca}
    \Box_{\mathbb{R}_+\times \mathrm{T^7}} H(y)& = & \frac{\mu}{4}\, e^{-2\,\mu \,
    U}\epsilon^{ijk\ell mnr} \,\partial_iY_{jk\ell} \,Y_{mnr}\label{2harmoniusca2}\\
\frac{1}{4!}\epsilon^{pqrijk\ell}\,\partial_iY_{jk\ell} &=& -\,\frac{\mu}{4} \,
Y_{pqr}\label{englerta}
\end{eqnarray}
The two equations admit the following index-free rewriting:
\begin{eqnarray}\label{FharmoniuscaF}
    \Box_{\mathbb{R}_+\times \mathrm{T^7}} H(y)& = & - \frac{3\,\mu^2}{2}\,
    e^{-2\,\mu \,U}\, \parallel\mathbf{Y}\parallel^2 \, \equiv \, J(y)\label{3harmoniusca3}\\
\star_{\,{\scriptsize\mathrm{T^7}}} \,\mathrm{d} \mathbf{Y}^{[3]} &=&
-\,\frac{\mu}{4} \, \mathbf{Y}^{[3]}\label{englertaF}
\end{eqnarray}
As we see eq.(\ref{englertaF}) is the generalization to a $7$-dimensional torus
of Beltrami equation on the three-dimensional one: it is just Englert equation
(\ref{englertus}) discussed in the introduction.
\section{The Killing spinor equation of M2-branes with Englert fluxes}
\label{kilspisection} In order to analyze the structure of the Killing spinor
equation in the background of the M2-branes with Englert fluxes, introduced in
the previous section, we need a basis of gamma matrices that is well-adapted to
the splitting of the 11-dimensional manifold, namely:
\begin{equation}\label{sparato}
    \mathcal{M}_{11} \, = \, \underbrace{\mathrm{Mink}_{1,2}}_{d=3\mbox{ brane world volume}}
    \times\underbrace{\left( \mathbb{R}_+\otimes
    \underbrace{\mathrm{T}^7}_{d=7}\right)}_{d=8 \mbox{ transverse space}}
\end{equation}
Such a well adapted basis is provided by the following nested hierarchy.
\subsection{Gamma matrices}
At the bottom of the hierarchy we have the Pauli matrices.
\paragraph{Pauli matrices.}
We use the following conventions:
\begin{equation}\label{paulisti}
\sigma_1 \,=\,\left(
\begin{array}{cc}
 0 & 1 \\
 1 & 0 \\
\end{array}
\right) \quad ; \quad \sigma_2 \,= \, \left(
\begin{array}{cc}
 0 & -i \\
 i & 0 \\
\end{array}
\right) \quad ; \quad \sigma_3 \, = \,\left(
\begin{array}{cc}
 1 & 0 \\
 0 & -1 \\
\end{array}
\right);
\end{equation}
\paragraph{Gamma matrices on the d=3 world-volume.} Next we construct the
 set of 2$\times $2 gamma matrices in d=3  in the following way
\begin{equation}\label{gammucci}
\left\{\gamma _{\underline{a}},\gamma _{\underline{b}} \right\} =2\eta
_{\underline{\text{ab}}} \quad ;\quad \gamma \, = \,\left\{\sigma
_2,\mathrm{i}\,\sigma _1,\mathrm{i}\,\sigma _3\right\}
\end{equation}
\paragraph{Gamma matrices in d=7} In d=7 we choose gamma matrices that are real and
antisymmetric and fulfill the following Clifford algebra:
\begin{equation}\label{taumatti}
\left\{\tau _i,\tau _j\right\} = -2\delta _{{ij}}
\end{equation}
The explicit basis utilized is that one where we express the $\tau $-matrices
in terms of $\phi_{{ijk}}$, namely of the $\mathrm{G_2}$-invariant
three-tensor:
\begin{eqnarray}
\left(\tau _i\right)_{{jk}} &=& \phi_{{ijk}} \nonumber\\
\left(\tau _i\right)_{{j8}} &=& \delta _{{ij}}\quad ; \quad \left(\tau
_i\right)_{8j} = - \delta _{{ij}} \label{taubasa}
\end{eqnarray}
The explicit form of the $\phi _{{ijk}}$ tensor will be given in
eq.(\ref{gorlandus})  and it is the one well-adapted to the immersion of the
discrete group which acts crystallographically on $\mathrm{T^7}$ into the
compact $\mathrm{G_2}$ Lie group, namely according to the canonical immersion
$\mathrm{L_{168}} \longrightarrow \mathrm{G_{2(-14)}}$.
\paragraph{Gamma matrices in d=8}
Because of our splitting $11=3\oplus 1\oplus7 $ we need also the gamma matrices
in d=8 corresponding to the transverse space to the M2-brane, namely
$\mathbb{R}_+\otimes \mathrm{T^7}$. We choose the following Clifford algebra:
\begin{equation}\label{d8gammi}
 \left\{T_I,T_J\right\} = -2 \delta _{{IJ}}
\end{equation}
and we utilize the following explicit realization:
 \begin{eqnarray}
  T_0 & = & \mathrm{i} \sigma _2\otimes  \pmb{1}_{8\times 8}  \nonumber \\
   T_i & = & \sigma _1\otimes  \tau_i  \nonumber\\
   T_9 &= & \sigma _3\otimes \pmb{1}_{8\times 8} \label{tmatre}
 \end{eqnarray}
The last matrix is the d=8 chirality operator which plays an important role in
the discussion of the Killing spinor equation.
\paragraph{Gamma matrices in d=11} At the top of the hierarchy we have the d=11 gamma
matrices, obeying the following Clifford algebra
\begin{equation}\label{d11gammi}
   \left\{\Gamma _a,\Gamma _b\right\}=2\eta _{\text{ab}}
\end{equation}
For them we utilize the following explicit realization:
\begin{eqnarray}
% \nonumber to remove numbering (before each equation)
 \Gamma _{\underline{a}}& = &\gamma_{\underline{a}}\otimes T_9\nonumber\\
\Gamma _I & = &\pmb{1}_{2\times 2}\otimes  T_I
\end{eqnarray}
With these choices the charge conjugation matrix, takes the following form:
\begin{eqnarray}
\mathfrak{C}&=& \mathrm{i} \sigma _2\otimes \pmb{1}_{2\times 2} \otimes \pmb{1}_{8\times 8} \nonumber\\
\mathfrak{C}\,\Gamma _a \mathfrak{C}^{ -1}& = &- \Gamma _a{}^T
\end{eqnarray}
Equipped with this set of properly chosen gamma matrices we can turn to the
investigation of the Killing spinor equation.
\subsection{The tensor structure of the Killing spinor equation}
The rheonomic solution of the d=11 Bianchi identities (see eq.(\ref{rheoFDA}))
allows us to write the Killing spinor equation in the following  general form:
\begin{equation}\label{spikillusequo}
\mathcal{D} \xi -\frac{\mathrm{i}}{3}\Gamma ^{\text{abc}}V^dF_{\text{abcd} }\xi
-\frac{\mathrm{i}}{24}\Gamma _{\text{abcdf}} F^{\text{abcd}}V^f\xi  = 0
\end{equation}
where
\begin{equation}\label{lorenzoderivo}
    \mathcal{D} \xi  \equiv \text{d$\xi $} - \frac{1}{4}\omega ^{\text{ab}}\Gamma _{\text{ab}} \xi
\end{equation}
is the Lorentz covariant differential in $d=11$.
\par
 Equation (\ref{spikillusequo}) can be usefully rewritten as follows:
 \begin{equation}\label{genCovar}
 \nabla \xi
\equiv \text{d$\xi $}+\Omega
 \xi  =0
 \end{equation}
where $\Omega $ is a generalized connection in the 32--dimensional spinor space,
defined as follows:
\begin{equation}\label{genOmme}
    \Omega  \equiv \Theta _L+\Theta _1^{[F]}+\Theta _2^{[F]}
\end{equation}
In the above equation we have  introduced the following definitions:
\begin{eqnarray}
 \Theta _L &\equiv& - \frac{1}{4}\omega ^{\text{ab}}\Gamma_{\text{ab}}\nonumber\\
 \Theta _1^{[F]} &\equiv& -\frac{\mathrm{i}}{3}\Gamma ^{\text{abc}}V^dF_{\text{abcd} }\nonumber \\
 \Theta _2^{[F]} &\equiv& -\frac{\mathrm{i}}{24}\Gamma _{\text{abcdf}} F^{\text{abcd}}V^f
\end{eqnarray}
Next let us make another splitting of the overall generalized connection:
\begin{equation}\label{omegasplittus}
   \Omega  = \Omega _H+ \Omega _Y
\end{equation}
where $\Omega _H$ depends only on the (inhomogeneous)-harmonic function H and
it is obtained from $\Omega $ by setting $Y_{{ijk}} \,\rightarrow \, 0$. Instead, the
other part $\Omega _Y$,  is just the difference and it depends linearly
on $Y_{ijk}$
%%%%%%%%%%%%%%%
\subsubsection{M2-branes without Englert fluxes: tensor structure of $\Omega_H$}
%%%%%%%%%%
Introducing the following operators:
\begin{eqnarray}
V\circ \gamma
 &= & V^{\underline{a}}\gamma _{\underline{a}}\label{oppo1}\\
 \mathbb{P}_{\pm } &=& \frac{1}{2}\left(\pmb{1}_{16} \pm  T_9\right) \label{oppo2}\\
 \partial H\circ T &=&\frac{1}{3}H ^{-\frac{7}{6}}\partial _IH T^I \label{oppo3}\\
 V\diamond \partial H\circ T & = &- \frac{1}{12} H ^{-\frac{7}{6}}
 V_{[I}\partial _{J]}H T^{\text{IJ}} \label{oppo4}\\
\text{$\mathbf{d}$H} & = &\frac{1}{6}H ^{-\frac{7}{6}} \sum _{I=1}^8 \partial
_IH V^I\label{oppo5}
\end{eqnarray}
we get that the H-part of the generalized connection has the following tensor
structure:
\begin{equation}\label{omegaH}
\Omega _H = V\circ \gamma \otimes  \partial H\circ T \,\mathbb{P}_- +
\pmb{\pmb{1}_2}\pmb{ }\otimes \text{ }V\diamond \partial H\circ T \,
\mathbb{P}_- +  \pmb{1}_2 \otimes\text{$\mathbf{d}$H}\,T_9
\end{equation}
From equation (\ref{omegaH}) one readily derives the form of the Killing
spinors for pure M2-brane solutions. Writing the 32 component Killing spinor as
a tensor product:
\begin{equation}\label{stensorokillo}
   \xi  = \epsilon \otimes \chi
\end{equation}
we find that, in the absence of Y-fields, the Killing spinor equation is
satisfied provided:
\begin{eqnarray}
 T_9\chi & = & \chi \quad \Rightarrow \quad \mathbb{P}_{-}\chi \, = \, 0\label{chiraltus}\\
 \chi  &=& H^{ -\frac{1}{6}} \, \chi_0
\end{eqnarray}
where $H$ is the (inhomogeneous)-harmonic function appearing in the metric
(\ref{colabrodo}) and $\chi _0$ is a constant spinor with commuting components.
Indeed, in view of our 2-brane interpretation of these backgrounds, we assume
that the two--component spinors $\epsilon $ are the anticommuting objects.
\par
Using the tensor structure of the d=8 T-matrices we set:
\begin{equation}\label{splittochi}
    \chi \, = \, \kappa \otimes \lambda
\end{equation}
where $\kappa $ is a two component spinor:
\begin{equation}\label{cappus}
\kappa =\left(
\begin{array}{c}
 \kappa _1 \\
 \kappa _2 \\
\end{array}
\right)
\end{equation}
with commuting components, while $\lambda $ is an eight--component spinor:
\begin{equation}\label{lambdus}
\lambda \, = \,\left(
\begin{array}{c}
\begin{array}{c}
\begin{array}{c}
 \lambda _1 \\
 \lambda _2 \\
\end{array}
 \\
 \lambda _3 \\
\end{array}
 \\
\begin{array}{c}
 \lambda _4 \\
\begin{array}{c}
 \lambda _5 \\
\begin{array}{c}
 \lambda _6 \\
\begin{array}{c}
 \lambda _7 \\
 \lambda _8 \\
\end{array}
 \\
\end{array}
 \\
\end{array}
 \\
\end{array}
 \\
\end{array}
\right)
\end{equation}
also with commuting components.
\par
In this language the most general 32--component spinor has the form:
\begin{equation}\label{tritensori}
    \xi  = \epsilon \otimes \kappa \otimes\lambda
\end{equation}
and  the general solution for the Killing spinor at $Y_{{ijk}}\,= \, 0$ is
obtained by setting:
\begin{equation}\label{crisaltus}
    \kappa _2 = 0 \quad ; \quad \kappa _1 = H^{ -\frac{1}{6}}
\end{equation}
This shows that the M2-branes without Englert-fluxes preserve 16
supersymmetries, namely $\frac{1}{2}$ of the total SUSY.
\subsubsection{M2-branes with Englert fluxes: tensor structure of $\Omega _Y$}
We come next to analyze the structure of the Y-part of the connection $\Omega
_Y$.
\par
We begin by introducing two  $d=7$ operators constructed with the Englert field
$Y_{{ijk}}$, the flat 8-dimensional  vielbein $\hat{V}^I\equiv dy^I$ and the
$\tau $-matrices:
\begin{equation}\label{operati}
    \mathcal{B} \equiv \tau_{{ijk}} \, {Y}_{{ijk}}\quad ; \quad\mathcal{T} = \hat{V}^i\tau _i
\end{equation}
In terms of these operators we get:
\begin{eqnarray}
   \Omega _Y &=& \frac{ \mu }{4}e^{-U \mu }H^{-2/3} \left( \frac{1}{3} \mathrm{i} V\circ \gamma \otimes
   \mathbb{P}_+\otimes
\mathcal{B}\right.\nonumber\\
&&\left.+\frac{1}{6} \mathrm{i} \pmb{1}\otimes \left(\sigma _1+\frac{\mathrm{i}
}{2}\sigma _2\right)\otimes [\mathcal{B},\mathcal{T}] +\frac{1}{6} \mathrm{i}
\pmb{1}\otimes \left(\frac{1}{2}\sigma _1+\mathrm{i} \sigma
_2\right)\otimes \{\mathcal{B},\mathcal{T}\}\right.\nonumber\\
 &&\left.+\frac{1}{3} \mathrm{i} \pmb{1}\otimes
\mathbb{P}_+ \otimes \mathcal{B} \hat{V}^0\right)\label{sodoma}
\end{eqnarray}
where $[\mathcal{B},\mathcal{T}]$ and $\{\mathcal{B},\mathcal{T}\}$
respectively denote the commutator and the anti-commutator of the two mentioned
operators. Observing the structure of the connection $\Omega _Y$ displayed in
eq.(\ref{sodoma}) we can rewrite it in the following more explicit form:
\begin{eqnarray}
   \Omega _Y &=&
\mathrm{i}\frac{1}{12}\mu e^{-U \mu }H^{-2/3}\times \nonumber\\
&& \left[V\circ \gamma
 \left(
\begin{array}{cc}
2 \mathcal{B} & 0
\\ 0 & 0 \\
\end{array}
\right) +\pmb{1}\otimes \left(
\begin{array}{cc}
\hat{V}^0\mathcal{B} &
0 \\ 0 & 0 \\
\end{array}
\right) +\frac{1}{2}\pmb{1}\otimes \left(
\begin{array}{cc}
0 & 3 \mathcal{B}
\mathcal{T} \\
-\mathcal{T} \mathcal{B} & 0
\\ \end{array}
\right) \right]\nonumber\\
\label{gomorra}
\end{eqnarray}
Eq.(\ref{gomorra}) reveals the mechanism behind the preservation of
supersymmetry by M2-branes with Englert fluxes. Writing the candidate Killing
spinor in the tensor product form (\ref{tritensori}) we see that the connection
$\Omega _Y$ annihilates it if $\kappa = \left(\begin{array}{c}
H^{-\ft 16} \\
0
\end{array}\right)$, as we already established from consideration of the H-part of the connection, and
if the 8-component $\lambda $ is a null-vector of $\mathcal{B}$:
\begin{equation}
   \mathcal{B} \, \lambda = 0\label{Binullo}
\end{equation}
This is the only possibility to integrate the Killing spinor equation. Indeed
the term with V$\circ \gamma $ which mixes the internal coordinateswith the world volume ones
has to vanish since it cannot be compensated in any other way. This implies
eq.(\ref{Binullo}). The magic thing is that because of the precise values of
the coefficients provided by the rheonomic solution of Bianchi identities in
d=11, the combinations of commutators and anticommutators appearing in
eq.(\ref{sodoma}) just produce  the structure in eq.(\ref{gomorra}). In this
way the condition (\ref{Binullo}) suffices to annihilate also the action of the
other terms in the connection.
\par
In conclusion M2-branes with Englert fluxes preserve part of the Killing
spinors existing in the case of $Y=0$ if and only if the operator $\mathcal{B}$
has a non trivial Null-Space, namely if the Rank of $\mathcal{B}$ is $<$ 8.
Every $\lambda $ satisfying (\ref{Binullo}) corresponds to a preserved
supersymmetry.
\section{Uplifting of $7$-dimensional Arnold-Beltrami flux $2$-branes to Englert  M2-branes}
\label{upliffo} In order to uplift the solutions of $d=7$ minimal supergravity
to the Englert M2-branes described in section. \ref{EngleM2T7} we split the
$7$-torus as it follows:
\begin{equation}\label{splittato}
     \mathrm{T^7}\,=\, \mathrm{T^3}\otimes \mathrm{T^4}
\end{equation}
and we name $\mathbf{x}_\shortparallel \, = \, x^{\mathfrak{i}}$,
($\mathfrak{i},\mathfrak{j},\mathfrak{k} \, = \, 1,2,3$) the coordinates of the
$3$-tours and $\mathbf{x}_\perp \, = \,x^\alpha$, ($\alpha,\beta,\gamma \, = \,
4,5,6,7$) the coordinates of the $4$-torus. The next step consists of setting:
\begin{eqnarray}\label{subartico}
    \mathbf{Y}^{[3]}\left[\mathbf{W}\right]& = & \mathbf{W}^{\Lambda}\wedge \mathbb{K}^{\Lambda}\nonumber\\
   \mathbb{K}^\Lambda &=& J^{\Lambda|-}_{\alpha\beta} dx^\alpha\wedge
    dx^\beta
\end{eqnarray}
where $\mathbf{W}^{\Lambda}$ denotes a triplet of Beltrami one-forms satisfying
Beltrami equation on the $3$-torus:
\begin{equation}\label{casinus}
    \star_{\,{\scriptsize\mathrm{T^3}}} \,\mathrm{d} \mathbf{W}^\Lambda \,=\, \mu \,\mathbf{W}^\Lambda
\end{equation}
while $J^{\Lambda|-}_{\alpha\beta}$ is the triplet of anti-self-dual 't Hooft
matrices, leading to:
\begin{equation}\label{Kapponi}
    \mathbb{K}^\Lambda \, = \,\left(
\begin{array}{c}
 2 \mathrm{d} {x^6}\wedge \mathrm{d} {x^7}-2 \mathrm{d} {x^4}\wedge \mathrm{d} {x^5} \\
 2 \mathrm{d} {x^4}\wedge \mathrm{d} {x^6}+2 \mathrm{d} {x^5}\wedge \mathrm{d} {x^7} \\
 2 \mathrm{d} {x^5}\wedge \mathrm{d} {x^6}-2 \mathrm{d} {x^4}\wedge \mathrm{d} {x^7} \\
\end{array}
\right)
\end{equation}
\par
Any triplet of Beltrami vector fields $\mathbf{W}^{\Lambda}$ satisfying
eq.(\ref{casinus}) produces an Englert 3-form
$\mathbf{Y}^{[3]}\left[\mathbf{W}\right]$ satisfying Englert equation
(\ref{englerta}) with the same $\mu$. Equation (\ref{FharmoniuscaF}) reduces to:
\begin{equation}
    \Box_{\mathbb{R}_+\times \mathrm{T^3}} H(U,\mathbf{x}_\shortparallel)\, = \, - 6 \,\mu^2\,
    e^{-2\,\mu \,U}\, \sum_{\Lambda=1}^3 \,\parallel \mathbf{W}^\Lambda \parallel^2 \label{4harmoniusca4}
\end{equation}
and the relations of the $d=7$ dilaton and metric with the $d=11$ metric are
the following ones:
\begin{eqnarray}
% \nonumber to remove numbering (before each equation)
  \phi &=& - \, \frac{2}{5} \, \log\left[ H(U,\mathbf{x}_\shortparallel)\right] \nonumber\\
  ds_{7}^2 &=& H(U,\mathbf{x}_\shortparallel)^{-\ft 2 5}\left(d\xi^\mu \otimes d\xi^\nu\,
    \eta_{\mu\nu}\right)\, - \,H(U,\mathbf{x}_\shortparallel)^{\ft 3 5}
    \left(dU^2 + d\mathbf{x}_\shortparallel^2\right) \nonumber  \\
 ds_{11}^2&=& \exp\left[ \ft 23 \, \phi\right] \,ds^{2}_{7} \, - \, \exp\left[- \ft 56 \, \phi\right]
  d\mathbf{x}_\perp^2 \nonumber\\
  \null &\Downarrow &\null \nonumber\\
 ds_{11}^2 & = & H(U,\mathbf{x}_\shortparallel)^{-\ft{2}{3}}\left(d\xi^\mu \otimes d\xi^\nu\,
    \eta_{\mu\nu}\right)\, - \,H(U,\mathbf{x}_\shortparallel)^{\ft{1}{3}}\left(dU^2 + d\mathbf{x}_\shortparallel^2
    +d\mathbf{x}_\perp^2\right)\,\label{baldogiovane}
\end{eqnarray}
\subsection{Uplifting of Killing spinors} We wonder whether the supersymmetry possibly preserved by the
Arnold-Beltrami flux $2$-brane in d=7 is preserved by its uplifting to an
M2-brane with Englert fluxes. The answer is clearly yes and actually the d=11
preserved SUSY is larger. In addition to the Killing $q$-charges preserved in
d=7 one has 8 further ones corresponding to  $\ft 12$ of the supersymmetries
that are truncated away in the consistent truncation to minimal d=7
supergravity. We illustrate this mechanism with the explicit consideration of
an example.
\subsubsection{Uplift of the Arnold Beltrami flux $2$-brane with $\mathcal{N}=2$ SUSY}
In \cite{Fre':2015lds} it was shown that the following triplet of Beltrami
vector fields:
\begin{equation}\label{beltramiN2}
  \mathbf{W}^\Lambda \, = \,  \left(
\begin{array}{c}
 \mathrm{d} {x^1}  \cos  2 \pi  {x^3} -\mathrm{d} {x^2}  \sin  2 \pi  {x^3}  \\
 \mathrm{d} {x^2}   -\cos  2 \pi  {x^3}  -\mathrm{d} {x^1}  \sin  2 \pi  {x^3}  \\
 0 \\
\end{array}
\right)
\end{equation}
leads to a d=7 supergravity solution preserving  $\mathcal{N}=2$ supersymmetry.
Inserted into eq.(\ref{subartico}), the vector field (\ref{beltramiN2})
produces an Englert field $Y_{ijk}[\mathbf{W}]$ which satisfies Englert
eq.(\ref{englertus}) with $\mu \, = \, 2\pi$ and admits, as solution of the
inhomogeneous harmonic equation (\ref{englerta}) the following function:
\begin{equation}\label{armoniosa}
   H(y)\, = \, 1-e^{-4 \pi  U}
\end{equation}
Inserted into eq.(\ref{operati}) the Englert field $Y_{ijk}[\mathbf{W}]$
produces the following $\mathcal{B}$-operator:
\begin{equation}\label{BdiW}
    \mathcal{B}\left[\mathbf{W}\right]\, = \, \left(
\begin{array}{cccccccc}
 0 & 0 & 0 & 0 & 0 & 0 & 0 & 0 \\
 0 & 0 & 0 & 0 & 0 & 0 & 0 & 0 \\
 0 & 0 & -4 \sin [2 \pi   {x^3}] & 0 & -4 \cos [2 \pi   {x^3}] & -4 \sin [2 \pi
    {x^3}] & 0 & -4 \cos [2 \pi   {x^3}] \\
 0 & 0 & 0 & 0 & 0 & 0 & 0 & 0 \\
 0 & 0 & -4 \cos [2 \pi   {x^3}] & 0 & 4 \sin [2 \pi   {x^3}] & -4 \cos [2 \pi
    {x^3}] & 0 & 4 \sin [2 \pi   {x^3}] \\
 0 & 0 & -4 \sin [2 \pi   {x^3}] & 0 & -4 \cos [2 \pi   {x^3}] & -4 \sin [2 \pi
    {x^3}] & 0 & -4 \cos [2 \pi   {x^3}] \\
 0 & 0 & 0 & 0 & 0 & 0 & 0 & 0 \\
 0 & 0 & -4 \cos [2 \pi   {x^3}] & 0 & 4 \sin [2 \pi   {x^3}] & -4 \cos [2 \pi
    {x^3}] & 0 & 4 \sin [2 \pi   {x^3}] \\
\end{array}
\right)
\end{equation}
The rank of the above matrix is $2$ and the most general vector in the
6-dimensional null-space has the following form:
\begin{equation}\label{zetaspinor}
    \zeta \, = \, \left(
\begin{array}{c}
 \zeta _6 \\
 \zeta _5 \\
 -\zeta _3 \\
 \zeta _4 \\
 -\zeta _1 \\
 \zeta _3 \\
 \zeta _2 \\
 \zeta _1 \\
\end{array}
\right) \quad ; \quad \mathcal{B}\left[\mathbf{W}\right]\, \zeta \, = \, 0
\end{equation}
This leads to the conclusion that:
\begin{equation}\label{surrocanto}
    \xi \, = \, \epsilon \otimes \left(\begin{array}{c}
                                   H^{-\ft 16} \\
                                   0
                                 \end{array}\right)\otimes \zeta
\end{equation}
is a Killing spinor for the constructed M2-brane solution, namely  we have a
total of 12 preserved supercharges. From a d=3 point of view it means that the
brane-volume theory has $\mathcal{N}=6$ supersymmetry. How to understand this
result we already outlined above.  Two out of the six are the supersymmetries
that are preserved in the theory obtained form d=7 supergravity. The other four
are the supersymmetries corresponding to the gravitinos that are truncated away
in d=7, when we reduce from maximal to minimal supergravity. From the point of
view of the d=3 world volume theory, the multiplets arising from d=7 minimal
supergravity are smaller since they are associated with the 4 transverse
coordinates of $\mathbb{R}_+\times \mathrm{T^3}$ and with their fermionic
superpartners. These multiplets   support $\mathcal{N}=2$ supersymmetry. The
multiplets arising from  the complete d=11 theory are bigger since they are
associated with the 8-transverse coordinates of $\mathbb{R}_+\times
\mathrm{T^3}\otimes \mathrm{T^4}$ and they support $\mathcal{N}=6$
supersymmetry.
\par
In conclusion we can establish the following general rule for the number of supersymmetries preserved by M2--branes with Englert fluxes that are uplift of the Arnold Beltrami $2$-branes in $d=7$:
\begin{equation}\label{rulla11da7}
  \mathcal{N}_{d=11}\, = \, \mathcal{N}_{d=7} \, + \, 4
\end{equation}
As already emphasized in the introduction, the above discussion of supersymmetry is quite relevant for the $d=3$ field theories on the brane world volume.
\begin{description}
  \item[A)] Pure M2-branes without Englert fluxes produce a $d=3$ theory with $\mathcal{N}=8$ supersymmetry.  This theory is essentially a free one containing 8 scalars and 8 fermions. In the case of $\mathrm{AdS_4 }\times \mathrm{S^7}$ compactifications it is the theory of the $\mathrm{OSp(8|4)}$-singleton \cite{embedtensor1}. Here the background geometry is different.
  \item[B)] Introducing Englert fluxes corresponds to the inclusion of non trivial interactions among the fields living on the brane that are just the transverse coordinates and their fermionic partners. The world volume fields now arrange into supermultiplets according with residual supersymmetry dictated by the number of Killing spinors.
  \item[C)] In the case of theories obtained from Arnold-Bletrami fluxes, the supersymmetry is a t least $\mathcal{N}=4$. However the $d=3$ theory can be consistently truncated by removing $4$ bosonic fields with their superpartners. In this way one is left with the $d=3$ theory one might directly construct from the $d=7$ 2-branes solution with Arnold Beltrami fluxes.
  \item[D)] We are interested in Englert fluxes which lead to $d=3$ theories having supersymmetry $\mathcal{N} \, \le \, 4$, in particular $\mathcal{N}=1$ or $\mathcal{N}=0$, where, possibly, no consistent truncation exists to smaller multiplets. For this reason we proceed to the consideration of the specially promising crystallographic group introduced in eq.(\ref{L168}).
\end{description}
\section{The simple group $\mathrm{L_{168}}=\mathrm{PSL(2,\mathbb{Z}_7)}$}
For the reasons outlined at the end of the previous section we turn next our attention to the simple group (\ref{L168}) and to its
crystallographic action in $d=7$. The Hurwitz simple group  $\mathrm{L_{168}}$
is abstractly presented as follows:
\begin{equation}\label{abstroL168}
\mathrm{L_{168}} \, = \, \left(R,S,T \,\parallel \, R^2 \, = \, S^3 \, = \, T^7
\, = \, RST \, = \, \left(TSR\right)^4 \, = \, \mathbf{e}\right)
\end{equation}
and, as its name implicitly advocates, it has order 168:
\begin{equation}\label{order168}
  \mid \mathrm{L_{168}} \mid \, = \, 168
\end{equation}
The elements of this  simple group are organized in six conjugacy classes
according to the scheme displayed below:
\begin{center}
\begin{tabular}{||c|c|c|c|c|c|c||}
\hline \hline
Conjugacy class &$\mathcal{C}_1$&$\mathcal{C}_2$&$\mathcal{C}_3$&$\mathcal{C}_4$&$\mathcal{C}_5$&$\mathcal{C}_6$\\
  \hline
  \hline
  representative of the class  & $\mathbf{e}$ & $R$ & $S$ &$TRS$ & $T$ & $SR$ \\
  \hline
  order of the elements in the class & 1 & 2 & 3 & 4 & 7 & 7 \\
  \hline
  number of elements in the class & 1 & 21 & 56 & 42 & 24 & 24  \\
  \hline
  \hline
\end{tabular}
\begin{equation}\label{coniugini}
\null
\end{equation}
\end{center}
As one sees from the above table (\ref{coniugini}) the group contains elements
of order $2$, $3$, $4$ and $7$ and there are two inequivalent conjugacy classes
of elements of the highest order. According to the general theory of finite
groups, there are $6$ different irreducible representations of dimensions
$1,6,7,8,3,3$, respectively. The character table of the group $\mathrm{L_{168}}$
can be found in the mathematical literature, for instance in the book
\cite{miobukko}. It reads as follows:
\begin{center}
\begin{tabular}{||c|c|c|c|c|c|c||}
\hline \hline
Representation &$\mathcal{C}_1$&$\mathcal{C}_2$&$\mathcal{C}_3$&$\mathcal{C}_4$&$\mathcal{C}_5$&$\mathcal{C}_6$\\
  \hline
 $\mathrm{D_1}\left[\mathrm{L_{168}}\right]$  & $1$ & $1$ & $1$ &$1$ & $1$ & $1$ \\
\hline
$\mathrm{D_6}\left[\mathrm{L_{168}}\right]$  & $6$ & $2$ & $0$ &$0$ & $-1$ & $-1$ \\
\hline
$\mathrm{D_7}\left[\mathrm{L_{168}}\right]$  & $7$ & $-1$ & $1$ &$-1$ & $0$ & $0$ \\
\hline
$\mathrm{D_8}\left[\mathrm{L_{168}}\right]$  & $8$ & $0$ & $-1$ &$0$ & $1$ & $1$ \\
\hline
$\mathrm{DA_{3}}\left[\mathrm{L_{168}}\right]$  & $3$ & $-1$ & $0$ &$1$ & $\ft 12
\left(-1+{\rm i}\sqrt{7}\right)$ & $\ft 12 \left(-1-{\rm i}\sqrt{7}\right)$ \\
\hline
$\mathrm{DB_{3}}\left[\mathrm{L_{168}}\right]$  & $3$ & $-1$ & $0$ &$1$ & $\ft 12
\left(-1-{\rm i}\sqrt{7}\right)$& $\ft 12 \left(-1+{\rm i}\sqrt{7}\right)$ \\
  \hline
  \hline
\end{tabular}
\begin{equation}\label{caratterini}
\null
\end{equation}
\end{center}
For our purposes the most interesting representation is the $7$ dimensional
one. Indeed its properties are the very reason to consider the group
$\mathrm{L_{168}}$ in the present context. The following three statements are
true:
\begin{enumerate}
  \item The $7$-dimensional irreducible representation  is crystallographic
  since all elements $\gamma\in \mathrm{L_{168}}$ are represented by integer valued
  matrices $D_7\left(\gamma\right)$ in a basis of vectors that span a lattice,
  namely the root lattice $\Lambda_{\mathrm{root}}$ of the $A_7$ simple Lie algebra.
  \item The $7$-dimensional irreducible representation provides an immersion
  $\mathrm{L_{168}} \hookrightarrow \mathrm{SO(7)}$ since its elements preserve
  the symmetric Cartan matrix of $A_7$:
      \begin{eqnarray}
        \forall \gamma \in \mathrm{L_{168}}\quad:\quad D_7^T\left(\gamma\right) \,
        \mathcal{C} \, D_7\left(\gamma\right)& = & \mathcal{C}\nonumber \\
         \mathcal{C}_{i,j} & = & \alpha_i \cdot \alpha_j \quad\quad \quad  (i,j \, =\,1 \,\dots ,7) \label{gomorito}
      \end{eqnarray}
     defined in terms of the simple roots $\alpha_i$ whose standard construction
     in terms of the unit vectors $\epsilon_i$ of $\mathbb{R}^8$ is recalled below:
      \begin{equation}\label{simplerutte}
      \begin{array}{cccccccccccc}
        \alpha_1 & = & \epsilon_1 -\epsilon_2 & ; & \alpha_2 & = &
        \epsilon_2-\epsilon_3 & = & ; &\alpha_3 & = & \epsilon_3 - \epsilon_4 \\
       \alpha_4 & = & \epsilon_4 -\epsilon_5 & ;
       & \alpha_5 & = & \epsilon_5-\epsilon_6 & = & ; &\alpha_6 & = & \epsilon_6 - \epsilon_7 \\
        \alpha_7 & = & \epsilon_7 -\epsilon_8 &\null & \null&\null & \null &\null &\null&\null &\null &\null \\
      \end{array}
      \end{equation}
  \item Actually the $7$-dimensional representation defines an
  embedding $\mathrm{L_{168}} \hookrightarrow \mathrm{G_2} \subset \mathrm{SO(7)}$ since there exists
  a three-index antisymmetric tensor $\phi_{ijk}$ satisfying the relations of
  octonionic structure constants that is preserved by all the matrices $D_7(\gamma)$:
      \begin{equation}\label{cantonus}
      \forall \gamma \in \mathrm{L_{168}} \quad : \quad
      D_7(\gamma)_{ii^\prime}\,D_7(\gamma)_{jj^\prime}\
      \,D_7(\gamma)_{kk^\prime}\,\phi_{i^\prime j^\prime k^\prime} \, = \, \phi_{ijk}
      \end{equation}
\end{enumerate}
\par
Let us prove the above statements. It suffices to write the explicit form of
the generators $R$, $S$ and $T$ in the crystallographic basis of the considered
root lattice:
\begin{equation}\label{rutello}
 \mathbf{ v} \, \in \, \Lambda_{\mathrm{root}}  \quad \Leftrightarrow
 \quad  \mathbf{ v} \, = \, n_i \, \alpha_i \quad n_i \in \mathbb{Z}
\end{equation}
Explicitly if  we set:
\begin{eqnarray}\label{rgen}
\mathcal{R} & = &  \left(
\begin{array}{ccccccc}
 0 & 0 & 0 & 0 & 0 & 0 & -1 \\
 0 & 0 & 0 & 0 & 0 & -1 & 0 \\
 0 & 0 & -1 & 1 & 0 & -1 & 0 \\
 0 & -1 & 0 & 1 & 0 & -1 & 0 \\
 0 & -1 & 0 & 1 & -1 & 0 & 0 \\
 0 & -1 & 0 & 0 & 0 & 0 & 0 \\
 -1 & 0 & 0 & 0 & 0 & 0 & 0 \\
\end{array}
\right) \quad ; \quad \mathcal{S} \, = \, \left(
\begin{array}{ccccccc}
 0 & 0 & 0 & 0 & 0 & 0 & -1 \\
 1 & 0 & 0 & 0 & 0 & 0 & -1 \\
 1 & 0 & 0 & -1 & 1 & 0 & -1 \\
 1 & 0 & -1 & 0 & 1 & 0 & -1 \\
 1 & 0 & -1 & 0 & 1 & -1 & 0 \\
 1 & 0 & -1 & 0 & 0 & 0 & 0 \\
 1 & -1 & 0 & 0 & 0 & 0 & 0 \\
\end{array}
\right) \nonumber\\
\mathcal{T} & = & \left(
\begin{array}{ccccccc}
 0 & 0 & 0 & 0 & 0 & -1 & 1 \\
 1 & 0 & 0 & 0 & 0 & -1 & 1 \\
 0 & 1 & 0 & 0 & 0 & -1 & 1 \\
 0 & 0 & 1 & 0 & 0 & -1 & 1 \\
 0 & 0 & 0 & 1 & 0 & -1 & 1 \\
 0 & 0 & 0 & 0 & 1 & -1 & 1 \\
 0 & 0 & 0 & 0 & 0 & 0 & 1 \\
\end{array}
\right)
\end{eqnarray}
we find that the defining relations of $\mathrm{L_{168}}$ are satisfied:
\begin{equation}\label{relationibus}
 \mathcal{R}^2 \, = \, \mathcal{S}^3 \, = \, \mathcal{T}^7 \,=\,
 \mathcal{RST} \, = \, (\mathcal{TSR})^4 \, = \, \mathbf{1}_{7\times7}
\end{equation}
and furthermore we have:
\begin{equation}\label{so7preservo}
  \mathcal{R}^T \mathcal{C} \mathcal{R} \,=\, \mathcal{S}^T
  \mathcal{C}\mathcal{ S} \, = \, \mathcal{T}^T \mathcal{C} \mathcal{T} \, = \, \mathcal{C}
\end{equation}
where the explicit form of the $A_7$ Cartan matrix is recalled below:
\begin{equation}\label{cartanschula}
\mathcal{C} \, = \,  \left(
\begin{array}{ccccccc}
 2 & -1 & 0 & 0 & 0 & 0 & 0 \\
 -1 & 2 & -1 & 0 & 0 & 0 & 0 \\
 0 & -1 & 2 & -1 & 0 & 0 & 0 \\
 0 & 0 & -1 & 2 & -1 & 0 & 0 \\
 0 & 0 & 0 & -1 & 2 & -1 & 0 \\
 0 & 0 & 0 & 0 & -1 & 2 & -1 \\
 0 & 0 & 0 & 0 & 0 & -1 & 2 \\
\end{array}
\right)
\end{equation}
This proves statements 1) and 2).
\par
In order to prove statement 3) we proceed as follows. In $\mathbb{R}^7$ we
consider the antisymmetric three-index tensor $\phi_{ABC}$ that, in the
standard orthonormal basis, has the following components:
\begin{eqnarray}\label{gorlandus}
  \begin{array}{ccc}
 \phi_{1,2,6} &=&\frac{1}{6} \\
 \phi_{1,3,4} &=& -\frac{1}{6} \\
 \phi_{1,5,7} &=& -\frac{1}{6} \\
  \phi_{2,3,7} &=& \frac{1}{6} \\
 \phi_{2,4,5} &=& \frac{1}{6} \\
  \phi_{3,5,6} &=& -\frac{1}{6} \\
  \phi_{4,6,7} &=& -\frac{1}{6} \\
\end{array} &; & \mbox{all other components vanish}
\end{eqnarray}
This tensor satisfies the algebraic relations of octonionic structure
constants, namely\footnote{In this equation the indices of the $G_2$-invariant tensor are denoted with
capital letter of the Latin alphabet, as it was the case in the quoted literature on weak $G_2$-structures. In the following we will use lower case latin letters to be consistent with our botation for the supergravity constructions, the upper Latin letters being reserved for $d=8$}:
\begin{eqnarray}\label{2colibri2}
  \phi_{ABM} \, \phi_{CDM} & = & \frac{1}{18} \delta^{AB}_{CD} \, +\, \frac{2}{3} \Phi_{ABCD} \\
  \phi_{ABC} & = &- \frac{1}{6} \epsilon_{ABCPQRS} \, \Phi_{ABCD}
\end{eqnarray}
and the subgroup of $\mathrm{SO(7)}$ which leaves $\phi_{ABC} $ invariant
is, by definition, the compact section $\mathrm{G_{(2,-14)}}$ of the complex
$\mathrm{G_2}$ Lie group (see for instance \cite{Fre':2006ut}). A particular
matrix that transforms the standard orthonormal basis of $\mathbb{R}^7$ into
the basis of simple roots $\alpha_i$ is the following one:
\begin{equation}\label{mgothica}
  \mathfrak{M} \, = \, \left(
\begin{array}{ccccccc}
 \sqrt{2} & -\frac{1}{\sqrt{2}} & 0 & 0 & 0 & 0 & 0 \\
 0 & -\frac{1}{\sqrt{2}} & \sqrt{2} & -\frac{1}{\sqrt{2}} & 0 & 0 &
   0 \\
 0 & 0 & 0 & -\frac{1}{\sqrt{2}} & \sqrt{2} & -\frac{1}{\sqrt{2}} &
   0 \\
 0 & 0 & 0 & 0 & 0 & -\frac{1}{\sqrt{2}} & \sqrt{2} \\
 0 & -\frac{1}{\sqrt{2}} & 0 & \frac{1}{\sqrt{2}} & 0 &
   -\frac{1}{\sqrt{2}} & 0 \\
 0 & 0 & 0 & -\frac{1}{\sqrt{2}} & 0 & 0 & 0 \\
 0 & \frac{1}{\sqrt{2}} & 0 & 0 & 0 & -\frac{1}{\sqrt{2}} & 0 \\
\end{array}
\right)
\end{equation}
since:
\begin{equation}\label{rattus}
  \mathfrak{M}^T \, \mathfrak{M} \, = \, \mathcal{C}
\end{equation}
Defining the transformed tensor:
\begin{equation}\label{perenospero}
  \varphi_{ijk} \, \equiv \, \left(\mathfrak{M}^{-1}\right)_i^{\phantom{i}I} \,
  \left(\mathfrak{M}^{-1}\right)_j^{\phantom{j}J}
   \,  \left(\mathfrak{M}^{-1}\right)_k^{\phantom{k}K} \, \phi_{IJK}
\end{equation}
we can explicitly verify that:
\begin{eqnarray}
\varphi_{ijk} &=& \left(\mathcal{R}\right)_i^{\phantom{i}p} \,
\left(\mathcal{R}\right)_j^{\phantom{i}q}
   \,  \left(\mathcal{R}\right)_k^{\phantom{i}r} \, \varphi_{pqr} \nonumber\\
 \varphi_{ijk} &=& \left(\mathcal{S}\right)_i^{\phantom{i}p} \, \left(\mathcal{S}\right)_j^{\phantom{i}q}
   \,  \left(\mathcal{S}\right)_k^{\phantom{i}r} \, \varphi_{pqr}\nonumber\\
\varphi_{ijk} &=& \left(\mathcal{T}\right)_i^{\phantom{i}p} \,
\left(\mathcal{T}\right)_j^{\phantom{i}q}
   \,  \left(\mathcal{T}\right)_k^{\phantom{i}r} \, \varphi_{pqr} \label{gargiulo}
\end{eqnarray}
Hence, being preserved by the three-generators $\mathcal{R}$,$\mathcal{S}$ and
$\mathcal{T}$, the antisymmetric tensor $\varphi_{ijk}$ is preserved by the entire
discrete group $\mathrm{L_{168}}$ which, henceforth, is  a subgroup of
$\mathrm{G_{(2,-14)}}\subset \mathrm{SO(7)}$, as it was shown by intrinsic
group theoretical arguments in \cite{kingus}.
\section{Classification of the proper subgroups $\mathrm{H}\subset\mathrm{L_{168}}$ and of the
$\mathrm{L_{168}}$-orbits in the weight lattice} We aim at the construction of
solutions of Englert equation (\ref{englerta}) on the crystallographic
$7$-torus:
\begin{equation}\label{7mucche}
 \mathrm{ T^7} \,  \equiv \, \frac{\mathbb{R}^7}{\Lambda_{\mathrm{root}}}
\end{equation}
where the root lattice is defined in eq.(\ref{rutello}). The first necessary
step is to introduce the dual weight lattice
\begin{equation}\label{roncisvaldo}
    \Lambda_w \, \ni \, \mathbf{w}\, = \, n_i \, \lambda^i \quad : \quad n^i \in \mathbb{Z}
\end{equation}
spanned by the simple weights that are implicitly defined by the relations:
\begin{equation}\label{dualabasata}
    \lambda^i \cdot \alpha_j \, = \, \delta^i_j \quad \Rightarrow \quad \lambda^i \, = \,
    \left(\mathcal{C}^{-1}\right)^{ij} \,
    \alpha_j
\end{equation}
Given the generators of the group $\mathrm{L_{168}}$ in the basis of simple
roots we obtain the same in the basis of simple weights through the following
transformation:
\begin{equation}\label{trasforlinzo}
    \mathcal{R}_w \, = \, \mathcal{C} \, \mathcal{R} \, \mathcal{C}^{-1} \quad ; \quad
    \mathcal{S}_w \, = \, \mathcal{C} \, \mathcal{S} \, \mathcal{C}^{-1}\quad ; \quad
    \mathcal{T}_w \, = \, \mathcal{C} \, \mathcal{T} \, \mathcal{C}^{-1}
\end{equation}
Explicitly we find:
\begin{eqnarray}
% \nonumber to remove numbering (before each equation)
  \mathcal{R}_w &=& \left(
\begin{array}{ccccccc}
 0 & 0 & 0 & 0 & 0 & 0 & -1 \\
 0 & 0 & 0 & -1 & -1 & -1 & 0 \\
 0 & 0 & -1 & 0 & 0 & 0 & 0 \\
 0 & 0 & 1 & 1 & 1 & 0 & 0 \\
 0 & 0 & 0 & 0 & -1 & 0 & 0 \\
 0 & -1 & -1 & -1 & 0 & 0 & 0 \\
 -1 & 0 & 0 & 0 & 0 & 0 & 0 \\
\end{array}
\right) \quad ; \quad   \mathcal{S}_w\, = \,\left(
\begin{array}{ccccccc}
 -1 & -1 & -1 & -1 & -1 & -1 & -1 \\
 1 & 1 & 1 & 1 & 0 & 0 & 0 \\
 0 & 0 & 0 & -1 & 0 & 0 & 0 \\
 0 & 0 & 0 & 1 & 1 & 1 & 0 \\
 0 & 0 & 0 & 0 & 0 & -1 & 0 \\
 0 & 0 & -1 & -1 & -1 & 0 & 0 \\
 0 & -1 & 0 & 0 & 0 & 0 & 0 \\
\end{array}
\right) \nonumber\\ \end{eqnarray}
\begin{eqnarray}
  \mathcal{T}_w &=& \left(
\begin{array}{ccccccc}
 -1 & -1 & -1 & -1 & -1 & -1 & 0 \\
 1 & 0 & 0 & 0 & 0 & 0 & 0 \\
 0 & 1 & 0 & 0 & 0 & 0 & 0 \\
 0 & 0 & 1 & 0 & 0 & 0 & 0 \\
 0 & 0 & 0 & 1 & 0 & 0 & 0 \\
 0 & 0 & 0 & 0 & 1 & 0 & 0 \\
 0 & 0 & 0 & 0 & 0 & 1 & 1 \\
\end{array}
\right)
\end{eqnarray}
Equipped with this result we can construct orbits of weight lattice vectors
under the action of the group $\mathrm{L_{168}}$.
\subsection{A first random exploration of the orbits}
\label{random} As a first orientation  exercise we resorted to random
calculations and we found that there are orbits $\mathcal{O}$ whose length
$\ell_\mathcal{O}$ takes  one among the following seven values:
\begin{equation}\label{kosterlitz}
  \ell_\mathcal{O} \, \in \, \{168,84,56,42,28,14,8\}
\end{equation}
In the next subsection we will retrieve these numbers from a rigorous and
exhaustive classification of all conjugacy classes of the proper subgroups
$\mathrm{H}\subset\mathrm{L_{168}}$. Indeed given a vector $\vec{v}_0 \in
\Lambda_w$, the $\mathrm{L_{168}}$-orbit $\mathcal{O}\left(\vec{v}_0\right)$ of
such a vector is isomorphic to the coset:
\begin{equation}\label{carducco}
    \mathrm{L_{168}}/\mathrm{H}^s_0
\end{equation}
where $\mathrm{H}^s_0$ is the stability subgroup of $\vec{v}_0$, that is:
\begin{equation}\label{stabillus}
   \forall \, \gamma \, \in \,\mathrm{H}^s_0 \quad : \quad\gamma\, \cdot \, \,\vec{v}_0 \, = \, \vec{v}_0
\end{equation}
Every other vector in the same orbit $\vec{v} \in
\mathcal{O}\left(\vec{v}_0\right)$ has a stability subgroup
$\mathrm{H}^s\subset\mathrm{L_{168}}$ which is conjugate to $\mathrm{H}_0$ via
the  group element $g\,\in \,\mathrm{L_{168}}$ which maps $\vec{v}_0$ into
$\vec{v}$. Hence the classification of all the possible orbits amounts to the
classification of the conjugacy classes of possible stability subgroups
$\mathrm{H}^s$ which certainly is included in the classification of conjugacy
classes of proper subgroups $\mathrm{H}\subset \mathrm{L_{168}}$. The latter
sentence is a \textit{caveat}. It might happen that a certain subgroup
$\mathrm{H}$ admits no fixed vector $v \in \Lambda_w$. In that case there is no
orbit with such a stability subgroup and the corresponding coset is not
isomorphic to any orbit.
\par
Before plunging into the above sketched systematics, let us note that some of
the numbers in eq.(\ref{kosterlitz})  correspond to the dimensions, of certain
irreducible representations of $\mathrm{SL(8,\mathbb{R})}$, other instead do
not correspond to the dimensions of any representation. In particular we have:
\begin{equation}\label{ciccio}
    \mbox{dim}\, \yng(2,1) \, = \, 168 \quad ; \quad \mbox{dim}\, \yng(1,1,1) \, = \, 56
    \quad ; \quad \mbox{dim}\, \yng(1,1) \, = \, 28\quad ; \quad \mbox{dim}\, \yng(1) \, = \, 8
\end{equation}
In the above cases we have verified that the set of weights of the
corresponding representation coincides with the $\mathrm{L_{168}}$-orbit of its
maximal  weight $\lambda_{max}$.
\par
Yet, as it will clearly appear from the exhaustive discussion of the next
section, this is just a curious coincidence but it is not the key to understand
the complete classification of orbits.
\subsection{Classification of conjugacy classes  of subgroups $\mathrm{H}\subset\mathrm{L_{168}}$}
An important mathematical result states that the simple group
$\mathrm{L_{168}}$ contains maximal subgroups only of index 8 and 7, namely of
order 21 and 24 \cite{turcomanno}. The order 21 subgroup $\mathrm{G_{21}}$ is the unique
non-abelian group of that order and abstractly it has the structure of the
semidirect product $\mathbb{Z}_3 \ltimes
 \mathbb{Z}_7$. Up to conjugation there is only one subgroup $\mathrm{G_{21}}$ as we have explicitly
verified with the computer. On the other hand, up to conjugation, there are two
different groups of order 24 that are both isomorphic to the octahedral group
$\mathrm{O_{24}}$.
\subsubsection{The maximal subgroup $\mathrm{G_{21}}$}
The group $\mathrm{G_{21}}$ has two generators $\mathcal{X}$ and $\mathcal{Y}$
that satisfy the following relations:
\begin{equation}
\mathcal{X}^3 \, = \, \mathcal{Y}^7 \, = \,\mathbf{ 1} \quad ; \quad
\mathcal{X}\mathcal{Y}  = \mathcal{Y}^2 \mathcal{X} \label{faloluna}
\end{equation}
The organization of the 21 group elements into conjugacy classes is displayed
below:
\begin{equation}\label{corsaro}
\begin{array}{|c|c|c|c|c|c|}
\hline \text{Conjugacy} \text{Class} & C_1 & C_2 & C_3 & C_4 & C_5 \\ \hline
\text{representative of the class} & e & \mathcal{Y} & \mathcal{X}^2
\mathcal{Y}\mathcal{X}\mathcal{Y}^2 & \mathcal{Y}\mathcal{X}^2 & \mathcal{X}
\\ \hline \text{order of the elements in the class} & 1 & 7 & 7 & 3
& 3 \\ \hline \text{number of elements in the class} & 1 & 3 & 3 & 7 & 7 \\
\hline
\end{array}
\end{equation}
As we see there are five conjugacy classes which implies that there should be
five irreducible representations the square of whose dimensions should sum up
to the group order 21. The solution of this problem is:
\begin{equation}\label{cavacchioli}
    21 \, = \, 1^2 + 1^2 + 1^2 + 3^2 + 3^2
\end{equation}
and the corresponding character table is mentioned below:
\begin{equation}\label{bruttocarattere21}
\begin{array}{|c|c|c|c|c|c|}
\hline 0 & e & \mathcal{Y} & \mathcal{X}^2 \mathcal{Y}\mathcal{X}\mathcal{Y}^2
& \mathcal{Y}\mathcal{X}^2 & \mathcal{X}
\\ \hline
\mathrm{D_1}\left[\mathrm{G_{21}}\right] & 1 & 1 & 1 &
1 & 1 \\
\hline \text{DX}_1\left[\mathrm{G_{21}}\right] & 1 & 1 & 1 & -(-1)^{1/3} &
(-1)^{2/3} \\
\hline \text{DY}_1\left[\mathrm{G_{21}}\right] & 1 & 1 & 1 & (-1)^{2/3} &
-(-1)^{1/3} \\
\hline \text{DA}_3\left[\mathrm{G_{21}}\right] & 3 & \frac{1}{2} i
\left(i+\sqrt{7}\right) & -\frac{1}{2} i \left(-i+\sqrt{7}\right)
& 0 & 0 \\
\hline \text{DB}_3\left[\mathrm{G_{21}}\right] & 3 & -\frac{1}{2} i
\left(-i+\sqrt{7}\right) & \frac{1}{2} i \left(i+\sqrt{7}\right)
& 0 & 0 \\
\hline
\end{array}
\end{equation}
In the weight-basis the two generators of the $\mathrm{G_{21}}$ subgroup of
$\mathrm{L_{168}}$ can be chosen to be the following matrices and this fixes our
representative of the unique conjugacy class:
\begin{equation}\label{XYgenerati}
    \mathcal{X}\, = \, \left(
\begin{array}{ccccccc}
 1 & 1 & 1 & 1 & 1 & 1 & 1 \\
 0 & 0 & 0 & 0 & 0 & 0 & -1 \\
 0 & -1 & -1 & -1 & -1 & -1 & 0 \\
 0 & 1 & 1 & 1 & 0 & 0 & 0 \\
 0 & 0 & -1 & -1 & 0 & 0 & 0 \\
 0 & 0 & 1 & 1 & 1 & 0 & 0 \\
 0 & 0 & 0 & -1 & -1 & 0 & 0 \\
\end{array}
\right) \quad  \mathcal{Y} \, = \,\left(
\begin{array}{ccccccc}
 0 & 1 & 1 & 0 & 0 & 0 & 0 \\
 0 & 0 & 0 & 1 & 1 & 1 & 1 \\
 0 & 0 & -1 & -1 & -1 & -1 & -1 \\
 0 & 0 & 1 & 1 & 0 & 0 & 0 \\
 -1 & -1 & -1 & -1 & 0 & 0 & 0 \\
 1 & 1 & 1 & 1 & 1 & 0 & 0 \\
 0 & 0 & 0 & 0 & 0 & 1 & 0 \\
\end{array}
\right)
\end{equation}
\subsubsection{The maximal subgroups $\mathrm{O_{24A}}$ and $\mathrm{O_{24B}}$} The octahedral group
$\mathrm{O_{24}}$ has two generators $S$ and $T$ that satisfy the following
relations:
\begin{equation}
S^2 \, = \, T^3\,  = \,(ST)^4 \, = \,  \mathbf{1}
\end{equation}
The 24 elements are organized in five conjugacy classes according to the scheme
displayed below:
\begin{equation}\label{o24coniugi}
  \begin{array}{|c|c|c|c|c|c|} \hline
 \text{Conjugacy Class} & C_1 & C_2 & C_3 & C_4 & C_5 \\
\hline
 \text{representative of the class} & e & T & STST  & S &  ST  \\
\hline
 \text{order of the elements in the class} & 1 & 3 & 2 & 2 & 4 \\
\hline
 \text{number of elements in the class} & 1 & 8 & 3 & 6 & 6 \\
\hline
\end{array}
\end{equation}
It follows that there are five irreducible representations that turn out to be
of dimensions, 1,1,2,3,3, according to the sum rule:
\begin{equation}\label{summarulla24}
    24 = 1^2 + 1^2 + 2^2 + 3^2+3^2
\end{equation}
The corresponding character table is the following one:
\begin{equation}\label{caratteri24}
\begin{array}{|c|c|c|c|c|c|}
\hline
 0 & e & T & STST  & S &  ST \\
\hline
 \mathrm{D_1}\left[\mathrm{O_{24}}\right] & 1 & 1 & 1 & 1 & 1 \\
\hline
 \text{DX}_1\left[\mathrm{O_{24}}\right] & 1 & 1 & 1 & -1 & -1 \\
\hline
 \mathrm{D_2}\left[\mathrm{O_{24}}\right] & 2 & -1 & 2 & 0 & 0 \\
\hline
 \mathrm{DA_3}\left[\mathrm{O_{24}}\right] & 3 & 0 & -1 & -1 & 1 \\
\hline
 \mathrm{DB_3}\left[\mathrm{O_{24}}\right] & 3 & 0 & -1 & 1 & -1 \\
\hline
\end{array}
\end{equation}
By computer calculations we have verified that there are just two
disjoint conjugacy classes of $\mathrm{O_{24}}$ maximal subgroups in
$\mathrm{L_{168}}$ that we have named A and B, respectively. We have chosen two
standard representatives, one for each conjugacy class, that we have named
$\mathrm{O_{24A}}$ and $\mathrm{O_{24B}}$ respectively. To fix these subgroups
it suffices to mention the explicit form of the their generators in the weight
basis.
\par
For the group $\mathrm{O_{24A}}$, we chose:
\begin{equation}\label{generatori24A}
    T_A \, = \, \left(
\begin{array}{ccccccc}
 1 & 1 & 1 & 1 & 1 & 1 & 1 \\
 0 & 0 & 0 & 0 & 0 & 0 & -1 \\
 0 & -1 & -1 & -1 & -1 & -1 & 0 \\
 0 & 1 & 1 & 1 & 0 & 0 & 0 \\
 0 & 0 & -1 & -1 & 0 & 0 & 0 \\
 0 & 0 & 1 & 1 & 1 & 0 & 0 \\
 0 & 0 & 0 & -1 & -1 & 0 & 0 \\
\end{array}
\right)    \quad  S_A \, = \,\left(
\begin{array}{ccccccc}
 0 & 0 & 0 & 1 & 1 & 1 & 0 \\
 0 & 0 & 0 & 0 & -1 & -1 & 0 \\
 -1 & -1 & -1 & -1 & 0 & 0 & 0 \\
 1 & 1 & 0 & 0 & 0 & 0 & 0 \\
 0 & 0 & 1 & 1 & 1 & 1 & 1 \\
 0 & -1 & -1 & -1 & -1 & -1 & -1 \\
 0 & 1 & 1 & 1 & 1 & 0 & 0 \\
\end{array}
\right)
\end{equation}
For the group $\mathrm{O_{24B}}$, we chose:
\begin{equation}\label{generatori24B}
    T_B \, = \, \left(
\begin{array}{ccccccc}
 1 & 1 & 1 & 1 & 0 & 0 & 0 \\
 0 & -1 & -1 & -1 & 0 & 0 & 0 \\
 0 & 1 & 1 & 1 & 1 & 0 & 0 \\
 0 & 0 & -1 & -1 & -1 & 0 & 0 \\
 0 & 0 & 1 & 1 & 1 & 1 & 0 \\
 0 & 0 & 0 & -1 & -1 & -1 & 0 \\
 0 & 0 & 0 & 1 & 1 & 1 & 1 \\
\end{array}
\right)    \quad  S_B \, = \,\left(
\begin{array}{ccccccc}
 0 & 0 & 1 & 1 & 1 & 0 & 0 \\
 -1 & -1 & -1 & -1 & -1 & 0 & 0 \\
 1 & 1 & 1 & 1 & 1 & 1 & 1 \\
 0 & 0 & 0 & 0 & 0 & 0 & -1 \\
 0 & -1 & -1 & -1 & -1 & -1 & 0 \\
 0 & 1 & 1 & 1 & 0 & 0 & 0 \\
 0 & 0 & 0 & -1 & 0 & 0 & 0 \\
\end{array}
\right)
\end{equation}
\subsubsection{The tetrahedral subgroup $\mathrm{T_{12}} \subset \mathrm{O_{24}}$}
Every octahedral group $\mathrm{O_{24}}$ has, up to
$\mathrm{O_{24}}$-conjugation, a unique tetrahedral subgroup $\mathrm{T_{12}}$
whose order is 12. The abstract description of the tetrahedral group is
provided by the following presentation in terms of two generators:
\begin{equation}\label{presentaT12}
  \mathrm{T_{12}} =\left(s,t\left|s^2\right. = t^3 = (st)^3 = 1\right)
\end{equation}
The 12 elements are organized into  four conjugacy classes as displayed below:
\begin{equation}\label{coniugatoT12}
\begin{array}{|c|c|c|c|c|}
\hline
 \text{Classes} & C_1 & C_2 & C_3 & C_4 \\
\hline
 \text{standard representative} & 1 & s & t & t^2s \\
\hline
 \text{order of the elements in the conjugacy class} & 1 & 2 & 3 & 3 \\
\hline
 \text{number of elements in the conjugacy class} & 1 & 3 & 4 & 4 \\
\hline
\end{array}
\end{equation}
We do not display the character table since we will not use it in the present
paper. We anticipate that the two tetrahedral subgroups $\mathrm{T_{12A}}
\subset \mathrm{O_{24A}}$ and $\mathrm{T_{12B}} \subset \mathrm{O_{24B}}$ are
not conjugate under the big group $\mathrm{L_{168}}$. Hence we have two
conjugacy classes of tetrahedral subgroups of $\mathrm{L_{168}}$, as we explain
in the sequel.
\subsubsection{The dihedral subgroup $\mathrm{Dih_{3}} \subset \mathrm{O_{24}}$}
Every octahedral group $\mathrm{O_{24}}$ has a dihedral group
$\mathrm{Dih_{3}}$ whose order is 6. The abstract description of the dihedral
group $\mathrm{Dih_{3}}$ is provided by the following presentation in terms of
two generators:
\begin{equation}\label{presentaDih3}
  \mathrm{Dih_{3}} =\left(A,B\left| A^3=B^2=(BA)^2=1\right.\right)
\end{equation}
The 6 elements are organized into  three conjugacy classes as displayed below:
\begin{equation}\label{coniugatoDih3}
\begin{array}{|c|c|c|c|}
\hline
 \text{Conjugacy} \text{Classes} & C_1 & C_2 & C_3 \\
\hline
 \text{standard representative of the class} & 1 & A & B \\
\hline
 \text{order of the elements in the class} & 1 & 3 & 2 \\
\hline
 \text{number of elements in the class} & 1 & 2 & 3 \\
\hline
\end{array}
\end{equation}
We do not display the character table since we will not use it in the present
paper. We anticipate that differently from the case of the tetrahedral
subgroups the two dihedral subgroups $\mathrm{Dih_{3A}} \subset
\mathrm{O_{24A}}$ and $\mathrm{Dih_{3B}} \subset \mathrm{O_{24B}}$ turn out to
be conjugate under the big group $\mathrm{L_{168}}$. Actually there is just one
$\mathrm{L_{168}}$-conjugacy class of dihedral subgroups $\mathrm{Dih_{3}}$.
\subsubsection{Enumeration of the possible subgroups and orbits}
Since the maximal subgroups of $\mathrm{L_{168}}$ are of index 7 or 8 we can
have subgroups $\mathrm{H} \subset \mathrm{L_{168}}$ that are either
$\mathrm{G_{21}}$ or $\mathrm{O_{24}}$ or subgroups thereof. Furthermore, as it
is well known, the order $|\mathrm{H}|$  of any subgroup $\mathrm{H}\subset
\mathrm{G}$ must be a divisor of $|\mathrm{G}|$. Hence we conclude that
\begin{equation}\label{ordinatini}
 |\mathrm{H}| \, \in \,   \{1,2,3,4,6,7,8,12,21,24\}
\end{equation}
Correspondingly we might have $\mathrm{L_{168}}$-orbits $\mathcal{O}$ in the
weight lattice $\Lambda_w$, whose length is one of the following nine numbers:
\begin{equation}\label{apprilunghi}
    \ell_{\mathcal{O}} \, \in \, \{168,84,56,42,28,24,21,14,8,7\}
\end{equation}
Comparing eq.(\ref{apprilunghi}) with the result of our numerical random
experiment displayed in eq.(\ref{kosterlitz}) we see that three orbit lengths are
excluded, namely $24$, $21$ and $7$. In the sequel of this subsection we will show
the reason for these exclusions.
\par
Combining the information about the possible group orders (\ref{ordinatini})
with the information that the maximal subgroups are of index 8 or 7, we arrive
at the following list of possible subgroups $\mathrm{H}$ (up to conjugation) of
the group $\mathrm{L_{168}}$:
\begin{description}
\item[Order 24)] Either H = $\mathrm{O_{24A}}$  or H = $\mathrm{O_{24B}}$.
\item[Order 21)] The only possibility is H = $\mathrm{G_{21}}$.
\item[Order 12)] The only possibilities are H = $\mathrm{T_{12A}}$ or H = $\mathrm{T_{12B}}$ where
$\mathrm{T_{12}}$ is the tetrahedral subgroup of the octahedral group
$\mathrm{O_{24}}$.
\item[Order 8)] Either H = $\mathbb{Z}_2\times\mathbb{Z}_2\times \mathbb{Z}_2$ or
H = $\mathbb{Z}_2\times\mathbb{Z}_4$.
\item[Order 7)] The only possibility is $\mathbb{Z}_7$.
\item [Order 6)] Either H = $\mathbb{Z}_2\times\mathbb{Z}_3$ or  H = $\text{Dih}_3$, where $\text{Dih}_3$
denotes the dihedral subgroup of index 3 of the octahedral group
$\mathrm{O_{24}}$.
\item[Order 4)] Either H = $\mathbb{Z}_2\times\mathbb{Z}_2$  or H = $\mathbb{Z}_4$.
\item[Order 3)] The only possibility is H = $\mathbb{Z}_3$
\item[Order 2)] The only possibility is H = $\mathbb{Z}_2$.
\end{description}
\subsubsection{Summary of our results for the subgroups and the orbits}
Let us summarize the results that we have derived by means of computer aided
calculations.
\begin{enumerate}
\item We have verified that there are no orbits with stability subgroups either $\mathrm{O_{24A}}$ or
$\mathrm{O_{24B}}$. Indeed the constraints imposed on a seven vector
$\mathbf{v}$ by the request that it should be an eigenstate of the generators
(\ref{generatori24A}) or (\ref{generatori24B}) admits the only solution
$\mathbf{v} \, = \,0$. \textit{This means that there are no orbits of length
7}.
\item On the contrary we have verified that there are orbits with stability subgroup $\mathrm{G_{21}}$.
These orbits have length $\ell_\mathcal{O} \, = \, 8$ and depend from a
\textit{ unique integer parameter} $n$. Indeed the most general vector
$\mathbf{v}_0$ invariant under $\mathrm{G_{21}}$ has the following form:
\begin{equation}\label{inva8}
    \mathbf{v}_0 \, = \, \{0,0,0,n,-n,0,0\}
\end{equation}
and the corresponding $\mathrm{L_{168}}$-orbit is displayed below:
\begin{equation}\label{orbita8}
 \mathcal{O}_8  \, = \, \left\{
\begin{array}{ccccccc}
 0 & 0 & 0 & 0 & 0 & 0 & n \\
 0 & 0 & 0 & 0 & 0 & n & -n \\
 0 & 0 & 0 & 0 & n & -n & 0 \\
 0 & 0 & 0 & n & -n & 0 & 0 \\
 0 & 0 & n & -n & 0 & 0 & 0 \\
 0 & n & -n & 0 & 0 & 0 & 0 \\
 -n & 0 & 0 & 0 & 0 & 0 & 0 \\
 n & -n & 0 & 0 & 0 & 0 & 0 \\
\end{array}
\right\}
\end{equation}
every line denoting the components of a 7-vector belonging to the orbit.
\item As we anticipated above, by means of computer calculations we have verified that there are
\textit{ two conjugacy classes} of \textit{ tetrahedral groups,
$\mathrm{T_{12A}}$ and $\mathrm{T_{12B}}$}. This implies that there are
\textit{two types} of orbits of  length 14, namely $\mathcal{O}_{\mathrm{14A}}$
and $\mathcal{O}_{\mathrm{14B }}$. Both of them   depend only on \textit{ one
integer parameter} $n$.
\begin{itemize}
  \item In the case of $\mathrm{T_{12A}}$, the two generators $s,t$ can be chosen as follows:
{\scriptsize  \begin{eqnarray}\label{stT12A}
    s_A & = &\left(
\begin{array}{ccccccc}
 0 & 0 & 1 & 1 & 1 & 1 & 1 \\
 -1 & -1 & -1 & -1 & -1 & -1 & -1 \\
 1 & 1 & 1 & 1 & 0 & 0 & 0 \\
 0 & 0 & 0 & -1 & 0 & 0 & 0 \\
 0 & 0 & 0 & 1 & 1 & 1 & 0 \\
 0 & 0 & 0 & 0 & 0 & -1 & 0 \\
 0 & -1 & -1 & -1 & -1 & 0 & 0 \\
\end{array}
\right)   \nonumber\\  t_A & = &\left(
\begin{array}{ccccccc}
 0 & -1 & -1 & -1 & -1 & 0 & 0 \\
 -1 & 0 & 0 & 0 & 0 & 0 & 0 \\
 1 & 1 & 1 & 1 & 0 & 0 & 0 \\
 0 & 0 & 0 & 0 & 1 & 1 & 1 \\
 0 & 0 & -1 & -1 & -1 & -1 & -1 \\
 0 & 0 & 1 & 1 & 1 & 1 & 0 \\
 0 & 0 & 0 & -1 & -1 & -1 & 0 \\
\end{array}
\right)\label{stT12A}
  \end{eqnarray}}
the most general vector $\mathbf{v}_0$ invariant under $\mathrm{T_{12A}}$ has
the following form: {\scriptsize
\begin{equation}\label{t12Ainva}
   \mathbf{v}_0\, = \, \{n, -n, n, 0, -n, 0, n\}
\end{equation}
} and the corresponding $\mathrm{L_{168}}$-orbit is displayed below:
{\scriptsize
\begin{equation}\label{orbita14A}
 \mathcal{O}_{\mathrm{14A}}  \, = \,  \left\{
\begin{array}{ccccccc}
 0 & 0 & -n & n & -n & 0 & 0 \\
 0 & 0 & n & -n & n & 0 & 0 \\
 0 & -n & 0 & 0 & n & -n & n \\
 0 & -n & n & -n & 0 & n & -n \\
 0 & n & 0 & 0 & -n & n & -n \\
 0 & n & -n & n & 0 & -n & n \\
 -n & 0 & 0 & n & -n & n & 0 \\
 -n & 0 & n & 0 & 0 & -n & 0 \\
 -n & n & 0 & -n & 0 & 0 & n \\
 -n & n & -n & 0 & n & 0 & -n \\
 n & 0 & 0 & -n & n & -n & 0 \\
 n & 0 & -n & 0 & 0 & n & 0 \\
 n & -n & 0 & n & 0 & 0 & -n \\
 n & -n & n & 0 & -n & 0 & n \\
\end{array}
\right\}
\end{equation}}
  \item In the case of $\mathrm{T_{12B}}$, the two generators $s,t$ can be chosen as follows:
{\scriptsize  \begin{eqnarray}\label{stT12A}
    s_B & = & \left(
\begin{array}{ccccccc}
 0 & 0 & 0 & 1 & 0 & 0 & 0 \\
 0 & 0 & 0 & 0 & 1 & 1 & 1 \\
 -1 & -1 & -1 & -1 & -1 & -1 & -1 \\
 1 & 0 & 0 & 0 & 0 & 0 & 0 \\
 0 & 1 & 1 & 1 & 1 & 1 & 0 \\
 0 & 0 & 0 & 0 & 0 & -1 & 0 \\
 0 & 0 & -1 & -1 & -1 & 0 & 0 \\
\end{array}
\right)  \nonumber\\  t_B & = & \left(
\begin{array}{ccccccc}
 0 & -1 & -1 & -1 & -1 & -1 & 0 \\
 0 & 1 & 1 & 1 & 1 & 1 & 1 \\
 -1 & -1 & -1 & -1 & -1 & -1 & -1 \\
 1 & 1 & 0 & 0 & 0 & 0 & 0 \\
 0 & 0 & 1 & 1 & 1 & 0 & 0 \\
 0 & 0 & 0 & -1 & -1 & 0 & 0 \\
 0 & 0 & 0 & 1 & 0 & 0 & 0 \\
\end{array}
\right)\label{stT12B}
  \end{eqnarray}}
the most general vector $\mathbf{v}_0$ invariant under $\mathrm{T_{12B}}$ has
the following form: {\scriptsize
\begin{equation}\label{t12Binva}
   \mathbf{v}_0\, = \, \{n, 0, -n, n, -n, 0, n\}
\end{equation}
} and the corresponding $\mathrm{L_{168}}$-orbit is displayed below:
{\scriptsize
\begin{equation}\label{orbita14B}
 \mathcal{O}_{\mathrm{14B}}  \, = \,   \left(
\begin{array}{ccccccc}
 0 & 0 & -n & 0 & n & -n & 0 \\
 0 & 0 & n & 0 & -n & n & 0 \\
 0 & -n & 0 & n & -n & n & -n \\
 0 & -n & n & -n & 0 & 0 & n \\
 0 & n & 0 & -n & n & -n & n \\
 0 & n & -n & n & 0 & 0 & -n \\
 -n & 0 & 0 & n & 0 & -n & n \\
 -n & 0 & n & -n & n & 0 & -n \\
 -n & n & 0 & 0 & -n & 0 & 0 \\
 -n & n & -n & 0 & 0 & n & 0 \\
 n & 0 & 0 & -n & 0 & n & -n \\
 n & 0 & -n & n & -n & 0 & n \\
 n & -n & 0 & 0 & n & 0 & 0 \\
 n & -n & n & 0 & 0 & -n & 0 \\
\end{array}
\right)
\end{equation}}
\end{itemize}
\item Next, as we anticipated above, we have verified that there is only
\textit{ one conjugacy class} of \textit{ dihedral groups  $\mathrm{Dih_3}$}.
This implies that there is \textit{only one type} of orbits of  length 28. They
depend only \textit{ on one integer parameter} $ n$. Indeed a choice of the two
generators $A,B$ introduced in eq.(\ref{presentaDih3}) is the following one:
{\scriptsize
\begin{eqnarray}\label{Dih3gene}
   A & = & \left(
\begin{array}{ccccccc}
 0 & -1 & -1 & -1 & -1 & 0 & 0 \\
 -1 & 0 & 0 & 0 & 0 & 0 & 0 \\
 1 & 1 & 1 & 1 & 0 & 0 & 0 \\
 0 & 0 & 0 & 0 & 1 & 1 & 1 \\
 0 & 0 & -1 & -1 & -1 & -1 & -1 \\
 0 & 0 & 1 & 1 & 1 & 1 & 0 \\
 0 & 0 & 0 & -1 & -1 & -1 & 0 \\
\end{array}
\right)\nonumber\\ B & = &\left(
\begin{array}{ccccccc}
 0 & 0 & 0 & 1 & 1 & 1 & 0 \\
 0 & 0 & 0 & 0 & -1 & -1 & 0 \\
 -1 & -1 & -1 & -1 & 0 & 0 & 0 \\
 1 & 1 & 0 & 0 & 0 & 0 & 0 \\
 0 & 0 & 1 & 1 & 1 & 1 & 1 \\
 0 & -1 & -1 & -1 & -1 & -1 & -1 \\
 0 & 1 & 1 & 1 & 1 & 0 & 0 \\
\end{array}
\right)
\end{eqnarray}
} and the vector $\mathbf{v}_0$ invariant  under the group they generate is the
following one: {\scriptsize
\begin{equation}\label{inva28}
    \mathbf{v}_0 \, = \,\{n, -n, 0, 0, 0, n, -n\}
\end{equation}
} The corresponding $\mathrm{L_{168}}$-orbit is displayed below{\scriptsize
\begin{equation}\label{orbit28}
 O_{28} \,= \,   \left\{
\begin{array}{ccccccc}
 0 & 0 & 0 & 0 & 0 & n & 0 \\
 0 & 0 & 0 & 0 & n & 0 & -n \\
 0 & 0 & 0 & 0 & n & -n & n \\
 0 & 0 & 0 & n & 0 & -n & 0 \\
 0 & 0 & 0 & n & -n & 0 & n \\
 0 & 0 & 0 & n & -n & n & -n \\
 0 & 0 & n & 0 & -n & 0 & 0 \\
 0 & 0 & n & -n & 0 & 0 & n \\
 0 & 0 & n & -n & 0 & n & -n \\
 0 & 0 & n & -n & n & -n & 0 \\
 0 & -n & 0 & 0 & 0 & 0 & 0 \\
 0 & n & 0 & -n & 0 & 0 & 0 \\
 0 & n & -n & 0 & 0 & 0 & n \\
 0 & n & -n & 0 & 0 & n & -n \\
 0 & n & -n & 0 & n & -n & 0 \\
 0 & n & -n & n & -n & 0 & 0 \\
 -n & 0 & 0 & 0 & 0 & 0 & n \\
 -n & 0 & 0 & 0 & 0 & n & -n \\
 -n & 0 & 0 & 0 & n & -n & 0 \\
 -n & 0 & 0 & n & -n & 0 & 0 \\
 -n & 0 & n & -n & 0 & 0 & 0 \\
 -n & n & -n & 0 & 0 & 0 & 0 \\
 n & 0 & -n & 0 & 0 & 0 & 0 \\
 n & -n & 0 & 0 & 0 & 0 & n \\
 n & -n & 0 & 0 & 0 & n & -n \\
 n & -n & 0 & 0 & n & -n & 0 \\
 n & -n & 0 & n & -n & 0 & 0 \\
 n & -n & n & -n & 0 & 0 & 0 \\
\end{array}
\right\}
\end{equation}
}
\item Next we have verified that there are two conjugacy classes of groups
H = $\mathbb{Z}_2\times\mathbb{Z}_2$. Yet any vector invariant with respect to
any  $\mathrm{H}\, = \,\mathbb{Z}_2\times \mathbb{Z}_2$ is automatically
invariant with respect to    the tetrahedral group $\mathrm{T_{12}}$ of which H
is a subgroup. This implies that we can obtain orbits of length 42 only from a
stability subgroup $ \mathrm{H} \, = \, \mathbb{Z}_4$. It also implies that
there cannot be orbits of length 21 with stability subgroup $\mathbb{Z}_2\times
\mathbb{Z}_2 \times \mathbb{Z}_2$.
\item Proceeding further we have verified  that there is a unique conjugacy class of $\mathbb{Z}_4$ subgroups.
Hence there is a unique type of orbits of length 42. They depend on a single
integer parameter $n$. Taking as generator of the $\mathbb{Z}_\mathrm{4}$ group
the following matrix: {\scriptsize
\begin{equation}\label{z4genero}
    g_{Z_4} \, = \, \left(
\begin{array}{ccccccc}
 -1 & -1 & -1 & -1 & -1 & -1 & -1 \\
 1 & 0 & 0 & 0 & 0 & 0 & 0 \\
 0 & 1 & 1 & 1 & 1 & 1 & 0 \\
 0 & 0 & 0 & 0 & 0 & -1 & 0 \\
 0 & 0 & 0 & -1 & -1 & 0 & 0 \\
 0 & 0 & 0 & 1 & 0 & 0 & 0 \\
 0 & 0 & -1 & -1 & 0 & 0 & 0 \\
\end{array}
\right)
\end{equation}}
the vector invariant under the action of such a $\mathbb{Z}_\mathrm{4}$ is the
following one: {\scriptsize
\begin{equation}\label{inva42}
  \mathbf{v}_0 \, = \,   \{0, 0, n, 0, 0, 0, -n\}
\end{equation}
 }
and the corresponding $\mathrm{L_{168}}$-orbit is displayed below: {\scriptsize
\begin{equation}\label{orbita42}
    \mathrm{O}_{42} \, = \, \left(
\begin{array}{ccccccc}
 0 & 0 & 0 & -n & 0 & 0 & 0 \\
 0 & 0 & 0 & n & 0 & 0 & 0 \\
 0 & 0 & -n & 0 & 0 & 0 & n \\
 0 & 0 & -n & 0 & 0 & n & -n \\
 0 & 0 & n & 0 & 0 & 0 & -n \\
 0 & 0 & n & 0 & 0 & -n & n \\
 0 & -n & 0 & 0 & 0 & n & 0 \\
 0 & -n & 0 & 0 & n & 0 & -n \\
 0 & -n & 0 & n & 0 & -n & 0 \\
 0 & -n & 0 & n & -n & 0 & n \\
 0 & -n & n & 0 & -n & 0 & 0 \\
 0 & -n & n & -n & n & -n & 0 \\
 0 & n & 0 & 0 & 0 & -n & 0 \\
 0 & n & 0 & 0 & -n & 0 & n \\
 0 & n & 0 & -n & 0 & n & 0 \\
 0 & n & 0 & -n & n & 0 & -n \\
 0 & n & -n & 0 & n & 0 & 0 \\
 0 & n & -n & n & -n & n & 0 \\
 -n & 0 & 0 & 0 & n & 0 & 0 \\
 -n & 0 & 0 & n & 0 & 0 & -n \\
 -n & 0 & n & 0 & -n & 0 & n \\
 -n & 0 & n & 0 & -n & n & -n \\
 -n & 0 & n & -n & 0 & n & 0 \\
 -n & 0 & n & -n & n & -n & n \\
 -n & n & 0 & -n & 0 & n & -n \\
 -n & n & 0 & -n & n & -n & 0 \\
 -n & n & -n & 0 & n & -n & n \\
 -n & n & -n & n & 0 & -n & 0 \\
 -n & n & -n & n & -n & 0 & n \\
 -n & n & -n & n & -n & n & -n \\
 n & 0 & 0 & 0 & -n & 0 & 0 \\
 n & 0 & 0 & -n & 0 & 0 & n \\
 n & 0 & -n & 0 & n & 0 & -n \\
 n & 0 & -n & 0 & n & -n & n \\
 n & 0 & -n & n & 0 & -n & 0 \\
 n & 0 & -n & n & -n & n & -n \\
 n & -n & 0 & n & 0 & -n & n \\
 n & -n & 0 & n & -n & n & 0 \\
 n & -n & n & 0 & -n & n & -n \\
 n & -n & n & -n & 0 & n & 0 \\
 n & -n & n & -n & n & 0 & -n \\
 n & -n & n & -n & n & -n & n \\
\end{array}
\right)
\end{equation}
}
\item In the next step we have verified that there is a
\textit{ unique conjugacy class of subgroups}  $\mathbb{Z}_3$.
This implies that there is a \textit{ unique type} of orbits of  length 56.
These depend on three integer parameters $ n,m,p.$ Indeed, taking as
$\mathbb{Z}_3$-generator the following element of $\mathrm{L_{168}}$:
{\scriptsize
\begin{equation}\label{z3generale}
    g_{Z3} \, = \, \left(
\begin{array}{ccccccc}
 0 & 1 & 0 & 0 & 0 & 0 & 0 \\
 -1 & -1 & 0 & 0 & 0 & 0 & 0 \\
 1 & 1 & 1 & 1 & 1 & 1 & 0 \\
 0 & 0 & 0 & 0 & -1 & -1 & 0 \\
 0 & 0 & 0 & 0 & 1 & 0 & 0 \\
 0 & 0 & 0 & 0 & 0 & 1 & 1 \\
 0 & 0 & 0 & -1 & -1 & -1 & -1 \\
\end{array}
\right)
\end{equation}
} the vector invariant under $\mathbb{Z}_3$ is the following one: {\scriptsize
\begin{equation}\label{inva56}
  v_0 \, = \   \{0,0,n,m,p,-m-p,0\}
\end{equation}
} and the corresponding orbit is displayed in fig.\ref{figura56}
\item Furthermore we verified that up to conjugation there is only  one $\mathbb{Z}_7$
subgroup of $\mathrm{L_{168}}$ and that any vector $\mathbf{v}$ that is
invariant with respect to this $\mathbb{Z}_7$ is also invariant with respect to
the $\mathrm{G_{21}}$ group  which contains it. \textit{Hence there are no
orbits of length 24}.
\item Furthermore we verified that \textit{there is no} $\mathbb{Z}_2\times \mathbb{Z}_4$ subgroup
of $\mathrm{O_{24}}$ and hence of $\mathrm{L_{168}}$. Hence \textit{orbits of
length 21 do not exist}.
\item
Finally, since in $\mathrm{L_{168}}$ there is a unique conjugacy class of
elements of order 2, it follows that there is a unique conjugacy class of
$\mathbb{Z}_2$ subgroups. Hence there is a unique type of orbits of length 84,
depending on 3 integer parameters  $ n,m,p$. To this effect it suffices to take
anyone of the 21
 elements belonging to the second conjugacy class of $\mathrm{L_{168}}$ as $\mathbb{Z}_2$--generator and the results
 follows. The orbit is shown in fig.\ref{figura84}.
\end{enumerate}
\begin{figure}[!hbt]
\begin{center}
\iffigs
\includegraphics[height=130mm]{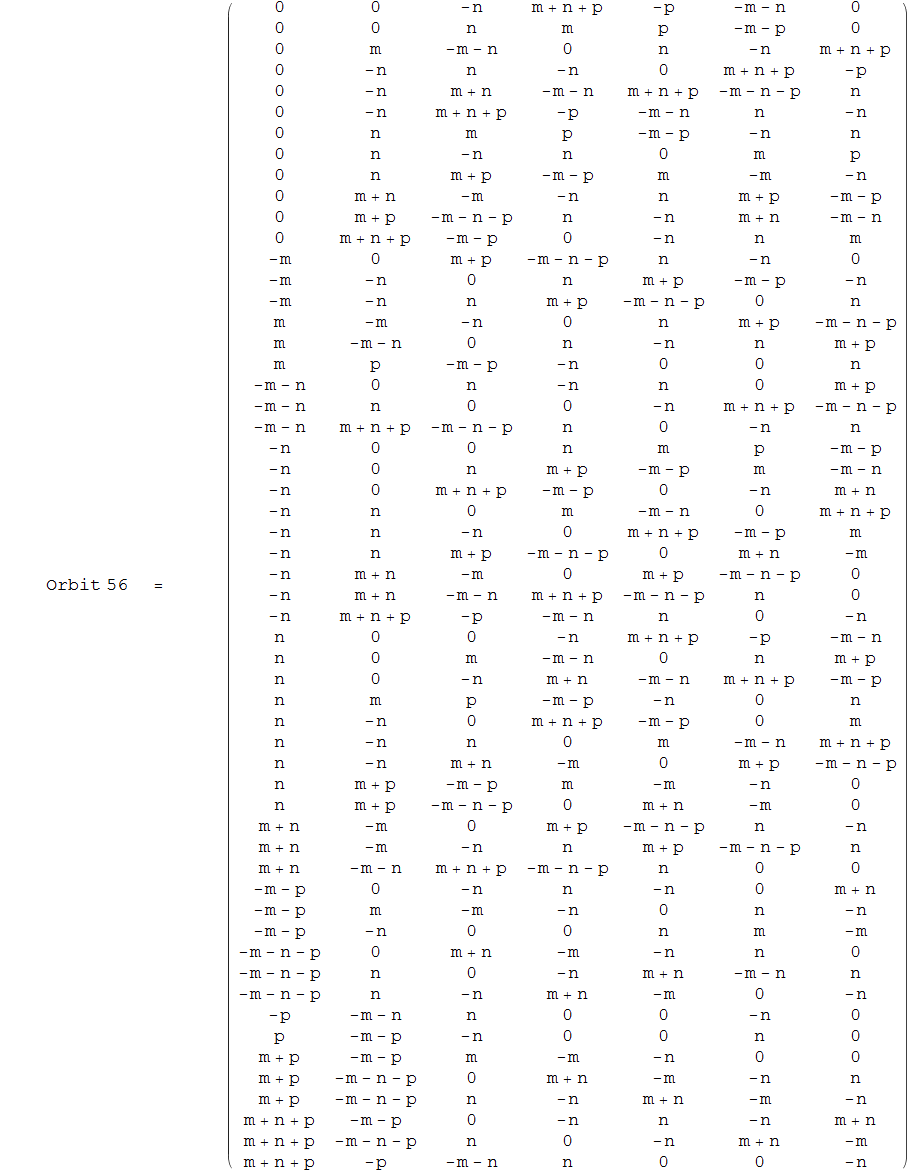}
\else
\end{center}
 \fi
\caption{\it The  vectors belonging to the length 56 orbit of
$\mathrm{L_{162}}$  in the A7 weight lattice\label{figura56}}
 \iffigs
 \hskip 1cm \unitlength=1.1mm
 \end{center}
  \fi
\end{figure}
\subsubsection{Synopsis of the $\mathrm{L_{168}}$ orbits in the weight lattice $\Lambda_w$}
Our findings about the available $\mathrm{L_{168}}$-orbits in the weight
lattice are summarized below
\begin{enumerate}
\item Orbits of length 8 (one parameter\pmb{  n}; stability subgroup $\mathrm{H}^s=\mathrm{G_{21}}$)
\item Orbits of length 14 (two types A $\&$ B) (one parameter\pmb{  n}; stability subgroup
$\mathrm{H}^s=\mathrm{T_{12A,B}}$)
\item Orbits of length 28 (one parameter\pmb{  n} ; stability subgroup
$\mathrm{H}^s=\mathrm{Dih_{3}}$)
\item Orbits of length 42 (one parameter\pmb{  n}; stability subgroup
$\mathrm{H}^s=\mathbb{Z}_4$) )
\item Orbits of length 56 (three parameters \pmb{ n,m,p}; stability subgroup
$\mathrm{H}^s=\mathbb{Z}_3$)
\item Orbits of length 84 (three parameters \pmb{ n,m,p}; stability subgroup
$\mathrm{H}^s=\mathbb{Z}_2$)
\item Generic orbits of length 168 (seven parameters ; stability subgroup $\mathrm{H}^s=\mathbf{1}$)
\end{enumerate}
\begin{figure}[!hbt]
\begin{center}
\iffigs
\includegraphics[height=130mm]{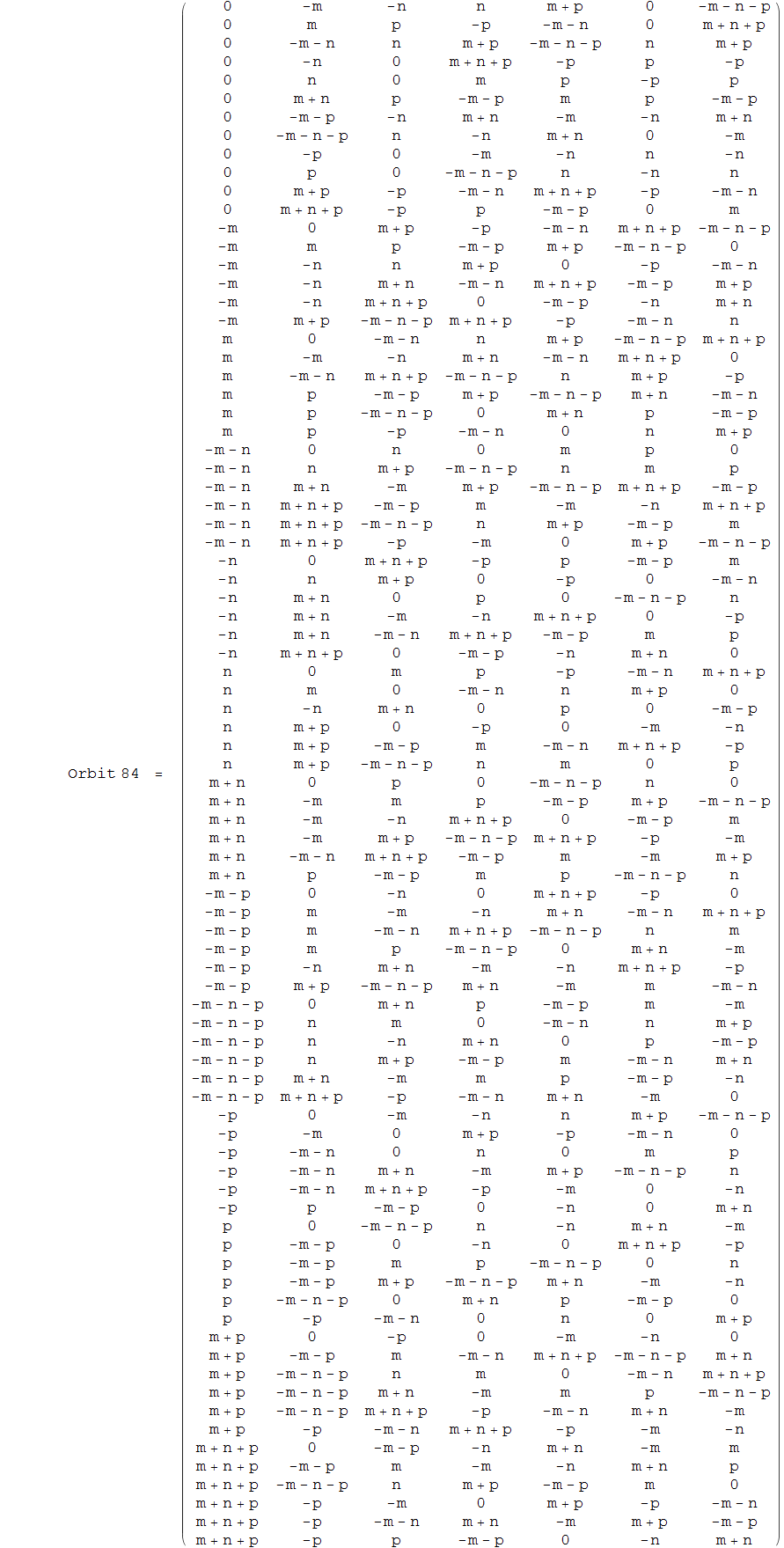}
\else
\end{center}
 \fi
\caption{\it The  vectors belonging to the length 84 orbit of
$\mathrm{L_{162}}$  in the A7 weight lattice\label{figura84}}
 \iffigs
 \hskip 1cm \unitlength=1.1mm
 \end{center}
  \fi
\end{figure}
%%%%%%%%%%%%%%%%%%%%%%%%%%%%%%%%%%%%%
\section{Solutions of Englert Equation associated with $\mathrm{L_{168}}$ orbits in the weight lattice}
Let us now generalize the solution algorithm introduced in \cite{arnolderie}
for Beltrami equation on the crystallographic $3$-torus:
\begin{equation}\label{7vitelli}
 \mathrm{ T^3} \,  \equiv \, \frac{\mathbb{R}^3}{\Lambda_{\mathrm{cubic}}}
\end{equation}
to the case of  Englert equation (\ref{englerta}) on the  crystallographic
$7$-torus eq.(\ref{7mucche}). To this effect, let us choose as line element on
the crystallographic torus (\ref{7mucche}) the following one:
\begin{equation}\label{lineus}
  ds^2_{\mathrm{T^7}} \, = \, \mathcal{C}_{ij} \, dX^i \otimes dX^j
\end{equation}
where $\mathcal{C}_{ij}$ is the Cartan matrix (\ref{cartanschula}). In relation
with the notations of section \ref{maxello} we have set:
\begin{equation}\label{caratazzo}
  x^i \, = \,  \alpha^i_{\phantom{i}j} \, X^j  \quad \Rightarrow \quad x \, = \, \mathfrak{M} \, X
\end{equation}
where $\alpha^i_{\phantom{i}j}$ denotes the $i$-th component, in an orthonormal
basis, of the $j$-th simple root, an explicit realization of which is indeed
provided by the corresponding  entry of the matrix $\mathfrak{M}$ introduced in
equation (\ref{mgothica}).
\par
Secondly let us introduce the following ansatz:
\begin{eqnarray}\label{unsuzzo}
  \mathbf{Y}^{[3]}& = & \sum_{\mathbf{k}\in \mathcal{O}}\,
  \left(v_{ijk}\left(\mathbf{k}\right)\, \cos\left[2\, \pi \,
  \mathbf{k} \cdot \mathbf{X}\right] \, + \, \omega_{ijk}\left(\mathbf{k}\right)\, \sin\left[2\, \pi \,
  \mathbf{k} \cdot \mathbf{X}\right] \right)\, dX^i \wedge dX^j \wedge dX^k \nonumber\\
  & \equiv & \mathcal{Y}_{ijk}(\mathbf{X})\, dX^i \wedge dX^j \wedge dX^k
  \, = \, {Y}_{ijk}(\mathbf{x})\, dx^i \wedge dx^j \wedge
  dx^k \label{carluccioB}
\end{eqnarray}
where $\mathcal{O}$ denotes some $\mathrm{L_{168}}$ orbit of momentum vectors
in the weight lattice $\Lambda_{\mathrm{weight}}$ and where:
 \begin{equation}\label{formichiere}
   \mathbf{k} \cdot \mathbf{X} \, \equiv \, k_\ell \, X^\ell
 \end{equation}
 Eq.
(\ref{unsuzzo}) defines also the relation between the tensors
${Y}_{ijk}(\mathbf{x})$ and $ \mathcal{Y}_{ijk}(\mathbf{X})$, namely:
\begin{equation}\label{casseruola}
{Y}_{ijk} \, = \, \left(\mathfrak{ M}^{-1}\right)_{i}^{\phantom{i}i^\prime}\,
\left(\mathfrak{ M}^{-1}\right)_{j}^{\phantom{j}j^\prime}\, \left(\mathfrak{
M}^{-1}\right)_{k}^{\phantom{k}k^\prime} \, \mathcal{Y}_{i^\prime j^\prime
k^\prime}
\end{equation}
 With such an ansatz, Englert equation (\ref{englerta}) is turned into the following  pair of algebraic equations for the coefficients:
\begin{eqnarray}
\sqrt{\mbox{det } \mathcal{C}} \,\,  \epsilon_{ijk}^{\phantom{ijk}\ell mnp}
\, k_\ell \, v_{mnp} &=& - \, \frac{6\,\mu}{\pi} \, \omega_{ijk} \\
\sqrt{\mbox{det } \mathcal{C}} \,\,   \epsilon_{ijk}^{\phantom{ijk}\ell mnp}\,
k_\ell \, \omega_{mnp} &=&  \, \frac{6\,\mu}{\pi} \, v_{ijk}\label{prosciutto}
\end{eqnarray}
where four   indices of the Levi-Civita epsilon symbol have being raised with
the inverse metric $\mathcal{C}^{-1}$. Substituting the first equation into the
second we obtain the following consistency conditions:
\begin{eqnarray}
\mu^2 &=& \pi^2
\parallel
\mathbf{k}
\parallel^2
\quad ; \quad
\parallel
\mathbf{k}
\parallel^2
\, = \, k_\ell \, k_m \, \left(\mathcal{C}^{-1}\right)^{\ell m}
\label{coppa}\\
0 & = & \left(\mathcal{C}^{-1}\right)^{\ell m} k_\ell \, v_{ijm} \,=\,
\left(\mathcal{C}^{-1}\right)^{\ell m} k_\ell \, \omega_{ijm} \label{bresaola}
\end{eqnarray}
It is easy to count the number of parameters in the general solution of
(\ref{prosciutto}). There are a priori $35+35=70$ parameters for each momentum
$\mathbf{k}$. Equation (\ref{bresaola}) imposes $21+21=42$ constraints.
 These latter are not all independent since the $7+7$ equations:
 \begin{equation}\label{lapizza}
   \left(\mathcal{C}^{-1}\right)^{\ell m} k_\ell \,
   \left(\mathcal{C}^{-1}\right)^{r n} \, k_r \, v_{inm} \, = \,
   \left(\mathcal{C}^{-1}\right)^{\ell m} k_\ell \,
   \left(\mathcal{C}^{-1}\right)^{r n} \, k_r \, \omega_{inm} \, = \, 0
 \end{equation}
 are automatically satisfied because of antisymmetry of the involved three-tensors.
 On their turn $1+1$ of the above equations
follow  from antisymmetry contracting once more with the momentum vector. In
conclusion the number of independent parameters for each momentum vector is
reduced by eq.(\ref{bresaola}) to
\begin{equation}\label{sinuhe}
   (35 -21 + 7 -1)\oplus (35 -21 + 7 -1) \, = \, 20 \oplus 20
\end{equation}
 Finally eq.(\ref{prosciutto}) halves this number so that for each momentum vector we have:
 \begin{equation}\label{egiziano}
  \# \, \mbox{parameters} \,= \,  20
\end{equation}
We conclude that the number of parameters in a solution of Englert equation
based on an orbit $\mathcal{O}$ is
\begin{equation}
np=20 \times |\mathcal{O}|
\end{equation}
where $|\mathcal{O}|$ denotes the number of different weights contained in the
considered orbit, counting weights that differ  by an overall sign only once.
The last specification is essential since cosine and sine are, respectively, an
even and an odd function  and we should not count $\cos\left[\pm 2\, \pi
\,\mathbf{k} \cdot \mathbf{X}\right]$ and $\sin\left[\pm 2\, \pi \,\mathbf{k}
\cdot \mathbf{X}\right]$ twice.
\par
In this way from each orbit $\mathcal{O}$ we obtain a $3$-form
\begin{equation}\label{solibrio}
  \mathbf{Y}^{[3]}_{\mathcal{O}}\left(\mathbf{X}|\mathbf{F}\right)
\end{equation}
which satisfies Englert equation and depends on a set of $np$-parameters $F_1 ,
\dots,F_{np}$ collectively denoted $\mathbf{F}$.
\par The
action of the crystallographic group $\mathrm{L_{168}}$ can be easily
transferred from the torus coordinates to the parameter space solving the
following linear equations:
\begin{equation}\label{pietradiavola}
\forall \,\gamma \in \mathrm{L_{168}} \quad : \quad
\mathbf{Y}^{[3]}_{\mathcal{O}}\left(\text{D}_7(\gamma) \cdot \mathbf{X}|
\mathbf{F}\right) \, = \, \mathbf{Y}^{[3]}_{\mathcal{O}}\left(
\mathbf{X}|\mathfrak{D}_{np}\left(\gamma\right) \cdot \mathbf{F}\right)
\end{equation}
 for the $np^2$ entries of the matrices
$\mathfrak{D}_{np}(\gamma)$. In this way one constructs a $np$-dimensional
linear representation of the group $\mathrm{L_{168}}$ which can always be
decomposed into irreps using the character table in eq.(\ref{caratterini}).
\section{A $\mathrm{L_{168}}$-invariant Englert $3$-form from the
orbit $\mathcal{O}_8$}
\label{168invariante}
As a first exemplification of the procedure  let us apply the described
algorithm to the case of the shortest orbit $\mathcal{O}_8$
displayed in eq.(\ref{orbita8}). Utilizing a Mathematica
code we produced the object:
\begin{equation}\label{oggetto}
  \mathbf{Y}^{[3]}_{\mathcal{O}_8}\left(\mathbf{X}|\mathbf{F}_{160}\right)
\end{equation} depending on $160$ real parameters and we
easily constructed the $160\times 160$ matrices of the
reducible representation $\mathfrak{D}_{160}$. Then we
calculated the traces of these matrices in each of the
conjugation classes ordered as in table \ref{coniugini} and
we obtained the following character:
\begin{equation}\label{brutcarat} \chi \, \equiv \,
\chi\left[\mathfrak{D}_{160}\right] \, = \,
\{160,0,4,0,-1,-1\} \end{equation} The multiplicities of
the 6 irreducible representation follow immediately from
the general formula: \begin{equation}\label{ragliata} a_\mu
\, = \, \langle \chi \, , \, \chi_\mu\rangle \end{equation}
where the scalar product in character space is defined as
follows:
\begin{eqnarray}\label{scalaruso} \langle \chi \, , \, \psi
\rangle & \equiv& \frac{1}{168} \sum_{i=1}^6 g_i \, \chi^i
\,
\psi^i \, = \, \chi^i \, \kappa_{ij} \, \psi^j\nonumber\\
\kappa & = & \left( \begin{array}{cccccc} \frac{1}{168} & 0 & 0 & 0 & 0 & 0 \\
0 & \frac{1}{8} & 0 & 0 & 0 & 0 \\ 0 & 0 &
\frac{1}{3} & 0 & 0 & 0 \\ 0 & 0 & 0 & \frac{1}{4} & 0 & 0 \\
0 & 0 & 0 & 0 & \frac{1}{7} & 0 \\ 0 & 0 & 0 & 0 & 0 & \frac{1}{7} \\
\end{array} \right) \end{eqnarray} the number $g_i$ denoting the length of the
conjugacy class $\mathcal{C}_i$. The result is the
following multiplicity vector:
\begin{equation}\label{multipiazza} a_{\mu} \, = \,
\{2,6,8,6,3,3\}
\end{equation} which corresponds to the following decomposition into irreps:
\begin{equation}\label{decompolo}
 \mathfrak{D}_\mathrm{160} \, = \,  2 \, \text{D}_1\oplus 6 \,\text{D}_6\oplus8
 \,\text{D}_7\oplus6\, \text{D}_8\oplus3\, \text{D}_{3a}
 \oplus 3 \,\text{D}_{3b}
\end{equation}
We conclude that there exists a $2$-parameter solution of
Englert equation, which is invariant under the full group
$\mathrm{L_{168}}$. It corresponds to the projection onto
the $\mathrm{D}_1$ representation in eq.(\ref{decompolo}).
\par
The projectors onto irreducible $\mathrm{D}_\mu$ can
be obtained by means of another classical formula of finite
group theory:
\begin{equation}\label{pruiutto} \mathbb{P}^{\mu} \, = \, \ft {1}{168}\sum_{i=1}^6 \chi^{\mu}_i \, \sum_{\gamma
\in \mathcal{C}_i} \, \mathfrak{D}_{160}\left(\gamma\right)
\end{equation} Applying $\mathbb{P}^{[D_1]}$ to the parameter vector
$\mathbf{F}_{160}$ we set to zero $158$ linear indipendent
combinations of the $F_i$ and the result is a
$2$-parameter three--form.  We have explicitly verified that it is
invariant under the full group $\mathrm{L_{168}}$ .
\par
In order to display the explcit form of this solution we introduce
the following set of 16 linearly indipendent trigonometric functions
\begin{equation}\label{casolare}
\mathfrak{f}_\alpha(\mathbf{X})  \, = \, \left(
\begin{array}{c}
 \cos (2 \pi  {X_1}) \\
 \cos (2 \pi  ({X_2}-{X_1})) \\
 \cos (2 \pi  ({X_3}-{X_2})) \\
 \cos (2 \pi  ({X_4}-{X_3})) \\
 \cos (2 \pi  ({X_5}-{X_4})) \\
 \cos (2 \pi  ({X_6}-{X_5})) \\
 \cos (2 \pi  {X_7}) \\
 \cos (2 \pi  ({X_7}-{X_6})) \\
 \sin (2 \pi  {X_1}) \\
 \sin (2 \pi  ({X_2}-{X_1})) \\
 \sin (2 \pi  ({X_3}-{X_2})) \\
 \sin (2 \pi  ({X_4}-{X_3})) \\
 \sin (2 \pi  ({X_5}-{X_4})) \\
 \sin (2 \pi  ({X_6}-{X_5})) \\
 -\sin (2 \pi  {X_7}) \\
 \sin (2 \pi  ({X_7}-{X_6})) \\
\end{array}
\right) \quad ; \quad \alpha = 1,\dots, 16
\end{equation}
and we organize  the 35 indipendent differentials of the integral coordinates into lexicographic order:
\begin{equation}\label{35deltoni}
\pmb{\Delta } = \left\{ \text{dX}_1\wedge \text{dX}_2\wedge
\text{dX}_3,\text{dX}_1\wedge \text{dX}_2\wedge
\text{dX}_4,\text{...}\text{...},\text{dX}_5\wedge
\text{dX}_6\wedge \text{dX}_7\right\}\, =\,\{\Delta
_{\mathfrak{q}}\} \,\, , \,\,
(\mathfrak{q}=1,\text{...},35)
\end{equation}
Then the $\mathrm{L_{168}}$-invariant Englert 3-form can be written as follows
\begin{equation}\label{168threeforma}
 \mathbf{Y}^{[168]}\pmb{(X|F)}\, = \, \sum_{\alpha=1}^{16} \sum_{\mathfrak{q}=1}^{35}
 \,\mathfrak{C}^{[168]}_{\mathfrak{q}\alpha}\left(\hat{\pmb{F}}\right) \Delta
_{\mathfrak{q}} \times \mathfrak{f}_{\alpha }(\pmb{X})
\end{equation}
%%%%%%%%%%%%%%
where
\begin{equation}\label{2modulini}
  \hat{\pmb{F}}\,= \, \{F_1,F_2\}
\end{equation}
and the the coefficients $\mathfrak{C}^{[168]}_{\mathfrak{q}\alpha}\left(\hat{\pmb{F}}\right)$ are displayed in appendix \ref{mostro}.
\par
We have carefully studied the $\mathcal{B}$-operator defined in eq.(\ref{operati}) when it is polarized on $\mathbf{Y}^{[168]}\pmb{(X|F)}$. To this effect one has to convert $\mathbf{Y}^{[168]}\pmb{(X|F)}$ to the orthonormal coordinates and then use its redefined components in eq.(\ref{operati}). We have verified that for no choice of the parameter $(F_1,F_2)$, the rank of $\mathcal{B}^{[168]}$ can be reduced to be smaller than 8. Hence at least from the orbit $\mathcal{O}_8$ no solution emerges of Englert equation that is $\mathrm{L_{168}}$-invariant and admits some residual supersymmetry.
%%%%%%%%%%%%%%%%%%%%%%%%%%%
\section{Construction of a $\mathrm{G_{21}}$-invariant solution of Englert equation}
Motivated by the results of the previous section we have made a new searching for an Englert solution with $\mathcal{N}=1$ supersymmetry and a reduced discrete symmetry $\mathrm{H}\subset \mathrm{L_{168}}$.
\par
An educated guess suggests that we should rather use the maximal subgroup $\mathrm{G_{21}}$. Indeed this latter contains a $\mathbb{Z}_7$ subgroup and no lattice of dimension less than seven can have a crystallographic realization of $\mathbb{Z}_7$. This encourages to think that any solution of Englert equation invariant under $\mathrm{G_{21}}$ is intrinsically 7-dimensional. Since we could not find an $\mathrm{L_{168}}$-invariant solution with $\mathcal{N}=1$ supersymmetry, the next obvious possibility is to try with the maximal subgroup $\mathrm{G_{21}}$.
\par
It turns out that our guess is correct and in this section we construct a $\mathrm{G_{21}}$-invariant  solution of Englert equation which admits $\mathcal{N}=1$ SUSY.
\subsection{The $\mathrm{G_{21}}$-orbit of length 7 in the weight lattice and
its associated solution of Englert equation} The orbit
$\mathcal{O}_7 \subset \Lambda _w$ of the group
$\mathrm{G_{21}}$ in the weight lattice that we consider is
made of 7 vectors and it is the displayed below:
\begin{equation}\label{orbita7nostra}
  \mathcal{O}_7=\left(
\begin{array}{ccccccc}
 -1 & 1 & 0 & 0 & 0 & 0 & 0 \\
 0 & -1 & 1 & 0 & 0 & 0 & 0 \\
 0 & 0 & -1 & 1 & 0 & 0 & 0 \\
 0 & 0 & 0 & 0 & -1 & 1 & 0 \\
 0 & 0 & 0 & 0 & 0 & -1 & 1 \\
 0 & 0 & 0 & 0 & 0 & 0 & -1 \\
 1 & 0 & 0 & 0 & 0 & 0 & 0 \\
\end{array}
\right)
\end{equation}
Using this orbit in the general construction algorithm (see
eq.(\ref{carluccioB}) and following ones) we construct a
solution of Englert equation (\ref{englertaF})  with
eigenvalue:
\begin{equation}\label{muvallo1}
  \mu  =\text{ }\frac{1}{2} \sqrt{\frac{7}{2}} \pi
\end{equation}
By computer calculation we find a solution solution that
depends on a set of 140 parameters:
\begin{equation}\label{effone}
\pmb{F} \,= \, \{F_1,\dots , F_{140}\}\quad ; \quad F_A
\,\,(A=1,\dots,140)
\end{equation}
and on a set of 14 independent trigonometric functions:
\begin{equation}\label{sakkopatat}
    \mathit{f}_{\alpha }(\pmb{X})\, = \, \left(
\begin{array}{c}
 \text{Cos}\left[2 \pi  X_1\right] \\
 \text{Cos}\left[2 \pi  \left(-X_1+X_2\right)\right] \\
 \text{Cos}\left[2 \pi  \left(-X_2+X_3\right)\right] \\
 \text{Cos}\left[2 \pi  \left(-X_3+X_4\right)\right] \\
 \text{Cos}\left[2 \pi  \left(-X_5+X_6\right)\right] \\
 \text{Cos}\left[2 \pi  X_7\right] \\
 \text{Cos}\left[2 \pi  \left(-X_6+X_7\right)\right] \\
 \text{Sin}\left[2 \pi  X_1\right] \\
 \text{Sin}\left[2 \pi  \left(-X_1+X_2\right)\right] \\
 \text{Sin}\left[2 \pi  \left(-X_2+X_3\right)\right] \\
 \text{Sin}\left[2 \pi  \left(-X_3+X_4\right)\right] \\
 \text{Sin}\left[2 \pi  \left(-X_5+X_6\right)\right] \\
 -\text{Sin}\left[2 \pi  X_7\right] \\
 \text{Sin}\left[2 \pi  \left(-X_6+X_7\right)\right] \\
\end{array}
\right)
\end{equation}
This solution of Englert equation can be written as follows:
\begin{equation}\label{Cgotico}
 \pmb{Y(X|F)}\, = \, \sum_{\alpha=1}^{14} \sum_{\mathfrak{q}=1}^{35}
 \,\mathfrak{C}_{\mathfrak{q}\alpha}\left(\pmb{F}\right) \Delta
_{\mathfrak{q}} \times \mathit{f}_{\alpha }(\pmb{X})
\end{equation}
where the differentials $\Delta_{\mathfrak{q}}$ were defined in eq.(\ref{35deltoni}) and
where the 35$\times $14=490 coefficients
$\mathfrak{C}_{\mathfrak{q}\alpha }(\pmb{ F})$ are linear
combinations of the the $F_A$ which we do not display for
obvious reasons of space, since they are very large
expressions.
\par
Next we have derived the 140-dimensional representation
$\mathfrak{D}_{140}$ of the group $\mathrm{G_{21}}\subset
\mathrm{L_{168}}$ induced on the parameter space by
the standard identity (see eq.(\ref{pietradiavola})):
\begin{equation}\label{consillo}
  \forall\gamma\in \mathrm{G_{21}}\quad : \quad
  \mathbf{Y}\left(\gamma\,\pmb{X}|\pmb{F}\right)\,=\,
\mathbf{Y}\left(\pmb{X}|{\mathfrak{D}}_{140}[\gamma]\pmb{F}\right)
\end{equation}
The character of this representation turns out to be:
\begin{equation}\label{carattere140}
 \chi ^{[140]}\text{ }=\text{ }\{140,0,0,2,2\}
\end{equation}
and by applying to $\mathfrak{D}_{140}$ the standard
group-theoretical formula that provides the multiplicity of
any irrep in a given reducible representation
\begin{equation}\label{multipilini}
    a^{\mu }\, =\, \ft{1}{21} \,\sum _{i=1}^5 g_i \bar{\chi }_i^{\mu }\chi _i^{[140]}
\end{equation}
using for $\bar{\chi }_i^{\mu }$ the character table in
eq.(\ref{bruttocarattere21}), we found the following
decomposition of $ \mathfrak{D}_{140}$ into irreducible
irreps of $\mathrm{G_{21}}$:
\begin{equation}\label{140decompo}
\mathfrak{D}_{140}\left[\mathrm{G_{21}}\right]\, =\,8 \,
\mathrm{D}_1\left[\mathrm{G_{21}}\right]\oplus 20 \,
\text{DA}_3\left[\mathrm{G_{21}}\right]\oplus 20 \,
\text{DB}_3\left[\mathrm{G_{21}}\right]\oplus 6\,
\text{DX}_1\left[\mathrm{G_{21}}\right]\oplus 6 \,
\text{DY}_1\left[\mathrm{G_{21}}\right]
\end{equation}
the notations being those of eq.(\ref{bruttocarattere21}).
\par
This means that from our constructed solution
\pmb{Y(X$|$F)} we can extract 8 singlets, namely an Englert
solution, invariant under $\mathrm{G_{21}}$ which depends
on 8 parameters. This solution can be easily obtained
utilizing
 the group-theoretical projection onto any
irreducible representation $\mathrm{D}^{\mu }$ encoded into
the following formula:
\begin{equation}\label{proiettokino}
    \mathbb{P}_{140}^{[\mu ]}\, =\, \frac{1}{21}\sum _{i=1}^5 \chi _i^{\mu }\sum _{\gamma \in \mathcal{C}_i}
\,\mathfrak{D}_{140}[\gamma ]
\end{equation}
where $\chi _i^{\mu }$ is the i-th compoment of the
$\mu$-character and $\mathcal{C}_i$ is the i-th conjugacy
class of group elements. The 132 dimensional null-space of
$\mathbb{P}_{140}^{\left[\mathrm{D_1}\right]}$ provides us
with 132 linear constraints on the parameters $F_A$ that
can be solved in terms of 8 parameters. We named these
parameters as follows:
\begin{equation}\label{felicione}
   \mathbf{\Psi}  =\left\{\psi_1,\dots,\psi_8\right\}
\end{equation}
The substitution of the solution of these constraints into
eq.(\ref{Cgotico}) produces a 3-form
$\mathbf{Y}_s(\mathbf{X}|\mathbf{\Psi})$ where the
subscript \textit{s} stands for singlet (with respect to
$\mathrm{G_{21}}$).
\subsection{Conversion of the $\mathrm{G_{21}}$-invariant
solution obtained in the root basis to the orthonormal
basis of coordinates} The next step in our construction is
the transcription of $\pmb{Y}_s(\pmb{X} |\pmb{\Psi}) $ that
was constructed in the integral coordinates \pmb{X},
adapted to the A7-root lattice, to the orthonormal
coordinates \pmb{x}, best suited to the construction of
supergravity solutions. Applying eq.(\ref{caratazzo}) we
obtain:
\begin{equation}\label{cazillo}
    \hat{\mathbf{Y}}_s\left(\pmb{x}|\pmb{\Psi}\right)\,\equiv \,
    \pmb{Y}_s\left(\mathfrak{M}^{-1}\mathbf{x}|\mathbf{\Psi}\right)
\end{equation}
and we can write a formula analogous to eq.(\ref{Cgotico})
\begin{equation}\label{bibbioconosco}
\pmb{\hat{Y}}_s(\pmb{x}|\pmb{\Psi})\,
=\,\sum_{\mathfrak{q}=1}^{35} \, \sum_{\alpha=1}^{14}
\mathfrak{C}_{\mathfrak{q}\alpha}(\pmb{\Psi })\,
\mathbf{d}_{\mathfrak{q}}\times
\hat{\mathit{f}}_{\alpha}(\pmb{x})
\end{equation}
where $\mathbf{d}_{\mathfrak{q}}$ denote the lexicographic
ordered differentials in the orthonormal basis:
\begin{equation}\label{curapalla}
\mathbf{d} = \left\{ \text{dx}_1\wedge \text{dx}_2\wedge
\text{dx}_3,\text{dx}_1\wedge \text{dx}_2\wedge
\text{dx}_4,\text{...}\text{...},\text{dx}_5\wedge
\text{dx}_6\wedge \text{dx}_7\right\}\, = \,
\left\{\mathbf{d}_{\mathfrak{q}}\right\} \quad
(\mathfrak{q}=1,\text{...},35)
\end{equation}
and $\hat{\mathit{f}}_{\alpha }(\pmb{x})$ are the basis
trigonometric functions in the same coordinates:
\begin{equation}\label{cospargolardo}
    \hat{\mathit{f}}_{\alpha }(\pmb{x})=\left(
\begin{array}{c}
 \text{Cos}\left[2 \pi  \left(\frac{x_1}{\sqrt{2}}-\frac{x_5}{2 \sqrt{2}}-\frac{x_6}{2 \sqrt{2}}+\frac{x_7}{2 \sqrt{2}}\right)\right] \\
 \text{Cos}\left[2 \pi  \left(-\frac{x_1}{\sqrt{2}}-\frac{x_5}{2 \sqrt{2}}-\frac{x_6}{2 \sqrt{2}}+\frac{x_7}{2 \sqrt{2}}\right)\right] \\
 \text{Cos}\left[2 \pi  \left(\frac{x_2}{\sqrt{2}}+\frac{x_5}{2 \sqrt{2}}-\frac{x_6}{2 \sqrt{2}}-\frac{x_7}{2 \sqrt{2}}\right)\right] \\
 \text{Cos}\left[2 \pi  \left(-\frac{x_2}{\sqrt{2}}+\frac{x_5}{2 \sqrt{2}}+\frac{3 x_6}{2 \sqrt{2}}-\sqrt{2} x_6-\frac{x_7}
 {2 \sqrt{2}}\right)\right]
\\
 \text{Cos}\left[2 \pi  \left(-\frac{x_3}{\sqrt{2}}-\frac{x_5}{2 \sqrt{2}}+\frac{x_6}{2 \sqrt{2}}-\frac{x_7}{2 \sqrt{2}}\right)\right] \\
 \text{Cos}\left[2 \pi  \left(\frac{x_4}{\sqrt{2}}-\frac{x_5}{2 \sqrt{2}}-\frac{x_6}{2 \sqrt{2}}-\frac{x_7}{2 \sqrt{2}}\right)\right] \\
 \text{Cos}\left[2 \pi  \left(\frac{x_4}{\sqrt{2}}+\frac{x_5}{2 \sqrt{2}}+\frac{x_6}{2 \sqrt{2}}+\frac{x_7}{2 \sqrt{2}}\right)\right] \\
 \text{Sin}\left[2 \pi  \left(\frac{x_1}{\sqrt{2}}-\frac{x_5}{2 \sqrt{2}}-\frac{x_6}{2 \sqrt{2}}+\frac{x_7}{2 \sqrt{2}}\right)\right] \\
 \text{Sin}\left[2 \pi  \left(-\frac{x_1}{\sqrt{2}}-\frac{x_5}{2 \sqrt{2}}-\frac{x_6}{2 \sqrt{2}}+\frac{x_7}{2 \sqrt{2}}\right)\right] \\
 \text{Sin}\left[2 \pi  \left(\frac{x_2}{\sqrt{2}}+\frac{x_5}{2 \sqrt{2}}-\frac{x_6}{2 \sqrt{2}}-\frac{x_7}{2 \sqrt{2}}\right)\right] \\
 \text{Sin}\left[2 \pi  \left(-\frac{x_2}{\sqrt{2}}+\frac{x_5}{2 \sqrt{2}}+\frac{3 x_6}{2 \sqrt{2}}-\sqrt{2} x_6
 -\frac{x_7}{2 \sqrt{2}}\right)\right]
\\
 \text{Sin}\left[2 \pi  \left(-\frac{x_3}{\sqrt{2}}-\frac{x_5}{2 \sqrt{2}}+\frac{x_6}{2 \sqrt{2}}-\frac{x_7}{2 \sqrt{2}}\right)\right] \\
 -\text{Sin}\left[2 \pi  \left(\frac{x_4}{\sqrt{2}}-\frac{x_5}{2 \sqrt{2}}-\frac{x_6}{2 \sqrt{2}}-\frac{x_7}{2 \sqrt{2}}\right)\right] \\
 \text{Sin}\left[2 \pi  \left(\frac{x_4}{\sqrt{2}}+\frac{x_5}{2 \sqrt{2}}+\frac{x_6}{2 \sqrt{2}}+\frac{x_7}{2 \sqrt{2}}\right)\right] \\
\end{array}
\right)\quad  ; \quad\alpha =1,\text{...},14
\end{equation}
The full structure of the general solution is encoded in
the coefficients $\hat{\mathfrak{C}}_{\mathfrak{q}\alpha }
(\mathbf{\Psi})$ which once again we do not display for
reasons of space, being they  very large objects.
\par
The components $Y_{{ijk}}^s(\mathbf{x}|\mathbf{\Psi})$ of
$\hat{\mathbf{Y}}_s (\pmb{x}|\mathbf{\Psi})$ satisfy
Englert equation in the orthonormal basis (see
eq.(\ref{englerta})) with the appropriate value of $\mu $,
namely
\begin{equation}\label{appropriomu}
    \mu =-4 \sqrt{\frac{7}{2}} \pi
\end{equation}
and they can be used to construct an exact M2-brane
solution of d=11 supergravity depending on the 8 moduli
$\psi _i$.
\subsection{Analysis of the $\mathcal{B}$-operator polarized on this solution}
Our aim is that of finding, if possible, a subspace of the
8-dimensional moduli space where the rank of the 8$\times$8
matrix:
\begin{equation}\label{dinuovoB}
   \mathcal{B}(\mathbf{x}|\mathbf{\Psi} ) =\sum _{{ijk}}
Y_{{ijk}}^s(\mathbf{x}|\mathbf{\Psi} ) \tau _{{ijk}}
\end{equation}
is reduced, leading to the existence of some preserved
supersymmetry. The strategy we have adopted for this task
is the following. We have expanded the
$\mathcal{B}(\mathbf{x}|\mathbf{\Psi})$ operator along the
basis functions:
\begin{equation}\label{Qbitti}
  \mathcal{B}(\mathbf{x}|\mathbf{\Psi} ) = \sum _{\alpha =1}^{14} \mathcal{Q}_{\alpha }(\mathbf{\Psi} )
\, \hat{\mathit{f}}_{\alpha }(\pmb{x})
\end{equation}
obtaining a set of 14 matrices $\mathcal{Q}_{\alpha
}(\mathbf{\Psi} )$ whose rank, for generic $\mathbf{\Psi}
$, is 8 for all of them. Next, in order to reduce the rank
we have considered the condition that one of the rows
(always the same) should be simultaneously zero for all the
14 matrices. Such conditions leads to a set of constraints
with no solutions for all the rows except for the last one, navely
the eight row. This row can be simultaneously annihilated
for all the 14 matrices by a constraint that has a solution
in terms of 4 parameters (\(\psi_1\),\(\psi _2\),\(\psi
_3\),\(\psi _5\)), namely:
\begin{equation}\label{riducopsi}
\begin{array}{rcl}
 \psi _4&=&-3 \psi _1-4 \psi _2-\psi _3-2 \psi _5 \\
 \psi _6&=&-2 \psi _2-2 \psi _5 \\
 \psi _7&=&-5 \psi _1+3 \psi _2-4 \psi _3+5 \psi _5 \\
 \psi _8&=&5 \psi _1+\psi _2+3 \psi _3-\psi _5 \\
\end{array}
\end{equation}
Under these conditions not only the 8-th row of the
operator $\mathcal{B}$(x$|\Psi)$ vanishes but also so it
does its 8-th column. In other words
$\mathcal{B}$(x$|\Psi)$ consistently reduces to a
7-dimensional operator in spinor space. Consequently the
spinor:
\begin{equation}\label{spinellochillo}
   \zeta =\left(
\begin{array}{c}
 0 \\
 0 \\
 0 \\
 0 \\
 0 \\
 0 \\
 0 \\
 \zeta _8 \\
\end{array}
\right)
\end{equation}
introduced in the tensor product formula (\ref{surrocanto})
leads to a Killing spinor of the considered 4-parameter
M2-brane solution of d=11 supergravity.
\par
Next, using the same procedure, we investigated whether the
remaining 7$\times $7 block of the $\mathcal{B}$-operator
could be further reduced in rank. The answer was negative.
Indeed it turns out that no condition can be imposed on the
remaining 4-moduli $\psi _i$ for which additional Killing
spinors pop up. The conclusion is that our construction
leads to an intrinsically $\mathcal{N}=1$ M2-brane solution
of d=11 supergravity. The operators associated with the
other 4-moduli (also $\mathrm{G_{21}}$-invariant)  break
supersymmetry.
\par
Let us name
\begin{equation}\label{craccopotto}
   \pmb{Y}^{\mathcal{N}=1}(\pmb{x}|\mathbf{\psi} )\, = \, \sum_{\alpha=1}^{14} \sum_{\mathfrak{q}=1}^{35}
 \,\mathfrak{C}^{\mathcal{N}=1}_{q\alpha}\left(\pmb{\psi }\right) \,
   \mathbf{d}_{\mathfrak{q}}\times\hat{\mathit{f}}_{\alpha}(\pmb{x})
\end{equation}
the Englert solution leading to $\mathcal{N}=1$
supersymmetry and dependent on the 4-parameters $\psi $
where, for simplicity,  in the last stage we have renamed,
$\psi _5\rightarrow \psi _4$. The unique shortest way of
displaying the result is that of displaying the 35$\times
$14 matrix \(\mathfrak{C}_{\mathfrak{q}\alpha
}{}^{N=1}\)($\psi $), dependent on the four moduli $\psi
_i$, $(i=1,\dots,4)$. The output is still ominously big,
yet due to the relevance of the final result for further
uses, we believe that it is worth showing it: furthermore
this is  the only way to give concreteness to the results
we have obtained. The matrix
$\mathfrak{C}^{\mathcal{N}=1}_{q\alpha}\left(\pmb{\psi
}\right)$ is displayed in appendix \ref{lagrandpap}.
\par
With some ingenuity we have also found that there are two
particulary nice points in the 4-dimensional moduli space
of this solution where the rank of all the 14 matrices
$\mathcal{Q}_{\alpha }(\mathbf{\psi} )$, coefficients of
the basis functions, reduces from 7 to 4, although these
matrices do not admit a common null-vector, leaving the
rank of the full operator $\mathcal{B}$(x$|\Psi $) equal to
7. These two special points in moduli space might have some
so far unknown deep significance and for this reason we
feel it important to mention them:
\begin{eqnarray}
 {\vec{\psi} _1}& \equiv & \left\{\psi
_1=1,\psi _2=-1,\psi _3=-2,\psi _4=1\right\}
\label{cosmopolitone1}\\
\vec{\psi _2}& \equiv & \left\{\psi _1=1,\psi _2=-1,\psi
_3=\frac{1}{2},\psi _4=1\right\} \label{cosmopolitone2}
\end{eqnarray}
\par
In the case of the point $\vec{\psi_1}$ in moduli space, the 35$\times $14 matrix encoding the
Englert solution takes the following relatively simple
look:
{\tiny
\begin{equation}\label{caspiterone}
\mathfrak{C}^{\mathcal{N}=1}_{q\alpha}\left(\pmb{\vec{\psi}_1
}\right)\, = \,\left(
\begin{array}{cccccccccccccc}
 0 & 0 & -3 & 0 & 9 & -15 & 9 & -\frac{24}{\sqrt{7}} & 0 & \frac{39}{\sqrt{7}} & 0 & \frac{27}{\sqrt{7}} & -3 \sqrt{7} & -3 \sqrt{7} \\
 -9 & -3 & 3 & -6 & 0 & 6 & 9 & 3 \sqrt{7} & \frac{15}{\sqrt{7}} & -\frac{15}{\sqrt{7}} & \frac{6}{\sqrt{7}} & 0 & -\frac{6}{\sqrt{7}} & -3 \sqrt{7}
\\
 9 & 0 & -6 & -6 & 12 & -9 & 0 & \frac{3}{\sqrt{7}} & 0 & \frac{6}{\sqrt{7}} & \frac{6}{\sqrt{7}} & -\frac{60}{\sqrt{7}} & \frac{45}{\sqrt{7}} &
0 \\
 3 & -3 & 6 & -6 & 15 & 3 & -18 & -\frac{15}{\sqrt{7}} & \frac{15}{\sqrt{7}} & -\frac{6}{\sqrt{7}} & \frac{6}{\sqrt{7}} & -\frac{3}{\sqrt{7}} & -\frac{39}{\sqrt{7}}
& 6 \sqrt{7} \\
 12 & -3 & -12 & 0 & -15 & 18 & 0 & -\frac{12}{\sqrt{7}} & \frac{15}{\sqrt{7}} & \frac{12}{\sqrt{7}} & 0 & \frac{3}{\sqrt{7}} & -\frac{18}{\sqrt{7}}
& 0 \\
 0 & -3 & -6 & 15 & 0 & -15 & 9 & -\frac{24}{\sqrt{7}} & \frac{39}{\sqrt{7}} & -6 \sqrt{7} & 3 \sqrt{7} & 0 & \frac{3}{\sqrt{7}} & \frac{3}{\sqrt{7}}
\\
 0 & -9 & 3 & 15 & -9 & -12 & 12 & 0 & -\frac{27}{\sqrt{7}} & \frac{33}{\sqrt{7}} & 3 \sqrt{7} & -\frac{27}{\sqrt{7}} & -\frac{12}{\sqrt{7}} & \frac{12}{\sqrt{7}}
\\
 6 & 6 & -3 & 15 & 0 & -6 & -18 & \frac{18}{\sqrt{7}} & -\frac{6}{\sqrt{7}} & -\frac{33}{\sqrt{7}} & 3 \sqrt{7} & 0 & \frac{30}{\sqrt{7}} & -\frac{30}{\sqrt{7}}
\\
 6 & -3 & 12 & 0 & 9 & -12 & -12 & \frac{18}{\sqrt{7}} & -\frac{33}{\sqrt{7}} & -\frac{12}{\sqrt{7}} & 0 & \frac{27}{\sqrt{7}} & -\frac{12}{\sqrt{7}}
& \frac{12}{\sqrt{7}} \\
 -9 & 12 & -18 & 18 & 0 & 3 & -6 & -\frac{3}{\sqrt{7}} & -\frac{12}{\sqrt{7}} & \frac{18}{\sqrt{7}} & -\frac{18}{\sqrt{7}} & 0 & -\frac{15}{\sqrt{7}}
& \frac{30}{\sqrt{7}} \\
 12 & -6 & -9 & 9 & 0 & -6 & 0 & -\frac{12}{\sqrt{7}} & \frac{6}{\sqrt{7}} & \frac{45}{\sqrt{7}} & -\frac{45}{\sqrt{7}} & 0 & \frac{30}{\sqrt{7}}
& -\frac{24}{\sqrt{7}} \\
 3 & 6 & -15 & -3 & 0 & 3 & 6 & -\frac{15}{\sqrt{7}} & -\frac{6}{\sqrt{7}} & \frac{3}{\sqrt{7}} & \frac{39}{\sqrt{7}} & 0 & -\frac{15}{\sqrt{7}}
& -\frac{6}{\sqrt{7}} \\
 -3 & 6 & 0 & -9 & 15 & -6 & -3 & -\frac{9}{\sqrt{7}} & -\frac{6}{\sqrt{7}} & 0 & -\frac{27}{\sqrt{7}} & -\frac{3}{\sqrt{7}} & \frac{30}{\sqrt{7}}
& \frac{15}{\sqrt{7}} \\
 -3 & 6 & 12 & -3 & -3 & 0 & -9 & -\frac{9}{\sqrt{7}} & -\frac{6}{\sqrt{7}} & -\frac{12}{\sqrt{7}} & \frac{39}{\sqrt{7}} & -\frac{57}{\sqrt{7}} &
0 & \frac{45}{\sqrt{7}} \\
 3 & -6 & -12 & -3 & 15 & 6 & -3 & \frac{9}{\sqrt{7}} & \frac{6}{\sqrt{7}} & \frac{12}{\sqrt{7}} & \frac{39}{\sqrt{7}} & -\frac{3}{\sqrt{7}} & -\frac{30}{\sqrt{7}}
& -\frac{33}{\sqrt{7}} \\
 9 & 6 & 0 & -15 & -9 & 0 & 9 & -3 \sqrt{7} & 6 \sqrt{7} & 0 & \frac{3}{\sqrt{7}} & -\frac{27}{\sqrt{7}} & 0 & \frac{3}{\sqrt{7}} \\
 -9 & 0 & -6 & 12 & 0 & 15 & -12 & -\frac{51}{\sqrt{7}} & 0 & \frac{6}{\sqrt{7}} & \frac{12}{\sqrt{7}} & 0 & 3 \sqrt{7} & \frac{12}{\sqrt{7}} \\
 6 & 6 & 3 & 3 & 0 & 0 & -18 & -\frac{30}{\sqrt{7}} & 6 \sqrt{7} & \frac{33}{\sqrt{7}} & -\frac{15}{\sqrt{7}} & 0 & 0 & -\frac{30}{\sqrt{7}} \\
 -3 & 6 & -9 & 9 & 0 & -15 & 12 & -\frac{33}{\sqrt{7}} & 6 \sqrt{7} & -\frac{27}{\sqrt{7}} & \frac{27}{\sqrt{7}} & 0 & -3 \sqrt{7} & \frac{12}{\sqrt{7}}
\\
 -18 & 15 & -3 & 3 & 3 & -6 & 6 & 6 \sqrt{7} & -\frac{3}{\sqrt{7}} & \frac{15}{\sqrt{7}} & -\frac{15}{\sqrt{7}} & -\frac{39}{\sqrt{7}} & \frac{6}{\sqrt{7}}
& -\frac{6}{\sqrt{7}} \\
 0 & -18 & -3 & -3 & 24 & 0 & 0 & 0 & \frac{18}{\sqrt{7}} & \frac{15}{\sqrt{7}} & \frac{15}{\sqrt{7}} & -\frac{24}{\sqrt{7}} & 0 & -\frac{24}{\sqrt{7}}
\\
 0 & -9 & 0 & 6 & 3 & 6 & -6 & 0 & \frac{45}{\sqrt{7}} & 0 & -\frac{30}{\sqrt{7}} & -\frac{39}{\sqrt{7}} & -\frac{6}{\sqrt{7}} & \frac{30}{\sqrt{7}}
\\
 12 & -15 & -6 & 3 & 0 & 3 & 3 & \frac{12}{\sqrt{7}} & \frac{3}{\sqrt{7}} & \frac{6}{\sqrt{7}} & -\frac{15}{\sqrt{7}} & 0 & -\frac{39}{\sqrt{7}}
& \frac{33}{\sqrt{7}} \\
 -6 & -15 & 6 & -3 & 0 & 9 & 9 & \frac{6}{\sqrt{7}} & \frac{3}{\sqrt{7}} & -\frac{6}{\sqrt{7}} & \frac{15}{\sqrt{7}} & 0 & \frac{27}{\sqrt{7}} &
-\frac{45}{\sqrt{7}} \\
 -18 & 9 & 6 & -3 & 0 & 3 & 3 & \frac{18}{\sqrt{7}} & \frac{27}{\sqrt{7}} & -\frac{6}{\sqrt{7}} & \frac{15}{\sqrt{7}} & 0 & -\frac{39}{\sqrt{7}}
& -\frac{15}{\sqrt{7}} \\
 -9 & 12 & 6 & -12 & 0 & -3 & 6 & -\frac{51}{\sqrt{7}} & -\frac{12}{\sqrt{7}} & 6 \sqrt{7} & -\frac{12}{\sqrt{7}} & 0 & \frac{15}{\sqrt{7}} & \frac{18}{\sqrt{7}}
\\
 3 & -3 & 6 & 12 & -9 & -9 & 0 & \frac{33}{\sqrt{7}} & -\frac{33}{\sqrt{7}} & 6 \sqrt{7} & \frac{12}{\sqrt{7}} & -\frac{27}{\sqrt{7}} & -\frac{27}{\sqrt{7}}
& 0 \\
 -6 & 3 & 0 & 6 & -9 & 12 & -6 & \frac{30}{\sqrt{7}} & \frac{33}{\sqrt{7}} & 0 & -\frac{30}{\sqrt{7}} & -\frac{27}{\sqrt{7}} & \frac{12}{\sqrt{7}}
& -\frac{18}{\sqrt{7}} \\
 3 & 15 & 3 & -12 & 0 & -6 & -3 & \frac{9}{\sqrt{7}} & 3 \sqrt{7} & \frac{33}{\sqrt{7}} & -\frac{12}{\sqrt{7}} & 0 & -6 \sqrt{7} & -\frac{9}{\sqrt{7}}
\\
 3 & -3 & -15 & -12 & 0 & 12 & 15 & \frac{9}{\sqrt{7}} & -\frac{33}{\sqrt{7}} & -3 \sqrt{7} & -\frac{12}{\sqrt{7}} & 0 & \frac{12}{\sqrt{7}} & \frac{45}{\sqrt{7}}
\\
 -15 & -3 & 3 & 6 & 0 & 12 & -3 & -\frac{45}{\sqrt{7}} & -\frac{33}{\sqrt{7}} & \frac{33}{\sqrt{7}} & 6 \sqrt{7} & 0 & \frac{12}{\sqrt{7}} & -\frac{9}{\sqrt{7}}
\\
 -6 & -12 & 9 & 0 & 3 & 3 & 3 & \frac{6}{\sqrt{7}} & \frac{12}{\sqrt{7}} & \frac{27}{\sqrt{7}} & 0 & -\frac{39}{\sqrt{7}} & -\frac{15}{\sqrt{7}}
& \frac{9}{\sqrt{7}} \\
 12 & 12 & -15 & -6 & 3 & -3 & -3 & \frac{12}{\sqrt{7}} & -\frac{12}{\sqrt{7}} & \frac{3}{\sqrt{7}} & \frac{30}{\sqrt{7}} & -\frac{39}{\sqrt{7}}
& \frac{15}{\sqrt{7}} & -\frac{9}{\sqrt{7}} \\
 -18 & 0 & -15 & 6 & 21 & 3 & 3 & \frac{18}{\sqrt{7}} & 0 & \frac{3}{\sqrt{7}} & -\frac{30}{\sqrt{7}} & \frac{15}{\sqrt{7}} & -\frac{15}{\sqrt{7}}
& \frac{9}{\sqrt{7}} \\
 6 & 12 & -12 & -6 & 0 & 6 & -6 & \frac{18}{\sqrt{7}} & -\frac{12}{\sqrt{7}} & \frac{12}{\sqrt{7}} & \frac{30}{\sqrt{7}} & 0 & -\frac{30}{\sqrt{7}}
& -\frac{18}{\sqrt{7}} \\
\end{array}
\right)
\end{equation}
}
\subsection{The inhomogeneous harmonic function}
The final step in our construction regards the explicit
form of the inhomogeneous harmonic function. To this effect
we have to construct the source term of eq.(\ref{3harmoniusca3})
starting from the explicit solution of Englert equation we
have derived.
\begin{equation}\label{JJsorgo}
    J(U,\pmb{x})\,\equiv\,\frac{3}{2} e^{-2 U \mu } \mu ^2 \sum _{{ijk}} Y_{{ijk}}^2 (\pmb{x})
\end{equation}
Focusing for simplicity on the case of the solution
$\pmb{Y}^{\mathcal{N}=1}\left(\pmb{x}|\vec{\psi}_1\right)$
we find that the source term has the following form:
\begin{equation}\label{caggofare}
  J=-1152\, e^{4 \sqrt{14} \,\pi  U}\, \pi ^2\, (-441\,+\, \mathfrak{W}(\pmb{x}))
\end{equation}
Where the function $\mathfrak{W}$(x) is the following
linear combination of 42 independent trigonometric
functions: {\tiny
\begin{eqnarray}
&&\mathfrak{W}(\pmb{x})\, = \, \nonumber\\
 && 21
\text{Cos}\left[2 \sqrt{2} \pi x_1\right]+21
\text{Cos}\left[2 \sqrt{2} \pi x_2\right]+21
\text{Cos}\left[2 \sqrt{2} \pi  x_4\right]+21
\text{Cos}\left[\sqrt{2} \pi
\left(x_1-x_4-x_5-x_6\right)\right]\nonumber\\
&&+21 \text{Cos}\left[\sqrt{2} \pi
\left(x_1+x_4-x_5-x_6\right)\right]+21
\text{Cos}\left[\sqrt{2} \pi
\left(x_2+x_3+x_5-x_6\right)\right] + 21
\text{Cos}\left[\sqrt{2} \pi
\left(x_2-x_3-x_5+x_6\right)\right]+21
\text{Cos}\left[\sqrt{2} \pi
\left(x_1-x_4+x_5+x_6\right)\right]\nonumber\\
&&+21 \text{Cos}\left[\sqrt{2} \pi
\left(x_1+x_4+x_5+x_6\right)\right]+21
\text{Cos}\left[\sqrt{2} \pi
\left(x_1-x_2+x_5-x_7\right)\right]+21
\text{Cos}\left[\sqrt{2} \pi
\left(x_1+x_2+x_5-x_7\right)\right]+21
\text{Cos}\left[\sqrt{2} \pi
\left(x_2-x_4-x_6-x_7\right)\right]\nonumber\\
&&+ 21 \text{Cos}\left[\sqrt{2} \pi
\left(x_2+x_4-x_6-x_7\right)\right]+21
\text{Cos}\left[\sqrt{2} \pi
\left(x_1-x_3+x_6-x_7\right)\right]+21
\text{Cos}\left[\sqrt{2} \pi
\left(x_1-x_2-x_5+x_7\right)\right]+21
\text{Cos}\left[\sqrt{2} \pi
\left(x_1+x_2-x_5+x_7\right)\right]\nonumber\\
&&+21 \text{Cos}\left[\sqrt{2} \pi
\left(x_3-x_4+x_5+x_7\right)\right]+21
\text{Cos}\left[\sqrt{2} \pi
\left(x_3+x_4+x_5+x_7\right)\right]+ 21
\text{Cos}\left[\sqrt{2} \pi
\left(x_1+x_3-x_6+x_7\right)\right]+21
\text{Cos}\left[\sqrt{2} \pi
\left(x_2-x_4+x_6+x_7\right)\right]\nonumber\\
&&+21 \text{Cos}\left[\sqrt{2} \pi
\left(x_2+x_4+x_6+x_7\right)\right]+\sqrt{7}
\text{Sin}\left[2 \sqrt{2} \pi  x_1\right]-\sqrt{7}
\text{Sin}\left[2 \sqrt{2} \pi x_2\right]-\sqrt{7}
\text{Sin}\left[2 \sqrt{2} \pi  x_4\right] -\sqrt{7}
\text{Sin}\left[\sqrt{2} \pi
\left(x_1-x_4-x_5-x_6\right)\right]\nonumber\\
&&+ \sqrt{7} \text{Sin}\left[\sqrt{2} \pi
\left(x_1+x_4-x_5-x_6\right)\right]-\sqrt{7}
\text{Sin}\left[\sqrt{2} \pi
\left(x_2+x_3+x_5-x_6\right)\right]-\sqrt{7}
\text{Sin}\left[\sqrt{2} \pi
\left(x_2-x_3-x_5+x_6\right)\right]\nonumber\\
&&+\sqrt{7} \text{Sin}\left[\sqrt{2} \pi
\left(x_1-x_4+x_5+x_6\right)\right]-\sqrt{7}
\text{Sin}\left[\sqrt{2} \pi
\left(x_1+x_4+x_5+x_6\right)\right]-\sqrt{7}
\text{Sin}\left[\sqrt{2} \pi
\left(x_1-x_2+x_5-x_7\right)\right]\nonumber\\
&& +\sqrt{7} \text{Sin}\left[\sqrt{2} \pi
\left(-x_1+x_2+x_5-x_7\right)\right]-\sqrt{7}
\text{Sin}\left[\sqrt{2} \pi
\left(x_1+x_2+x_5-x_7\right)\right]+\sqrt{7}
\text{Sin}\left[\sqrt{2} \pi
\left(x_2-x_4-x_6-x_7\right)\right]\nonumber\\
&&+\sqrt{7} \text{Sin}\left[\sqrt{2} \pi
\left(x_2+x_4-x_6-x_7\right)\right]+\sqrt{7}
\text{Sin}\left[\sqrt{2} \pi
\left(x_1-x_3+x_6-x_7\right)\right]\nonumber\\
&& + \sqrt{7} \text{Sin}\left[\sqrt{2} \pi
\left(x_1+x_2-x_5+x_7\right)\right]+\sqrt{7}
\text{Sin}\left[\sqrt{2} \pi
\left(x_3-x_4+x_5+x_7\right)\right]+\sqrt{7}
\text{Sin}\left[\sqrt{2} \pi
\left(x_3+x_4+x_5+x_7\right)\right]\nonumber\\
&&-\sqrt{7} \text{Sin}\left[\sqrt{2} \pi
\left(x_1+x_3-x_6+x_7\right)\right]-\sqrt{7}
\text{Sin}\left[\sqrt{2} \pi
\left(x_2-x_4+x_6+x_7\right)\right]+\sqrt{7}
\text{Sin}\left[\sqrt{2} \pi
\left(x_2+x_4+x_6+x_7\right)\right]
\end{eqnarray}
} The function $\mathfrak{W}$(x) is an eigenstate of the
Laplacian on the seven torus:
\begin{equation}\label{autfunzionka}
    \square_{\mathrm{T^7}} \,\mathfrak{W}(x) = - 8 \pi
    ^2\mathfrak{W}(x)
\end{equation}
Hence the equation
\begin{equation}\label{armonicelli}
  \square _{\mathbb{R}_+\otimes \mathrm{T^7}}H(U,x)+J(U,\pmb{x}) = 0
\end{equation}
admits the following solution:
\begin{equation}\label{cuscolotto}
  H(U,x)\,=\, \alpha \, -\, 2268\, e^{4 \sqrt{14} \pi  U}\, + \, \frac{16}{3}e^{4 \sqrt{14} \pi
U}\, \mathfrak{W}(x)
\end{equation}
where $\alpha $ is an arbitrary constant that can be fixed
by boundary conditions.
\par
In this way we have completed the derivation of an M2-brane solution of $d=11$ supergravity that preserves $\mathcal{N}=1$ supersymmetry in $d=3$ counting and has an exact $\mathrm{G_{21}}$ discrete symmetry. It should be noted that the metric has a quite non trivial dependence on all the coordinates of the space $\mathbb{R}_+ \times T^7$, transverse to the brane volume. This together with the structure of the Englert fields implies that the fields of the $d=3$ brane model are all strongly, non linearly interacting.
\section{Conclusions}
In this paper we have shown that, contrary to what happens in compactifications of the type:
\begin{equation}\label{ads4m7}
  \mathcal{M}_{11}\, = \, \mathrm{AdS_4}\times \mathcal{M}_7
\end{equation}
there exist M2-brane supersymmetric solutions of $d=11$ supergravity with internal fluxes governed by the Englert equation. In the case of eq.(\ref{ads4m7}) Englert solutions exist but they always break supersymmetry. We have identified Englert equation as the proper generalization to $7$-manifolds of Beltrami equation defined on $3$-manifolds. We have shown a general procedure to construct M2-brane solutions with Englert fluxes and we have defined a simple and algorithmic criterion to determine the numebr of supersymmetries preserved by such backgrounds.
\par
Building on our experience with the torus $\mathrm{T^3}$ and the use of its crystallographic point group for the construction of solutions of Beltrami equation, we have spotted the simple group $\mathrm{L_{168}} \equiv \mathrm{PSL(2,\mathbb{Z}_7)}$ as a very much challenging crystallographic point--group in $7$-dimensions, the corresponding lattice being the A7-root lattice. Relying on this we have defined an algorithm to construct solutions of Englert equation associated with orbits of $\mathrm{L_{168}}$ and of its subgroups in the weight-lattice of A7.
\par
In this framework we have constructed a very non trivial M2-brane solution with $\mathcal{N}=1$--supersymmetry and a large non-abelian discrete symmetry, namely $\mathrm{G_{21}} \,\equiv \,\mathbb{Z}_3\ltimes \mathbb{Z}_7$.
\par
The next obvious step is the analysis of the $d=3$ theories on the world volume that are dual to  $d=11$ supergravity localized on the considered backgrounds. We plan to address this problem in future publications.
\par
We note in passing that, as a by-product of our main investigation, we have classified all the orbits of
$\mathrm{PSL(2,\mathbb{Z}_7)}$ in the A7-weight lattice. This mathematical result might prove useful in different contests both mathematical and physical.
\vskip 2cm
\section*{Aknowledgments}
Very important and illuminating discussions during the entire development of this work are aknowledged with author's good friends P.A. Grassi, A. Sorin and M.Trigiante. Also important discussions occurred the recent visit of S. Ferrara to Moscow. Finally the author would like to thank D. Luest for his useful comments during a short visit of the author to Munich.
\newpage
\appendix
\section{The explicit form of the $\mathrm{L_{168}}$-invariant Englert solution}
\label{mostro}
{\small In this appendix we display display the explicit
form of the 2-moduli dependent coefficients $\mathfrak{C}^{[168]}_{\mathfrak{q}\alpha}\left(\hat{\pmb{F}}\right)$ that define the $\mathrm{L_{168}}$-invariant  solution of Englert equation discussed in section \ref{168invariante}. Because of the large form of the output the entries of the matrix
 are organized in 4 tables containing the  columns from 1to 4, from 5 to 8, from 8 to 12 and finally from
 13 to 16.}
{\scriptsize
\begin{equation}\label{1tavola168}
  \left(
\begin{array}{c|cccc}
 0 & 1 & 2 & 3 & 4 \\
 \hline
 1 & 6 F_1 & 6 F_1 & -6 F_1 & -6 F_1 \\
 2 & 6 F_2 & 6 F_2 & 6 \left(2 F_1+F_2\right) & -6 \left(4 F_1+F_2\right) \\
 3 & -6 \left(3 F_1+2 F_2\right) & -6 \left(3 F_1+2 F_2\right) & -6 \left(3 F_1+2 F_2\right) & 6 \left(3 F_1+2 F_2\right) \\
 4 & 6 \left(F_1+F_2\right) & 6 \left(3 F_1+F_2\right) & 6 \left(3 F_1+F_2\right) & -6 \left(F_1+3 F_2\right) \\
 5 & 12 F_1 & -12 F_1 & 0 & 36 F_1+24 F_2 \\
 6 & -6 \left(F_1+F_2\right) & 6 \left(F_1-F_2\right) & 6 \left(F_1+F_2\right) & 6 \left(3 F_1+F_2\right) \\
 7 & 6 \left(5 F_1+2 F_2\right) & 6 \left(F_1+2 F_2\right) & -6 \left(3 F_1+2 F_2\right) & -6 \left(3 F_1+2 F_2\right) \\
 8 & -6 \left(3 F_1+2 F_2\right) & -6 \left(3 F_1+2 F_2\right) & 6 \left(5 F_1+2 F_2\right) & 6 \left(F_1+2 F_2\right) \\
 9 & 6 \left(F_1+2 F_2\right) & 6 \left(5 F_1+2 F_2\right) & -6 \left(3 F_1+2 F_2\right) & -6 \left(3 F_1+2 F_2\right) \\
 10 & -18 F_1 & 18 F_1 & 6 \left(3 F_1+2 F_2\right) & -6 \left(3 F_1+2 F_2\right) \\
 11 & 6 \left(5 F_1+2 F_2\right) & 6 \left(F_1+2 F_2\right) & -6 \left(3 F_1+2 F_2\right) & 6 \left(5 F_1+2 F_2\right) \\
 12 & -18 \left(2 F_1+F_2\right) & -18 \left(2 F_1+F_2\right) & 6 \left(F_2-2 F_1\right) & 6 \left(F_2-2 F_1\right) \\
 13 & -6 \left(4 F_1+F_2\right) & -6 \left(2 F_1+F_2\right) & 6 \left(F_2-2 F_1\right) & -6 \left(8 F_1+3 F_2\right) \\
 14 & 18 F_1 & 18 F_1 & 42 F_1 & 42 F_1 \\
 15 & 6 F_2-6 F_1 & 6 \left(F_1+F_2\right) & 6 \left(F_1+3 F_2\right) & -6 \left(5 F_1+F_2\right) \\
 16 & 6 \left(3 F_1+2 F_2\right) & 6 F_1 & 6 F_1 & -6 F_1 \\
 17 & -6 \left(8 F_1+3 F_2\right) & 6 F_2 & 6 F_2 & 6 \left(2 F_1+F_2\right) \\
 18 & 6 \left(3 F_1+2 F_2\right) & -6 \left(3 F_1+2 F_2\right) & -6 \left(3 F_1+2 F_2\right) & -6 \left(3 F_1+2 F_2\right) \\
 19 & -6 \left(F_1+3 F_2\right) & 6 \left(3 F_1+F_2\right) & 6 \left(F_1+F_2\right) & 6 \left(3 F_1+F_2\right) \\
 20 & 6 \left(5 F_1+F_2\right) & -6 \left(F_1+F_2\right) & 6 \left(F_1-F_2\right) & 6 \left(F_1+F_2\right) \\
 21 & -6 \left(3 F_1+2 F_2\right) & 6 \left(5 F_1+2 F_2\right) & 6 \left(F_1+2 F_2\right) & -6 \left(3 F_1+2 F_2\right) \\
 22 & 36 F_1+24 F_2 & -12 F_1 & 12 F_1 & 12 F_1 \\
 23 & 6 \left(3 F_1+2 F_2\right) & -18 F_1 & 18 F_1 & 6 \left(3 F_1+2 F_2\right) \\
 24 & -6 \left(5 F_1+F_2\right) & -6 \left(F_1+F_2\right) & -6 \left(5 F_1+F_2\right) & -6 \left(5 F_1+F_2\right) \\
 25 & 6 \left(5 F_1+F_2\right) & -6 \left(F_1+F_2\right) & 6 \left(F_1-F_2\right) & 6 \left(5 F_1+F_2\right) \\
 26 & -6 \left(3 F_1+2 F_2\right) & 6 \left(3 F_1+2 F_2\right) & 6 F_1 & 6 F_1 \\
 27 & 6 \left(F_2-2 F_1\right) & -6 \left(8 F_1+3 F_2\right) & 6 F_2 & 6 F_2 \\
 28 & 12 F_1-6 F_2 & 12 F_1-6 F_2 & -6 F_2 & -6 \left(2 F_1+F_2\right) \\
 29 & -6 \left(F_1+3 F_2\right) & 6 \left(5 F_1+F_2\right) & -6 \left(F_1+F_2\right) & 6 \left(F_1-F_2\right) \\
 30 & 6 \left(3 F_1+2 F_2\right) & 6 \left(3 F_1+2 F_2\right) & 6 \left(3 F_1+2 F_2\right) & 6 \left(3 F_1+2 F_2\right) \\
 31 & -6 \left(8 F_1+3 F_2\right) & 6 \left(F_2-2 F_1\right) & -6 \left(4 F_1+F_2\right) & -6 \left(2 F_1+F_2\right) \\
 32 & 6 \left(3 F_1+2 F_2\right) & -6 \left(3 F_1+2 F_2\right) & 6 \left(3 F_1+2 F_2\right) & 6 F_1 \\
 33 & 0 & 0 & -12 \left(3 F_1+2 F_2\right) & 0 \\
 34 & 12 F_1-6 F_2 & 12 F_1-6 F_2 & 6 \left(8 F_1+3 F_2\right) & 6 \left(2 F_1+F_2\right) \\
 35 & 6 \left(3 F_1+2 F_2\right) & 6 \left(3 F_1+2 F_2\right) & -6 \left(3 F_1+2 F_2\right) & -6 \left(3 F_1+2 F_2\right) \\
\end{array}
\right)
\end{equation}
}
{\scriptsize
\begin{equation}\label{2tavola168}
 \left(
\begin{array}{c|cccc}
 0 & 5 & 6 & 7 & 8 \\
 \hline
 1 & 6 \left(3 F_1+2 F_2\right) & -6 \left(3 F_1+2 F_2\right) & -6 \left(3 F_1+2 F_2\right) & 6 \left(3 F_1+2 F_2\right) \\
 2 & -6 \left(2 F_1+F_2\right) & 6 \left(F_2-2 F_1\right) & 6 \left(F_2-2 F_1\right) & -6 \left(8 F_1+3 F_2\right) \\
 3 & -18 F_1 & 18 F_1 & 6 \left(3 F_1+2 F_2\right) & 6 \left(3 F_1+2 F_2\right) \\
 4 & 6 \left(5 F_1+F_2\right) & -6 \left(F_1+F_2\right) & 6 \left(5 F_1+F_2\right) & 6 \left(F_1-F_2\right) \\
 5 & 0 & 0 & -12 \left(2 F_1+F_2\right) & -12 \left(F_1+F_2\right) \\
 6 & 6 \left(3 F_1+F_2\right) & -6 \left(F_1+3 F_2\right) & 6 \left(5 F_1+F_2\right) & 6 \left(5 F_1+F_2\right) \\
 7 & 6 \left(5 F_1+2 F_2\right) & 6 \left(F_1+2 F_2\right) & -6 \left(3 F_1+2 F_2\right) & -6 \left(3 F_1+2 F_2\right) \\
 8 & -6 \left(3 F_1+2 F_2\right) & 6 \left(5 F_1+2 F_2\right) & -6 \left(3 F_1+2 F_2\right) & 6 \left(F_1+2 F_2\right) \\
 9 & -6 \left(3 F_1+2 F_2\right) & -6 \left(3 F_1+2 F_2\right) & 6 \left(F_1+2 F_2\right) & 6 \left(5 F_1+2 F_2\right) \\
 10 & -6 \left(3 F_1+2 F_2\right) & -6 \left(3 F_1+2 F_2\right) & 6 \left(3 F_1+2 F_2\right) & 6 \left(3 F_1+2 F_2\right) \\
 11 & 6 \left(F_1+2 F_2\right) & -6 \left(3 F_1+2 F_2\right) & -6 \left(3 F_1+2 F_2\right) & -6 \left(3 F_1+2 F_2\right) \\
 12 & 6 \left(2 F_1+F_2\right) & 6 \left(F_2-2 F_1\right) & 6 \left(2 F_1+F_2\right) & 6 \left(F_2-2 F_1\right) \\
 13 & 6 F_2 & 6 F_2 & 6 \left(F_2-2 F_1\right) & 6 \left(2 F_1+F_2\right) \\
 14 & 18 F_1 & 18 F_1 & 18 F_1 & 18 F_1 \\
 15 & -6 \left(5 F_1+F_2\right) & -6 \left(F_1+F_2\right) & -6 \left(3 F_1+F_2\right) & -6 \left(3 F_1+F_2\right) \\
 16 & -6 F_1 & 6 \left(3 F_1+2 F_2\right) & -6 \left(3 F_1+2 F_2\right) & -6 \left(3 F_1+2 F_2\right) \\
 17 & -6 \left(4 F_1+F_2\right) & -6 \left(2 F_1+F_2\right) & 6 \left(F_2-2 F_1\right) & 6 \left(F_2-2 F_1\right) \\
 18 & 6 \left(3 F_1+2 F_2\right) & -18 F_1 & 6 \left(3 F_1+2 F_2\right) & 18 F_1 \\
 19 & 6 \left(5 F_1+F_2\right) & 6 \left(5 F_1+F_2\right) & 6 \left(F_1-F_2\right) & -6 \left(F_1+F_2\right) \\
 20 & 6 \left(3 F_1+F_2\right) & 6 \left(3 F_1+F_2\right) & 6 \left(5 F_1+F_2\right) & -6 \left(F_1+3 F_2\right) \\
 21 & -6 \left(3 F_1+2 F_2\right) & 6 \left(5 F_1+2 F_2\right) & -6 \left(3 F_1+2 F_2\right) & 6 \left(F_1+2 F_2\right) \\
 22 & -12 F_1 & -12 \left(3 F_1+2 F_2\right) & -12 F_1 & 12 F_1 \\
 23 & -6 \left(3 F_1+2 F_2\right) & -6 \left(3 F_1+2 F_2\right) & 6 \left(3 F_1+2 F_2\right) & -6 \left(3 F_1+2 F_2\right) \\
 24 & 18 \left(F_1+F_2\right) & 18 \left(F_1+F_2\right) & -6 \left(F_1+F_2\right) & -6 \left(5 F_1+F_2\right) \\
 25 & -6 \left(F_1+3 F_2\right) & 6 \left(3 F_1+F_2\right) & 6 \left(F_1+F_2\right) & 6 \left(3 F_1+F_2\right) \\
 26 & -6 F_1 & -6 F_1 & -6 \left(3 F_1+2 F_2\right) & 6 \left(3 F_1+2 F_2\right) \\
 27 & 6 \left(2 F_1+F_2\right) & -6 \left(4 F_1+F_2\right) & 6 \left(F_2-2 F_1\right) & -6 \left(2 F_1+F_2\right) \\
 28 & -6 F_2 & 6 \left(8 F_1+3 F_2\right) & 6 \left(4 F_1+F_2\right) & 6 \left(2 F_1+F_2\right) \\
 29 & 6 \left(F_1+F_2\right) & 6 \left(3 F_1+F_2\right) & 6 \left(5 F_1+F_2\right) & 6 \left(3 F_1+F_2\right) \\
 30 & -6 \left(F_1+2 F_2\right) & -6 \left(5 F_1+2 F_2\right) & -6 \left(5 F_1+2 F_2\right) & -6 \left(F_1+2 F_2\right) \\
 31 & 6 \left(F_2-2 F_1\right) & 6 F_2 & 6 \left(2 F_1+F_2\right) & 6 F_2 \\
 32 & 6 F_1 & -6 F_1 & -6 \left(3 F_1+2 F_2\right) & -6 F_1 \\
 33 & -12 F_1 & 12 F_1 & 12 \left(F_1+F_2\right) & 12 \left(2 F_1+F_2\right) \\
 34 & 6 \left(4 F_1+F_2\right) & -6 F_2 & -6 F_2 & -6 \left(2 F_1+F_2\right) \\
 35 & 6 F_1 & -6 F_1 & -6 F_1 & 6 F_1 \\
\end{array}
\right)
\end{equation}
}
{\scriptsize
\begin{equation}\label{3tavola168}
  \left(
\begin{array}{c|cccc}
 0 & 9 & 10 & 11 & 12 \\
 \hline
 1 & -\frac{6 \left(3 F_1+2 F_2\right)}{\sqrt{7}} & -\frac{6 \left(3 F_1+2 F_2\right)}{\sqrt{7}} & \frac{6 \left(3 F_1+2 F_2\right)}{\sqrt{7}}
   & \frac{6 \left(3 F_1+2 F_2\right)}{\sqrt{7}} \\
 2 & \frac{6 \left(8 F_1+3 F_2\right)}{\sqrt{7}} & \frac{6 \left(8 F_1+3 F_2\right)}{\sqrt{7}} & \frac{6 \left(2 F_1-F_2\right)}{\sqrt{7}} &
   \frac{6 \left(4 F_1+5 F_2\right)}{\sqrt{7}} \\
 3 & -6 \sqrt{7} F_1 & -6 \sqrt{7} F_1 & -6 \sqrt{7} F_1 & 6 \sqrt{7} F_1 \\
 4 & \frac{6 \left(5 F_1+F_2\right)}{\sqrt{7}} & -\frac{6 \left(F_1+3 F_2\right)}{\sqrt{7}} & -\frac{6 \left(F_1+3 F_2\right)}{\sqrt{7}} & -6
   \sqrt{7} \left(3 F_1+F_2\right) \\
 5 & -\frac{12 \left(3 F_1+2 F_2\right)}{\sqrt{7}} & \frac{12 \left(3 F_1+2 F_2\right)}{\sqrt{7}} & 0 & 12 \sqrt{7} F_1 \\
 6 & -\frac{6 \left(5 F_1+F_2\right)}{\sqrt{7}} & -\frac{6 \left(11 F_1+5 F_2\right)}{\sqrt{7}} & \frac{6 \left(5 F_1+F_2\right)}{\sqrt{7}} &
   -\frac{6 \left(F_1+3 F_2\right)}{\sqrt{7}} \\
 7 & \frac{6 \left(F_1-4 F_2\right)}{\sqrt{7}} & \frac{6 \left(13 F_1+4 F_2\right)}{\sqrt{7}} & -6 \sqrt{7} F_1 & -6 \sqrt{7} F_1 \\
 8 & -6 \sqrt{7} F_1 & -6 \sqrt{7} F_1 & \frac{6 \left(F_1-4 F_2\right)}{\sqrt{7}} & \frac{6 \left(13 F_1+4 F_2\right)}{\sqrt{7}} \\
 9 & \frac{6 \left(13 F_1+4 F_2\right)}{\sqrt{7}} & \frac{6 \left(F_1-4 F_2\right)}{\sqrt{7}} & -6 \sqrt{7} F_1 & -6 \sqrt{7} F_1 \\
 10 & \frac{18 \left(3 F_1+2 F_2\right)}{\sqrt{7}} & -\frac{18 \left(3 F_1+2 F_2\right)}{\sqrt{7}} & 6 \sqrt{7} F_1 & -6 \sqrt{7} F_1 \\
 11 & \frac{6 \left(F_1-4 F_2\right)}{\sqrt{7}} & \frac{6 \left(13 F_1+4 F_2\right)}{\sqrt{7}} & -6 \sqrt{7} F_1 & \frac{6 \left(F_1-4
   F_2\right)}{\sqrt{7}} \\
 12 & \frac{18 \left(F_2-2 F_1\right)}{\sqrt{7}} & \frac{18 \left(F_2-2 F_1\right)}{\sqrt{7}} & 6 \sqrt{7} \left(2 F_1+F_2\right) & 6 \sqrt{7}
   \left(2 F_1+F_2\right) \\
 13 & \frac{6 \left(4 F_1+5 F_2\right)}{\sqrt{7}} & \frac{6 \left(F_2-2 F_1\right)}{\sqrt{7}} & 6 \sqrt{7} \left(2 F_1+F_2\right) & 6 \sqrt{7}
   F_2 \\
 14 & -\frac{18 \left(3 F_1+2 F_2\right)}{\sqrt{7}} & -\frac{18 \left(3 F_1+2 F_2\right)}{\sqrt{7}} & -6 \sqrt{7} \left(3 F_1+2 F_2\right) &
   -6 \sqrt{7} \left(3 F_1+2 F_2\right) \\
 15 & \frac{6 \left(11 F_1+5 F_2\right)}{\sqrt{7}} & \frac{6 \left(5 F_1+F_2\right)}{\sqrt{7}} & 6 \sqrt{7} \left(3 F_1+F_2\right) & 6
   \sqrt{7} \left(F_1+F_2\right) \\
 16 & 6 \sqrt{7} F_1 & -\frac{6 \left(3 F_1+2 F_2\right)}{\sqrt{7}} & -\frac{6 \left(3 F_1+2 F_2\right)}{\sqrt{7}} & \frac{6 \left(3 F_1+2
   F_2\right)}{\sqrt{7}} \\
 17 & 6 \sqrt{7} F_2 & \frac{6 \left(8 F_1+3 F_2\right)}{\sqrt{7}} & \frac{6 \left(8 F_1+3 F_2\right)}{\sqrt{7}} & \frac{6 \left(2
   F_1-F_2\right)}{\sqrt{7}} \\
 18 & 6 \sqrt{7} F_1 & -6 \sqrt{7} F_1 & -6 \sqrt{7} F_1 & -6 \sqrt{7} F_1 \\
 19 & -6 \sqrt{7} \left(3 F_1+F_2\right) & -\frac{6 \left(F_1+3 F_2\right)}{\sqrt{7}} & \frac{6 \left(5 F_1+F_2\right)}{\sqrt{7}} & -\frac{6
   \left(F_1+3 F_2\right)}{\sqrt{7}} \\
 20 & -6 \sqrt{7} \left(F_1+F_2\right) & -\frac{6 \left(5 F_1+F_2\right)}{\sqrt{7}} & -\frac{6 \left(11 F_1+5 F_2\right)}{\sqrt{7}} & \frac{6
   \left(5 F_1+F_2\right)}{\sqrt{7}} \\
 21 & -6 \sqrt{7} F_1 & \frac{6 \left(F_1-4 F_2\right)}{\sqrt{7}} & \frac{6 \left(13 F_1+4 F_2\right)}{\sqrt{7}} & -6 \sqrt{7} F_1 \\
 22 & 12 \sqrt{7} F_1 & \frac{12 \left(3 F_1+2 F_2\right)}{\sqrt{7}} & -\frac{12 \left(3 F_1+2 F_2\right)}{\sqrt{7}} & -\frac{12 \left(3 F_1+2
   F_2\right)}{\sqrt{7}} \\
 23 & 6 \sqrt{7} F_1 & \frac{18 \left(3 F_1+2 F_2\right)}{\sqrt{7}} & -\frac{18 \left(3 F_1+2 F_2\right)}{\sqrt{7}} & 6 \sqrt{7} F_1 \\
 24 & 6 \sqrt{7} \left(F_1+F_2\right) & -\frac{6 \left(5 F_1+F_2\right)}{\sqrt{7}} & 6 \sqrt{7} \left(F_1+F_2\right) & 6 \sqrt{7}
   \left(F_1+F_2\right) \\
 25 & -6 \sqrt{7} \left(F_1+F_2\right) & -\frac{6 \left(5 F_1+F_2\right)}{\sqrt{7}} & -\frac{6 \left(11 F_1+5 F_2\right)}{\sqrt{7}} & -6
   \sqrt{7} \left(F_1+F_2\right) \\
 26 & -6 \sqrt{7} F_1 & 6 \sqrt{7} F_1 & -\frac{6 \left(3 F_1+2 F_2\right)}{\sqrt{7}} & -\frac{6 \left(3 F_1+2 F_2\right)}{\sqrt{7}} \\
 27 & 6 \sqrt{7} \left(2 F_1+F_2\right) & 6 \sqrt{7} F_2 & \frac{6 \left(8 F_1+3 F_2\right)}{\sqrt{7}} & \frac{6 \left(8 F_1+3
   F_2\right)}{\sqrt{7}} \\
 28 & -6 \sqrt{7} \left(2 F_1+F_2\right) & -6 \sqrt{7} \left(2 F_1+F_2\right) & -\frac{6 \left(8 F_1+3 F_2\right)}{\sqrt{7}} & \frac{6
   \left(F_2-2 F_1\right)}{\sqrt{7}} \\
 29 & -6 \sqrt{7} \left(3 F_1+F_2\right) & -6 \sqrt{7} \left(F_1+F_2\right) & -\frac{6 \left(5 F_1+F_2\right)}{\sqrt{7}} & -\frac{6 \left(11
   F_1+5 F_2\right)}{\sqrt{7}} \\
 30 & 6 \sqrt{7} F_1 & 6 \sqrt{7} F_1 & 6 \sqrt{7} F_1 & 6 \sqrt{7} F_1 \\
 31 & 6 \sqrt{7} F_2 & 6 \sqrt{7} \left(2 F_1+F_2\right) & \frac{6 \left(4 F_1+5 F_2\right)}{\sqrt{7}} & \frac{6 \left(F_2-2
   F_1\right)}{\sqrt{7}} \\
 32 & 6 \sqrt{7} F_1 & -6 \sqrt{7} F_1 & 6 \sqrt{7} F_1 & -\frac{6 \left(3 F_1+2 F_2\right)}{\sqrt{7}} \\
 33 & 0 & 0 & -12 \sqrt{7} F_1 & 0 \\
 34 & -6 \sqrt{7} \left(2 F_1+F_2\right) & -6 \sqrt{7} \left(2 F_1+F_2\right) & -6 \sqrt{7} F_2 & \frac{6 \left(2 F_1-F_2\right)}{\sqrt{7}} \\
 35 & 6 \sqrt{7} F_1 & 6 \sqrt{7} F_1 & -6 \sqrt{7} F_1 & -6 \sqrt{7} F_1 \\
\end{array}
\right)
\end{equation}
}
{\scriptsize
\begin{equation}\label{4tavola168}
  \left(
\begin{array}{ccccc}
 0 & 13 & 14 & 15 & 16 \\
 1 & 6 \sqrt{7} F_1 & -6 \sqrt{7} F_1 & -6 \sqrt{7} F_1 & 6 \sqrt{7} F_1 \\
 2 & \frac{6 \left(F_2-2 F_1\right)}{\sqrt{7}} & 6 \sqrt{7} \left(2 F_1+F_2\right) & 6 \sqrt{7} \left(2 F_1+F_2\right) & 6 \sqrt{7} F_2 \\
 3 & \frac{18 \left(3 F_1+2 F_2\right)}{\sqrt{7}} & -\frac{18 \left(3 F_1+2 F_2\right)}{\sqrt{7}} & 6 \sqrt{7} F_1 & 6 \sqrt{7} F_1 \\
 4 & -6 \sqrt{7} \left(F_1+F_2\right) & -\frac{6 \left(5 F_1+F_2\right)}{\sqrt{7}} & -6 \sqrt{7} \left(F_1+F_2\right) & -\frac{6 \left(11
   F_1+5 F_2\right)}{\sqrt{7}} \\
 5 & 0 & 0 & \frac{12 \left(F_2-2 F_1\right)}{\sqrt{7}} & -\frac{12 \left(5 F_1+F_2\right)}{\sqrt{7}} \\
 6 & -\frac{6 \left(F_1+3 F_2\right)}{\sqrt{7}} & -6 \sqrt{7} \left(3 F_1+F_2\right) & -6 \sqrt{7} \left(F_1+F_2\right) & -6 \sqrt{7}
   \left(F_1+F_2\right) \\
 7 & \frac{6 \left(F_1-4 F_2\right)}{\sqrt{7}} & \frac{6 \left(13 F_1+4 F_2\right)}{\sqrt{7}} & -6 \sqrt{7} F_1 & -6 \sqrt{7} F_1 \\
 8 & -6 \sqrt{7} F_1 & \frac{6 \left(F_1-4 F_2\right)}{\sqrt{7}} & -6 \sqrt{7} F_1 & \frac{6 \left(13 F_1+4 F_2\right)}{\sqrt{7}} \\
 9 & -6 \sqrt{7} F_1 & -6 \sqrt{7} F_1 & \frac{6 \left(13 F_1+4 F_2\right)}{\sqrt{7}} & \frac{6 \left(F_1-4 F_2\right)}{\sqrt{7}} \\
 10 & -6 \sqrt{7} F_1 & -6 \sqrt{7} F_1 & 6 \sqrt{7} F_1 & 6 \sqrt{7} F_1 \\
 11 & \frac{6 \left(13 F_1+4 F_2\right)}{\sqrt{7}} & -6 \sqrt{7} F_1 & -6 \sqrt{7} F_1 & -6 \sqrt{7} F_1 \\
 12 & \frac{6 \left(2 F_1-F_2\right)}{\sqrt{7}} & 6 \sqrt{7} \left(2 F_1+F_2\right) & \frac{6 \left(2 F_1-F_2\right)}{\sqrt{7}} & 6 \sqrt{7}
   \left(2 F_1+F_2\right) \\
 13 & \frac{6 \left(8 F_1+3 F_2\right)}{\sqrt{7}} & \frac{6 \left(8 F_1+3 F_2\right)}{\sqrt{7}} & 6 \sqrt{7} \left(2 F_1+F_2\right) & \frac{6
   \left(2 F_1-F_2\right)}{\sqrt{7}} \\
 14 & -\frac{18 \left(3 F_1+2 F_2\right)}{\sqrt{7}} & -\frac{18 \left(3 F_1+2 F_2\right)}{\sqrt{7}} & -\frac{18 \left(3 F_1+2
   F_2\right)}{\sqrt{7}} & -\frac{18 \left(3 F_1+2 F_2\right)}{\sqrt{7}} \\
 15 & 6 \sqrt{7} \left(F_1+F_2\right) & -\frac{6 \left(5 F_1+F_2\right)}{\sqrt{7}} & \frac{6 \left(F_1+3 F_2\right)}{\sqrt{7}} & \frac{6
   \left(F_1+3 F_2\right)}{\sqrt{7}} \\
 16 & \frac{6 \left(3 F_1+2 F_2\right)}{\sqrt{7}} & 6 \sqrt{7} F_1 & -6 \sqrt{7} F_1 & -6 \sqrt{7} F_1 \\
 17 & \frac{6 \left(4 F_1+5 F_2\right)}{\sqrt{7}} & \frac{6 \left(F_2-2 F_1\right)}{\sqrt{7}} & 6 \sqrt{7} \left(2 F_1+F_2\right) & 6 \sqrt{7}
   \left(2 F_1+F_2\right) \\
 18 & 6 \sqrt{7} F_1 & \frac{18 \left(3 F_1+2 F_2\right)}{\sqrt{7}} & 6 \sqrt{7} F_1 & -\frac{18 \left(3 F_1+2 F_2\right)}{\sqrt{7}} \\
 19 & -6 \sqrt{7} \left(F_1+F_2\right) & -6 \sqrt{7} \left(F_1+F_2\right) & -\frac{6 \left(11 F_1+5 F_2\right)}{\sqrt{7}} & -\frac{6 \left(5
   F_1+F_2\right)}{\sqrt{7}} \\
 20 & -\frac{6 \left(F_1+3 F_2\right)}{\sqrt{7}} & -\frac{6 \left(F_1+3 F_2\right)}{\sqrt{7}} & -6 \sqrt{7} \left(F_1+F_2\right) & -6 \sqrt{7}
   \left(3 F_1+F_2\right) \\
 21 & -6 \sqrt{7} F_1 & \frac{6 \left(F_1-4 F_2\right)}{\sqrt{7}} & -6 \sqrt{7} F_1 & \frac{6 \left(13 F_1+4 F_2\right)}{\sqrt{7}} \\
 22 & \frac{12 \left(3 F_1+2 F_2\right)}{\sqrt{7}} & -12 \sqrt{7} F_1 & \frac{12 \left(3 F_1+2 F_2\right)}{\sqrt{7}} & -\frac{12 \left(3 F_1+2
   F_2\right)}{\sqrt{7}} \\
 23 & -6 \sqrt{7} F_1 & -6 \sqrt{7} F_1 & 6 \sqrt{7} F_1 & -6 \sqrt{7} F_1 \\
 24 & \frac{18 \left(5 F_1+F_2\right)}{\sqrt{7}} & \frac{18 \left(5 F_1+F_2\right)}{\sqrt{7}} & -\frac{6 \left(5 F_1+F_2\right)}{\sqrt{7}} & 6
   \sqrt{7} \left(F_1+F_2\right) \\
 25 & -6 \sqrt{7} \left(3 F_1+F_2\right) & -\frac{6 \left(F_1+3 F_2\right)}{\sqrt{7}} & \frac{6 \left(5 F_1+F_2\right)}{\sqrt{7}} & -\frac{6
   \left(F_1+3 F_2\right)}{\sqrt{7}} \\
 26 & \frac{6 \left(3 F_1+2 F_2\right)}{\sqrt{7}} & \frac{6 \left(3 F_1+2 F_2\right)}{\sqrt{7}} & -6 \sqrt{7} F_1 & 6 \sqrt{7} F_1 \\
 27 & \frac{6 \left(2 F_1-F_2\right)}{\sqrt{7}} & \frac{6 \left(4 F_1+5 F_2\right)}{\sqrt{7}} & 6 \sqrt{7} \left(2 F_1+F_2\right) & \frac{6
   \left(F_2-2 F_1\right)}{\sqrt{7}} \\
 28 & -\frac{6 \left(8 F_1+3 F_2\right)}{\sqrt{7}} & -6 \sqrt{7} F_2 & -\frac{6 \left(4 F_1+5 F_2\right)}{\sqrt{7}} & \frac{6 \left(2
   F_1-F_2\right)}{\sqrt{7}} \\
 29 & \frac{6 \left(5 F_1+F_2\right)}{\sqrt{7}} & -\frac{6 \left(F_1+3 F_2\right)}{\sqrt{7}} & -6 \sqrt{7} \left(F_1+F_2\right) & -\frac{6
   \left(F_1+3 F_2\right)}{\sqrt{7}} \\
 30 & -\frac{6 \left(13 F_1+4 F_2\right)}{\sqrt{7}} & -\frac{6 \left(F_1-4 F_2\right)}{\sqrt{7}} & -\frac{6 \left(F_1-4 F_2\right)}{\sqrt{7}}
   & -\frac{6 \left(13 F_1+4 F_2\right)}{\sqrt{7}} \\
 31 & 6 \sqrt{7} \left(2 F_1+F_2\right) & \frac{6 \left(8 F_1+3 F_2\right)}{\sqrt{7}} & \frac{6 \left(2 F_1-F_2\right)}{\sqrt{7}} & \frac{6
   \left(8 F_1+3 F_2\right)}{\sqrt{7}} \\
 32 & -\frac{6 \left(3 F_1+2 F_2\right)}{\sqrt{7}} & \frac{6 \left(3 F_1+2 F_2\right)}{\sqrt{7}} & -6 \sqrt{7} F_1 & \frac{6 \left(3 F_1+2
   F_2\right)}{\sqrt{7}} \\
 33 & \frac{12 \left(3 F_1+2 F_2\right)}{\sqrt{7}} & -\frac{12 \left(3 F_1+2 F_2\right)}{\sqrt{7}} & \frac{12 \left(5
   F_1+F_2\right)}{\sqrt{7}} & \frac{12 \left(2 F_1-F_2\right)}{\sqrt{7}} \\
 34 & -\frac{6 \left(4 F_1+5 F_2\right)}{\sqrt{7}} & -\frac{6 \left(8 F_1+3 F_2\right)}{\sqrt{7}} & -\frac{6 \left(8 F_1+3
   F_2\right)}{\sqrt{7}} & \frac{6 \left(F_2-2 F_1\right)}{\sqrt{7}} \\
 35 & -\frac{6 \left(3 F_1+2 F_2\right)}{\sqrt{7}} & \frac{6 \left(3 F_1+2 F_2\right)}{\sqrt{7}} & \frac{6 \left(3 F_1+2 F_2\right)}{\sqrt{7}}
   & -\frac{6 \left(3 F_1+2 F_2\right)}{\sqrt{7}} \\
\end{array}
\right)
\end{equation}
}
\section{The Explicit form of the $35\times14$ coefficient matrix
$\mathfrak{C}_{\mathfrak{q}\alpha}(\psi)$
 for the $\mathcal{N}=1$ Englert 3-form}
 \label{lagrandpap}
{\small In this appendix we display display the explicit
form of the 4-moduli dependent coefficients that define our
 $4$-parameter dependent solution of Englert equation leading to an $N=1$ supergravity solution invariant
 under the discrete group $\mathrm{G_{21}}$. Because of the large form of the output the entries of the matrix
 are organized in 4 tables containing the first columns from 1to 4, from 5 to 8, from 8 to 12 and finally from
 13 to 14.}
\vskip 0.5cm
{\tiny
\begin{doublespace}
\noindent\(\pmb{
\begin{array}{|c|c|c|c|c|}
\hline
 0 & 1 & 2 & 3 & 4 \\
\hline
 1 & \frac{3 \left(6 \psi _1+\psi _2+2 \psi _3-\psi _4\right)}{\sqrt{2}} & -3 \sqrt{2} \left(\psi _2+\psi _4\right) & -\frac{3 \left(9 \psi _1+2
\psi _2+2 \psi _3-\psi _4\right)}{\sqrt{2}} & -\frac{3 \left(\psi _1+5 \psi _2+4 \psi _4\right)}{\sqrt{2}} \\
\hline
 2 & -\frac{3 \left(4 \psi _1+3 \psi _2-2 \psi _3+\psi _4\right)}{\sqrt{2}} & -3 \sqrt{2} \left(2 \psi _1+3 \psi _2+2 \psi _4\right) & \frac{3 \left(5
\psi _1+6 \psi _2+3 \psi _4\right)}{\sqrt{2}} & \frac{3 \left(3 \psi _1+\psi _2+2 \psi _3-2 \psi _4\right)}{\sqrt{2}} \\
\hline
 3 & -\frac{3 \left(6 \psi _1+3 \psi _2+4 \psi _3-\psi _4\right)}{\sqrt{2}} & -3 \sqrt{2} \left(\psi _1+\psi _2\right) & \frac{3 \left(\psi _1+2
\psi _2+2 \psi _3+\psi _4\right)}{\sqrt{2}} & \frac{3 \left(\psi _1-\psi _2+2 \psi _3-2 \psi _4\right)}{\sqrt{2}} \\
\hline
 4 & 3 \sqrt{2} \left(2 \psi _1+3 \psi _2+2 \psi _4\right) & -3 \sqrt{2} \psi _1 & -3 \sqrt{2} \left(\psi _1-\psi _2+\psi _3-2 \psi _4\right) & -3
\sqrt{2} \left(\psi _2-\psi _3+\psi _4\right) \\
\hline
 5 & -\frac{3 \left(2 \psi _1-3 \psi _2+4 \psi _3-5 \psi _4\right)}{\sqrt{2}} & -3 \sqrt{2} \left(2 \psi _1+\psi _2\right) & \frac{3 \left(3 \psi
_1+4 \psi _3-3 \psi _4\right)}{\sqrt{2}} & \frac{3 \left(\psi _1+\psi _2\right)}{\sqrt{2}} \\
\hline
 6 & \frac{3 \left(8 \psi _1+5 \psi _2+2 \psi _3+\psi _4\right)}{\sqrt{2}} & -3 \sqrt{2} \left(3 \psi _1-\psi _2+\psi _3-\psi _4\right) & \frac{3
\left(15 \psi _1+6 \psi _2+6 \psi _3-\psi _4\right)}{\sqrt{2}} & -\frac{3 \left(13 \psi _1+\psi _2+8 \psi _3-6 \psi _4\right)}{\sqrt{2}} \\
\hline
 7 & 3 \sqrt{2} \left(\psi _2+\psi _4\right) & 3 \sqrt{2} \left(5 \psi _1-\psi _2+3 \psi _3-3 \psi _4\right) & -3 \sqrt{2} \left(5 \psi _1+2 \psi
_3-2 \psi _4\right) & -3 \sqrt{2} \left(4 \psi _1-5 \psi _2+4 \psi _3-6 \psi _4\right) \\
\hline
 8 & -\frac{3 \left(6 \psi _1+\psi _2+4 \psi _3-\psi _4\right)}{\sqrt{2}} & 3 \sqrt{2} \left(\psi _2-\psi _3+\psi _4\right) & \frac{3 \left(9 \psi
_1+2 \psi _2+4 \psi _3-\psi _4\right)}{\sqrt{2}} & -\frac{3 \left(13 \psi _1+\psi _2+8 \psi _3-6 \psi _4\right)}{\sqrt{2}} \\
\hline
 9 & \frac{3 \left(-6 \psi _1+\psi _2-4 \psi _3+3 \psi _4\right)}{\sqrt{2}} & 3 \sqrt{2} \left(5 \psi _1+2 \psi _3-2 \psi _4\right) & -\frac{3 \left(\psi
_1-2 \psi _2+4 \psi _3-3 \psi _4\right)}{\sqrt{2}} & \frac{3 \left(3 \psi _1+\psi _2-2 \psi _4\right)}{\sqrt{2}} \\
\hline
 10 & \frac{3 \left(4 \psi _1+\psi _2+4 \psi _3-\psi _4\right)}{\sqrt{2}} & -3 \sqrt{2} \left(\psi _1-2 \psi _2+2 \psi _3-3 \psi _4\right) & \frac{3
\left(\psi _1-2 \psi _2+6 \psi _3-3 \psi _4\right)}{\sqrt{2}} & -\frac{3 \left(5 \psi _1-3 \psi _2+6 \psi _3-8 \psi _4\right)}{\sqrt{2}} \\
\hline
 11 & \frac{3 \left(5 \psi _2-4 \psi _3+5 \psi _4\right)}{\sqrt{2}} & 3 \sqrt{2} \left(\psi _1-\psi _2+\psi _3-2 \psi _4\right) & -\frac{3 \left(11
\psi _1+12 \psi _2+7 \psi _4\right)}{\sqrt{2}} & \frac{3 \left(7 \psi _1-\psi _2-2 \psi _4\right)}{\sqrt{2}} \\
\hline
 12 & 3 \sqrt{2} \left(2 \psi _1+3 \psi _2+2 \psi _4\right) & 3 \sqrt{2} \left(\psi _2-\psi _3+\psi _4\right) & 3 \sqrt{2} \left(\psi _1-\psi _2+3
\psi _3-\psi _4\right) & -3 \sqrt{2} \left(3 \psi _1-3 \psi _2+\psi _3-3 \psi _4\right) \\
\hline
 13 & \frac{6 \psi _1-3 \left(\psi _2-2 \psi _3+\psi _4\right)}{\sqrt{2}} & -3 \sqrt{2} \left(\psi _1+\psi _2+\psi _3\right) & \frac{9 \psi _1+6
\psi _2-3 \psi _4}{\sqrt{2}} & \frac{3 \left(9 \psi _1-\psi _2+6 \psi _3-4 \psi _4\right)}{\sqrt{2}} \\
\hline
 14 & \frac{6 \psi _1-3 \left(\psi _2-2 \psi _3+\psi _4\right)}{\sqrt{2}} & -3 \sqrt{2} \left(\psi _1+\psi _2+\psi _3\right) & -\frac{3 \left(5 \psi
_1+6 \psi _2+4 \psi _3+\psi _4\right)}{\sqrt{2}} & -\frac{3 \left(7 \psi _1+\psi _2+2 \psi _3\right)}{\sqrt{2}} \\
\hline
 15 & \frac{3 \left(-2 \psi _1+\psi _2-2 \psi _3+\psi _4\right)}{\sqrt{2}} & 3 \sqrt{2} \left(\psi _1+\psi _2+\psi _3\right) & \frac{3 \left(\psi
_1-6 \psi _2+4 \psi _3-7 \psi _4\right)}{\sqrt{2}} & -\frac{3 \left(9 \psi _1+3 \psi _2+2 \psi _3\right)}{\sqrt{2}} \\
\hline
 16 & -\frac{3 \left(5 \psi _2+2 \psi _3+3 \psi _4\right)}{\sqrt{2}} & -3 \sqrt{2} \left(5 \psi _1+\psi _2+3 \psi _3\right) & \frac{3 \left(\psi
_1-2 \psi _2-3 \psi _4\right)}{\sqrt{2}} & \frac{3 \left(5 \psi _1-\psi _2+6 \psi _3-4 \psi _4\right)}{\sqrt{2}} \\
\hline
 17 & \frac{3 \left(18 \psi _1+\psi _2+8 \psi _3-7 \psi _4\right)}{\sqrt{2}} & -3 \sqrt{2} \left(\psi _1+2 \psi _2+\psi _4\right) & \frac{9 \psi
_1+6 \psi _2+6 \psi _3-3 \psi _4}{\sqrt{2}} & -\frac{3 \left(5 \psi _1-3 \psi _2+6 \psi _3-4 \psi _4\right)}{\sqrt{2}} \\
\hline
 18 & \frac{3 \left(2 \psi _1-\psi _2+\psi _4\right)}{\sqrt{2}} & -3 \sqrt{2} \left(7 \psi _1+\psi _2+3 \psi _3-2 \psi _4\right) & -\frac{3 \left(7
\psi _1-4 \psi _2+4 \psi _3-5 \psi _4\right)}{\sqrt{2}} & \frac{3 \left(5 \psi _1+5 \psi _2+2 \psi _4\right)}{\sqrt{2}} \\
\hline
 19 & 3 \sqrt{2} \left(4 \psi _1-\psi _2+2 \psi _3-2 \psi _4\right) & -3 \sqrt{2} \left(8 \psi _1+5 \psi _2+3 \psi _3+\psi _4\right) & 3 \sqrt{2}
\left(5 \psi _1-\psi _2+3 \psi _3-3 \psi _4\right) & -3 \sqrt{2} \left(5 \psi _1+\psi _2+3 \psi _3-\psi _4\right) \\
\hline
 20 & -3 \sqrt{2} \left(2 \psi _1+\psi _2-2 \psi _3+\psi _4\right) & -3 \sqrt{2} \left(\psi _1-5 \psi _2+3 \psi _3-5 \psi _4\right) & -3 \sqrt{2}
\psi _1 & 3 \sqrt{2} \left(2 \psi _1+3 \psi _2+2 \psi _4\right) \\
\hline
 21 & \frac{3 \left(\psi _2+\psi _4\right)}{\sqrt{2}} & -9 \sqrt{2} \left(\psi _2-\psi _3+\psi _4\right) & -\frac{3 \left(3 \psi _1+2 \psi _2+\psi
_4\right)}{\sqrt{2}} & \frac{-9 \psi _1+3 \psi _2+6 \psi _4}{\sqrt{2}} \\
\hline
 22 & \frac{3 \left(4 \psi _1+5 \psi _2+\psi _4\right)}{\sqrt{2}} & -3 \sqrt{2} \left(5 \psi _1+4 \psi _2+2 \psi _4\right) & \frac{3 \left(\psi _1+2
\psi _2+\psi _4\right)}{\sqrt{2}} & \frac{3 \left(7 \psi _1+5 \psi _2+2 \psi _4\right)}{\sqrt{2}} \\
\hline
 23 & -\frac{3 \left(10 \psi _1+3 \psi _2+6 \psi _3-3 \psi _4\right)}{\sqrt{2}} & 3 \sqrt{2} \left(3 \psi _1+\psi _2+3 \psi _3-\psi _4\right) & \frac{3
\left(\psi _1-2 \psi _2+2 \psi _3-3 \psi _4\right)}{\sqrt{2}} & \frac{9 \psi _1+9 \psi _2+6 \psi _4}{\sqrt{2}} \\
\hline
 24 & \frac{3 \left(2 \psi _1+3 \psi _2+2 \psi _3+\psi _4\right)}{\sqrt{2}} & 3 \sqrt{2} \left(\psi _1-\psi _2+3 \psi _3-\psi _4\right) & -\frac{3
\left(\psi _1-2 \psi _2+2 \psi _3-3 \psi _4\right)}{\sqrt{2}} & -\frac{3 \left(3 \psi _1+3 \psi _2+2 \psi _4\right)}{\sqrt{2}} \\
\hline
 25 & \frac{3 \left(2 \psi _1-9 \psi _2+6 \psi _3-11 \psi _4\right)}{\sqrt{2}} & -3 \sqrt{2} \left(5 \psi _1+\psi _2+3 \psi _3-\psi _4\right) & -\frac{3
\left(\psi _1-2 \psi _2+2 \psi _3-3 \psi _4\right)}{\sqrt{2}} & -\frac{3 \left(3 \psi _1+3 \psi _2+2 \psi _4\right)}{\sqrt{2}} \\
\hline
 26 & \frac{3 \left(20 \psi _1+7 \psi _2+8 \psi _3-3 \psi _4\right)}{\sqrt{2}} & -3 \sqrt{2} \left(\psi _1-\psi _2+2 \psi _3-2 \psi _4\right) & -\frac{3
\left(15 \psi _1+6 \psi _2+6 \psi _3-\psi _4\right)}{\sqrt{2}} & \frac{3 \left(7 \psi _1-3 \psi _2+6 \psi _3-6 \psi _4\right)}{\sqrt{2}} \\
\hline
 27 & -3 \sqrt{2} \left(4 \psi _1+\psi _2+2 \psi _3\right) & 3 \sqrt{2} \left(3 \psi _1-4 \psi _2+2 \psi _3-4 \psi _4\right) & -3 \sqrt{2} \left(5
\psi _1+\psi _2+3 \psi _3\right) & -3 \sqrt{2} \left(4 \psi _1+\psi _2+3 \psi _3-\psi _4\right) \\
\hline
 28 & -\frac{3 \left(4 \psi _1+5 \left(\psi _2+\psi _4\right)\right)}{\sqrt{2}} & -3 \sqrt{2} \left(4 \psi _1+\psi _2+2 \psi _3\right) & -\frac{3
\left(3 \psi _1+8 \psi _2+5 \psi _4\right)}{\sqrt{2}} & \frac{3 \left(5 \psi _1+\psi _2\right)}{\sqrt{2}} \\
\hline
 29 & -\frac{3 \left(2 \psi _1-7 \psi _2+2 \psi _3-7 \psi _4\right)}{\sqrt{2}} & -3 \sqrt{2} \left(6 \psi _1-\psi _2+4 \psi _3-4 \psi _4\right) &
-\frac{3 \left(7 \psi _1-6 \psi _2+4 \psi _3-7 \psi _4\right)}{\sqrt{2}} &
\frac{3 \left(5 \psi _1-\psi _2+6 \psi _3-2 \psi _4\right)}{\sqrt{2}}
\\
\hline
 30 & \frac{-6 \psi _1+9 \psi _2-6 \psi _3+9 \psi _4}{\sqrt{2}} & 3 \sqrt{2} \left(4 \psi _1-\psi _2+2 \psi _3-2 \psi _4\right) & \frac{3 \left(13
\psi _1-2 \psi _2+8 \psi _3-9 \psi _4\right)}{\sqrt{2}} & \frac{3 \left(5 \psi _1-5 \psi _2+6 \psi _3-6 \psi _4\right)}{\sqrt{2}} \\
\hline
 31 & \frac{3 \left(14 \psi _1-5 \psi _2+10 \psi _3-9 \psi _4\right)}{\sqrt{2}} & 3 \sqrt{2} \left(6 \psi _1+\psi _2+2 \psi _3-2 \psi _4\right) &
-\frac{3 \left(7 \psi _1-2 \psi _2+4 \psi _3-3 \psi _4\right)}{\sqrt{2}} &
-\frac{3 \left(15 \psi _1+9 \psi _2+6 \psi _3+2 \psi _4\right)}{\sqrt{2}}
\\
\hline
 32 & -\frac{3 \left(\psi _2-2 \psi _3+\psi _4\right)}{\sqrt{2}} & 3 \sqrt{2} \left(2 \psi _1-\psi _2+2 \psi _3-3 \psi _4\right) & -\frac{3 \left(9
\psi _1-2 \psi _2+6 \psi _3-5 \psi _4\right)}{\sqrt{2}} & -\frac{3 \left(3 \psi _1+7 \psi _2+4 \psi _4\right)}{\sqrt{2}} \\
\hline
 33 & -\frac{3 \left(8 \psi _1+3 \psi _2+6 \psi _3-\psi _4\right)}{\sqrt{2}} & 3 \sqrt{2} \left(3 \psi _2-2 \psi _3+3 \psi _4\right) & \frac{3 \left(5
\psi _1-2 \psi _2+6 \psi _3-5 \psi _4\right)}{\sqrt{2}} & -\frac{3 \left(5 \psi _1+5 \psi _2+4 \psi _4\right)}{\sqrt{2}} \\
\hline
 34 & -\frac{9 \left(3 \psi _2-2 \psi _3+3 \psi _4\right)}{\sqrt{2}} & -3 \sqrt{2} \left(\psi _2+\psi _4\right) & \frac{3 \left(3 \psi _1-6 \psi
_2+6 \psi _3-7 \psi _4\right)}{\sqrt{2}} & \frac{3 \left(9 \psi _1+5 \psi _2\right)}{\sqrt{2}} \\
\hline
 35 & -3 \sqrt{2} \left(2 \psi _1-\psi _2+2 \psi _3-\psi _4\right) & -6 \sqrt{2} \left(\psi _1+\psi _2+\psi _3\right) & 3 \sqrt{2} \left(\psi _1-2
\psi _2+2 \psi _3-3 \psi _4\right) & -3 \sqrt{2} \left(3 \psi _1+3 \psi _2+2 \psi _4\right) \\
\hline
\end{array}
}\)
\end{doublespace}
} {\tiny
\begin{doublespace}
\noindent\(\pmb{
\begin{array}{|c|c|c|c|c|}
\hline
 0 & 5 & 6 & 7 & 8 \\
\hline
 1 & \frac{3 \left(-8 \psi _1+\psi _2-6 \psi _3+3 \psi _4\right)}{\sqrt{2}} & 3 \sqrt{2} \left(4 \psi _1-5 \psi _2+4 \psi _3-6 \psi _4\right) & \frac{3
\left(4 \psi _1+\psi _2-2 \psi _3-\psi _4\right)}{\sqrt{2}} & \frac{3 \left(8 \psi _1-5 \psi _2+10 \psi _3-9 \psi _4\right)}{\sqrt{14}} \\
\hline
 2 & \frac{3 \left(2 \psi _1+3 \psi _2+\psi _4\right)}{\sqrt{2}} & -3 \sqrt{2} \left(\psi _1-\psi _2+\psi _3-2 \psi _4\right) & -\frac{3 \left(3
\psi _2+2 \psi _3+\psi _4\right)}{\sqrt{2}} & \frac{-54 \psi _1+45 \psi _2-42 \psi _3+57 \psi _4}{\sqrt{14}} \\
\hline
 3 & \frac{3 \left(12 \psi _1+11 \psi _2+7 \psi _4\right)}{\sqrt{2}} & -9 \sqrt{2} \psi _1 & \frac{3 \left(\psi _2+\psi _4\right)}{\sqrt{2}} & \frac{3
\left(12 \psi _1+7 \psi _2+4 \psi _3+5 \psi _4\right)}{\sqrt{14}} \\
\hline
 4 & -3 \sqrt{2} \left(\psi _1-5 \psi _2+3 \psi _3-5 \psi _4\right) & 3 \sqrt{2} \left(3 \psi _1+\psi _2+\psi _3+\psi _4\right) & -3 \sqrt{2} \left(2
\psi _1+\psi _2-2 \psi _3+\psi _4\right) & 3 \sqrt{\frac{2}{7}} \left(8 \psi _1+\psi _2+4 \psi _3-4 \psi _4\right) \\
\hline
 5 & \frac{6 \psi _1-9 \left(\psi _2-2 \psi _3+\psi _4\right)}{\sqrt{2}} & -3 \sqrt{2} \left(2 \psi _1-\psi _2+3 \psi _3-3 \psi _4\right) & \frac{3
\left(4 \psi _1+7 \psi _2+3 \psi _4\right)}{\sqrt{2}} & \frac{3 \left(28 \psi _1+9 \psi _2+12 \psi _3-3 \psi _4\right)}{\sqrt{14}} \\
\hline
 6 & \frac{9 \left(\psi _2+\psi _4\right)}{\sqrt{2}} & 3 \sqrt{2} \left(\psi _1-3 \left(\psi _2-\psi _3+\psi _4\right)\right) & -\frac{3 \left(6
\psi _1+\psi _2+4 \psi _3-3 \psi _4\right)}{\sqrt{2}} & \frac{3 \left(10 \psi _1-9 \psi _2+10 \psi _3-15 \psi _4\right)}{\sqrt{14}} \\
\hline
 7 & 3 \sqrt{2} \left(5 \psi _1+\psi _2+3 \psi _3-\psi _4\right) & 3 \sqrt{2} \left(2 \psi _1-\psi _2+3 \psi _3-\psi _4\right) & -3 \sqrt{2} \left(3
\psi _1-3 \psi _2+3 \psi _3-4 \psi _4\right) & 3 \sqrt{\frac{2}{7}} \left(2 \psi _1+3 \psi _2+\psi _4\right) \\
\hline
 8 & -\frac{3 \left(2 \psi _1+3 \psi _2+\psi _4\right)}{\sqrt{2}} & -3 \sqrt{2} \left(3 \psi _1+2 \psi _2+\psi _4\right) & \frac{3 \left(18 \psi
_1+5 \left(\psi _2+2 \psi _3-\psi _4\right)\right)}{\sqrt{2}} & \frac{3 \left(-3 \psi _2-4 \psi _3+\psi _4\right)}{\sqrt{14}} \\
\hline
 9 & -\frac{3 \left(12 \psi _1+5 \psi _2+6 \psi _3-\psi _4\right)}{\sqrt{2}} & 3 \sqrt{2} \left(3 \psi _1-3 \psi _2+3 \psi _3-4 \psi _4\right) &
-\frac{3 \left(9 \psi _2-4 \psi _3+9 \psi _4\right)}{\sqrt{2}} & \frac{3 \left(4 \psi _1+3 \psi _2-4 \psi _3+3 \psi _4\right)}{\sqrt{14}} \\
\hline
 10 & \frac{3 \left(2 \psi _1+5 \psi _2+3 \psi _4\right)}{\sqrt{2}} & 3 \sqrt{2} \left(2 \psi _1+3 \psi _2+2 \psi _4\right) & -\frac{3 \left(4 \psi
_1+\psi _2+\psi _4\right)}{\sqrt{2}} & \frac{3 \left(-10 \psi _1+3 \psi _2-4 \psi _3+3 \psi _4\right)}{\sqrt{14}} \\
\hline
 11 & \frac{3 \left(2 \psi _1+\psi _2-\psi _4\right)}{\sqrt{2}} & -3 \sqrt{2} \left(3 \psi _1+3 \psi _2+2 \psi _4\right) & \frac{3 \left(6 \psi _1-\psi
_2+2 \psi _3-3 \psi _4\right)}{\sqrt{2}} & \frac{78 \psi _1-3 \psi _2+36 \psi _3-33 \psi _4}{\sqrt{14}} \\
\hline
 12 & 3 \sqrt{2} \left(\psi _2+\psi _4\right) & 3 \sqrt{2} \psi _1 & -3 \sqrt{2} \left(\psi _1-\psi _2+\psi _3-2 \psi _4\right) & 3 \sqrt{\frac{2}{7}}
\left(8 \psi _1+\psi _2+4 \psi _3-4 \psi _4\right) \\
\hline
 13 & \frac{-6 \psi _1+9 \left(\psi _2-2 \psi _3+\psi _4\right)}{\sqrt{2}} & -3 \sqrt{2} \left(4 \psi _1+5 \psi _2+3 \psi _4\right) & \frac{3 \left(-2
\psi _1+\psi _2+\psi _4\right)}{\sqrt{2}} & \frac{3 \left(5 \psi _2+2 \psi _3+3 \psi _4\right)}{\sqrt{14}} \\
\hline
 14 & \frac{3 \left(18 \psi _1+7 \psi _2+6 \psi _3-\psi _4\right)}{\sqrt{2}} & -3 \sqrt{2} \left(\psi _2+\psi _4\right) & -\frac{9 \left(2 \psi _1+\psi
_2+\psi _4\right)}{\sqrt{2}} & \frac{3 \left(5 \psi _2+2 \psi _3+3 \psi _4\right)}{\sqrt{14}} \\
\hline
 15 & -\frac{3 \left(6 \psi _1+3 \psi _2+6 \psi _3-\psi _4\right)}{\sqrt{2}} & 3 \sqrt{2} \left(2 \psi _1+\psi _2+\psi _4\right) & \frac{3 \left(10
\psi _1-\psi _2+4 \psi _3-5 \psi _4\right)}{\sqrt{2}} & -\frac{3 \left(5 \psi _2+2 \psi _3+3 \psi _4\right)}{\sqrt{14}} \\
\hline
 16 & \frac{3 \left(8 \psi _1-5 \psi _2+6 \psi _3-7 \psi _4\right)}{\sqrt{2}} & -3 \sqrt{2} \left(\psi _2+\psi _4\right) & \frac{3 \left(-4 \psi
_1+\psi _2-4 \psi _3+3 \psi _4\right)}{\sqrt{2}} & 3 \sqrt{\frac{7}{2}} \left(2 \psi _1-\psi _2+2 \psi _3-\psi _4\right) \\
\hline
 17 & \frac{3 \left(\psi _2+\psi _4\right)}{\sqrt{2}} & -3 \sqrt{2} \left(6 \psi _1+\psi _2+4 \psi _3-2 \psi _4\right) & -\frac{3 \left(2 \psi _1+9
\psi _2-4 \psi _3+7 \psi _4\right)}{\sqrt{2}} & -\frac{3 \left(4 \psi _1+29 \psi _2-16 \psi _3+27 \psi _4\right)}{\sqrt{14}} \\
\hline
 18 & -\frac{3 \left(\psi _2+\psi _4\right)}{\sqrt{2}} & -3 \sqrt{2} \left(\psi _1+\psi _2\right) & \frac{3 \left(16 \psi _1+5 \psi _2+10 \psi _3-3
\psi _4\right)}{\sqrt{2}} & \frac{3 \left(24 \psi _1+13 \psi _2+16 \psi _3+\psi _4\right)}{\sqrt{14}} \\
\hline
 19 & -3 \sqrt{2} \left(\psi _2+\psi _4\right) & 3 \sqrt{2} \left(7 \psi _1+4 \psi _3-4 \psi _4\right) & -3 \sqrt{2} \left(3 \psi _1-\psi _2+3 \psi
_3-2 \psi _4\right) & 3 \sqrt{\frac{2}{7}} \left(2 \psi _1-3 \psi _2+6 \psi _3-4 \psi _4\right) \\
\hline
 20 & 3 \sqrt{2} \left(3 \psi _1+\psi _2+\psi _3+\psi _4\right) & -3 \sqrt{2} \left(\psi _2-\psi _3+\psi _4\right) & -3 \sqrt{2} \left(\psi _1-\psi
_2+\psi _3-2 \psi _4\right) & 3 \sqrt{\frac{2}{7}} \left(-20 \psi _1+5 \psi _2-14 \psi _3+11 \psi _4\right) \\
\hline
 21 & -\frac{3 \left(4 \psi _1-7 \psi _2+8 \psi _3-11 \psi _4\right)}{\sqrt{2}} & 3 \sqrt{2} \left(\psi _1+2 \psi _2+\psi _4\right) & \frac{3 \left(8
\psi _1+3 \psi _2+2 \psi _3-\psi _4\right)}{\sqrt{2}} & \frac{3 \left(2 \psi _1+3 \psi _2+\psi _4\right)}{\sqrt{14}} \\
\hline
 22 & \frac{3 \left(6 \psi _1-5 \psi _2+2 \psi _3-5 \psi _4\right)}{\sqrt{2}} & -3 \sqrt{2} \left(\psi _1+\psi _2+\psi _3\right) & -\frac{3 \left(6
\psi _1+5 \psi _2+3 \psi _4\right)}{\sqrt{2}} & -\frac{3 \left(2 \psi _1+17 \psi _2+15 \psi _4\right)}{\sqrt{14}} \\
\hline
 23 & \frac{3 \left(2 \psi _1+3 \psi _2+\psi _4\right)}{\sqrt{2}} & 3 \sqrt{2} \left(5 \psi _1+\psi _2+\psi _3-\psi _4\right) & -\frac{3 \left(12
\psi _1+5 \psi _2+4 \psi _3-\psi _4\right)}{\sqrt{2}} & \frac{48 \psi _1+45 \psi _2+6 \psi _3+33 \psi _4}{\sqrt{14}} \\
\hline
 24 & \frac{3 \left(2 \psi _1+5 \psi _2+3 \psi _4\right)}{\sqrt{2}} & 3 \sqrt{2} \left(-5 \psi _1+\psi _2-3 \psi _3+3 \psi _4\right) & \frac{3 \left(8
\psi _1+9 \psi _2+7 \psi _4\right)}{\sqrt{2}} & \frac{3 \left(-8 \psi _1+\psi _2-6 \psi _3+\psi _4\right)}{\sqrt{14}} \\
\hline
 25 & \frac{3 \left(2 \psi _1+\psi _2-\psi _4\right)}{\sqrt{2}} & 3 \sqrt{2} \left(3 \psi _1-3 \psi _2+\psi _3-3 \psi _4\right) & \frac{3 \left(4
\psi _1+\psi _2-\psi _4\right)}{\sqrt{2}} & -\frac{3 \left(48 \psi _1+19 \psi _2+18 \psi _3-5 \psi _4\right)}{\sqrt{14}} \\
\hline
 26 & -\frac{3 \left(2 \psi _1+3 \psi _2+\psi _4\right)}{\sqrt{2}} & -3 \sqrt{2} \psi _1 & -\frac{3 \left(6 \psi _1-3 \psi _2+4 \psi _3-5 \psi _4\right)}{\sqrt{2}}
& \frac{6 \psi _1-81 \psi _2+48 \psi _3-93 \psi _4}{\sqrt{14}} \\
\hline
 27 & 3 \sqrt{2} \left(5 \psi _1-\psi _2+3 \psi _3-3 \psi _4\right) & 3 \sqrt{2} \left(5 \psi _1+\psi _2+3 \psi _3-\psi _4\right) & -3 \sqrt{2} \left(\psi
_2+\psi _4\right) & -3 \sqrt{\frac{2}{7}} \left(6 \psi _1+3 \psi _2+6 \psi _3-2 \psi _4\right) \\
\hline
 28 & \frac{3 \left(12 \psi _1+\psi _2+6 \psi _3-5 \psi _4\right)}{\sqrt{2}} & -3 \sqrt{2} \left(4 \psi _1+\psi _2+3 \psi _3-\psi _4\right) & \frac{3
\left(6 \psi _1-\psi _2+4 \psi _3-3 \psi _4\right)}{\sqrt{2}} & -\frac{3 \left(30 \psi _1+15 \psi _2+16 \psi _3-3 \psi _4\right)}{\sqrt{14}} \\
\hline
 29 & \frac{3 \left(\psi _2+\psi _4\right)}{\sqrt{2}} & 3 \sqrt{2} \left(7 \psi _1+3 \left(\psi _2+\psi _3\right)\right) & \frac{6 \psi _1+9 \psi
_2+6 \psi _3+9 \psi _4}{\sqrt{2}} & \frac{36 \psi _1+39 \psi _2-6 \psi _3+9 \psi _4}{\sqrt{14}} \\
\hline
 30 & \frac{3 \left(\psi _2+\psi _4\right)}{\sqrt{2}} & -3 \sqrt{2} \left(5 \psi _1+3 \left(\psi _2+\psi _3\right)\right) & \frac{-42 \psi _1+9 \psi
_2-30 \psi _3+21 \psi _4}{\sqrt{2}} & \frac{3 \left(4 \psi _1+\psi _2-2 \psi _3-\psi _4\right)}{\sqrt{14}} \\
\hline
 31 & \frac{3 \left(\psi _2+\psi _4\right)}{\sqrt{2}} & -3 \sqrt{2} \left(3 \psi _1-\psi _2+3 \psi _3-2 \psi _4\right) & \frac{6 \psi _1-3 \left(\psi
_2-2 \psi _3+\psi _4\right)}{\sqrt{2}} & -\frac{3 \left(12 \psi _1+7 \psi _2+5 \left(-2 \psi _3+\psi _4\right)\right)}{\sqrt{14}} \\
\hline
 32 & \frac{3 \left(6 \psi _1-5 \psi _2+2 \psi _3-5 \psi _4\right)}{\sqrt{2}} & 3 \sqrt{2} \left(2 \psi _1+\psi _2\right) & \frac{-6 \psi _1+9 \psi
_2-6 \psi _3+9 \psi _4}{\sqrt{2}} & -\frac{3 \left(10 \psi _1-5 \psi _2+6 \psi _3-7 \psi _4\right)}{\sqrt{14}} \\
\hline
 33 & \frac{3 \left(10 \psi _1+\psi _2+2 \psi _3-3 \psi _4\right)}{\sqrt{2}} & -3 \sqrt{2} \left(2 \psi _1+\psi _2\right) & \frac{6 \psi _1-9 \psi
_2+6 \psi _3-9 \psi _4}{\sqrt{2}} & \frac{3 \left(10 \psi _1-\psi _2+2 \psi _3+\psi _4\right)}{\sqrt{14}} \\
\hline
 34 & -\frac{3 \left(14 \psi _1+\psi _2+10 \psi _3-7 \psi _4\right)}{\sqrt{2}} & 3 \sqrt{2} \left(2 \psi _1+\psi _2\right) & \frac{-6 \psi _1+9 \psi
_2-6 \psi _3+9 \psi _4}{\sqrt{2}} & -\frac{9 \left(14 \psi _1+\psi _2+6 \psi _3-5 \psi _4\right)}{\sqrt{14}} \\
\hline
 35 & -3 \sqrt{2} \left(\psi _2+\psi _4\right) & 6 \sqrt{2} \left(2 \psi _1+\psi _2\right) & 3 \sqrt{2} \left(2 \psi _1-3 \psi _2+2 \psi _3-3 \psi
_4\right) & -3 \sqrt{\frac{2}{7}} \left(5 \psi _2+2 \psi _3+3 \psi _4\right) \\
\hline
\end{array}
}\)
\end{doublespace}
} {\tiny
\begin{doublespace}
\noindent\(\pmb{
\begin{array}{|c|c|c|c|c|}
\hline
 0 & 9 & 10 & 11 & 12 \\
\hline
 1 & -3 \sqrt{\frac{2}{7}} \left(2 \psi _1+3 \psi _2+\psi _4\right) & \frac{-51 \psi _1+30 \psi _2-54 \psi _3+51 \psi _4}{\sqrt{14}} & -3 \sqrt{\frac{7}{2}}
\left(\psi _1+\psi _2\right) & \frac{3 \left(2 \psi _1-5 \psi _2-6 \psi _3-\psi _4\right)}{\sqrt{14}} \\
\hline
 2 & -3 \sqrt{\frac{2}{7}} \left(8 \psi _1+\psi _2+4 \psi _3-4 \psi _4\right) & \frac{3 \left(13 \psi _1-6 \psi _2+8 \psi _3-13 \psi _4\right)}{\sqrt{14}}
& -\frac{3 \left(15 \psi _1+13 \psi _2+6 \left(\psi _3+\psi _4\right)\right)}{\sqrt{14}} & -3 \sqrt{\frac{7}{2}} \left(\psi _2+\psi _4\right) \\
\hline
 3 & 3 \sqrt{\frac{2}{7}} \left(\psi _1+5 \psi _2+4 \psi _4\right) & -\frac{3 \left(7 \psi _1-6 \psi _2+6 \psi _3-5 \psi _4\right)}{\sqrt{14}} &
-\frac{3 \left(13 \psi _1+3 \psi _2+6 \psi _3-2 \psi _4\right)}{\sqrt{14}} &
\frac{3 \left(50 \psi _1+\psi _2+32 \psi _3-25 \psi _4\right)}{\sqrt{14}}
\\
\hline
 4 & -3 \sqrt{\frac{2}{7}} \left(5 \psi _1+4 \psi _3-2 \psi _4\right) & 3 \sqrt{\frac{2}{7}} \left(9 \psi _1+7 \psi _2+3 \psi _3+2 \psi _4\right)
& 3 \sqrt{\frac{2}{7}} \left(-6 \psi _1+\psi _2-3 \psi _3+3 \psi _4\right) & 3
\sqrt{\frac{2}{7}} \left(17 \psi _1+3 \psi _2+5 \psi _3-5 \psi _4\right)
\\
\hline
 5 & -3 \sqrt{\frac{2}{7}} \left(4 \psi _1-5 \psi _2+4 \psi _3-6 \psi _4\right) & -\frac{3 \left(25 \psi _1+8 \psi _2+12 \psi _3-\psi _4\right)}{\sqrt{14}}
& -\frac{3 \left(\psi _1+5 \psi _2+4 \psi _4\right)}{\sqrt{14}} & \frac{-60 \psi _1+45 \psi _2-30 \psi _3+51 \psi _4}{\sqrt{14}} \\
\hline
 6 & -3 \sqrt{\frac{2}{7}} \left(9 \psi _1+\psi _2+9 \psi _3-3 \psi _4\right) & 3 \sqrt{\frac{7}{2}} \left(\psi _1-2 \psi _2+2 \psi _3-3 \psi _4\right)
& 3 \sqrt{\frac{7}{2}} \left(3 \psi _1+3 \psi _2+2 \psi _4\right) & \frac{9 \left(2 \psi _1+3 \psi _2+\psi _4\right)}{\sqrt{14}} \\
\hline
 7 & -3 \sqrt{\frac{2}{7}} \left(5 \psi _1+7 \psi _2-3 \psi _3+5 \psi _4\right) & -3 \sqrt{\frac{2}{7}} \left(\psi _1-8 \psi _2+6 \psi _3-8 \psi
_4\right) & 3 \sqrt{14} \left(2 \psi _1+\psi _2\right) & -3 \sqrt{\frac{2}{7}} \left(\psi _1+\psi _2-3 \psi _3+3 \psi _4\right) \\
\hline
 8 & 3 \sqrt{\frac{2}{7}} \left(6 \psi _1-\psi _2+3 \psi _3-3 \psi _4\right) & \frac{3 \left(9 \psi _1-2 \psi _2+12 \psi _3-9 \psi _4\right)}{\sqrt{14}}
& 3 \sqrt{\frac{7}{2}} \left(3 \psi _1+3 \psi _2+2 \psi _4\right) & 3 \sqrt{\frac{7}{2}} \left(\psi _2+\psi _4\right) \\
\hline
 9 & 3 \sqrt{\frac{2}{7}} \left(\psi _1-8 \psi _2+6 \psi _3-8 \psi _4\right) & \frac{3 \left(23 \psi _1-2 \psi _2+12 \psi _3-9 \psi _4\right)}{\sqrt{14}}
& -3 \sqrt{\frac{7}{2}} \left(\psi _1+3 \psi _2+2 \psi _4\right) & \frac{3 \left(2 \psi _1+9 \psi _2-6 \psi _3+13 \psi _4\right)}{\sqrt{14}} \\
\hline
 10 & 3 \sqrt{\frac{2}{7}} \left(15 \psi _1+6 \psi _2+6 \psi _3-\psi _4\right) & \frac{-93 \psi _1+30 \psi _2-54 \psi _3+51 \psi _4}{\sqrt{14}} &
\frac{3 \left(45 \psi _1+25 \psi _2+18 \psi _3+4 \psi _4\right)}{\sqrt{14}} & \frac{3 \left(4 \psi _1-\psi _2-5 \psi _4\right)}{\sqrt{14}} \\
\hline
 11 & -3 \sqrt{\frac{2}{7}} \left(9 \psi _1+7 \psi _2+3 \psi _3+2 \psi _4\right) & \frac{3 \left(-39 \psi _1+4 \psi _2-24 \psi _3+25 \psi _4\right)}{\sqrt{14}}
& \frac{75 \psi _1-33 \psi _2+72 \psi _3-54 \psi _4}{\sqrt{14}} & -\frac{3 \left(4 \psi _1+13 \psi _2+9 \psi _4\right)}{\sqrt{14}} \\
\hline
 12 & 3 \sqrt{\frac{2}{7}} \left(6 \psi _1-\psi _2+3 \psi _3-3 \psi _4\right) & -3 \sqrt{\frac{2}{7}} \left(9 \psi _1-9 \psi _2+5 \psi _3-9 \psi
_4\right) & -3 \sqrt{\frac{2}{7}} \left(5 \psi _1-5 \psi _2+9 \psi _3-5 \psi
_4\right) & 3 \sqrt{\frac{2}{7}} \left(2 \psi _1+3 \psi _2+\psi _4\right)
\\
\hline
 13 & 3 \sqrt{\frac{2}{7}} \left(5 \psi _1+\psi _2+3 \psi _3\right) & -\frac{3 \left(5 \psi _1+18 \psi _2+13 \psi _4\right)}{\sqrt{14}} & -\frac{3
\left(5 \psi _1+3 \psi _2-6 \psi _3+4 \psi _4\right)}{\sqrt{14}} & \frac{60 \psi _1-45 \psi _2+30 \psi _3-51 \psi _4}{\sqrt{14}} \\
\hline
 14 & 3 \sqrt{\frac{2}{7}} \left(5 \psi _1+\psi _2+3 \psi _3\right) & \frac{57 \psi _1+9 \left(2 \psi _2+4 \psi _3+\psi _4\right)}{\sqrt{14}} & -\frac{3
\left(21 \psi _1+3 \psi _2+18 \psi _3-8 \psi _4\right)}{\sqrt{14}} & \frac{48 \psi _1-57 \psi _2+66 \psi _3-87 \psi _4}{\sqrt{14}} \\
\hline
 15 & -3 \sqrt{\frac{2}{7}} \left(5 \psi _1+\psi _2+3 \psi _3\right) & -\frac{3 \left(31 \psi _1+10 \psi _2+12 \psi _3-5 \psi _4\right)}{\sqrt{14}}
& \frac{3 \left(-19 \psi _1+7 \psi _2-18 \psi _3+16 \psi _4\right)}{\sqrt{14}} & \frac{60 \psi _1-3 \psi _2+30 \psi _3-9 \psi _4}{\sqrt{14}} \\
\hline
 16 & -3 \sqrt{14} \left(\psi _1+\psi _2+\psi _3\right) & -3 \sqrt{\frac{7}{2}} \left(\psi _1+2 \psi _2+\psi _4\right) & -\frac{3 \left(25 \psi _1+3
\psi _2+10 \psi _3-4 \psi _4\right)}{\sqrt{14}} & -\frac{3 \left(10 \psi _1+7 \psi _2-6 \psi _3+3 \psi _4\right)}{\sqrt{14}} \\
\hline
 17 & -3 \sqrt{\frac{2}{7}} \left(\psi _1-2 \psi _2-3 \psi _4\right) & -\frac{3 \left(13 \psi _1+10 \psi _2+6 \psi _3+5 \psi _4\right)}{\sqrt{14}}
& \frac{3 \left(13 \psi _1-7 \psi _2+2 \psi _3-8 \psi _4\right)}{\sqrt{14}} & \frac{3 \left(2 \psi _1+3 \psi _2+\psi _4\right)}{\sqrt{14}} \\
\hline
 18 & -3 \sqrt{\frac{2}{7}} \left(\psi _1-9 \psi _2+7 \psi _3-10 \psi _4\right) & \frac{3 \left(-3 \psi _1+4 \psi _2-12 \psi _3+5 \psi _4\right)}{\sqrt{14}}
& \frac{3 \left(11 \psi _1-9 \psi _2+8 \psi _3-14 \psi _4\right)}{\sqrt{14}} & -\frac{3 \left(2 \psi _1+3 \psi _2+\psi _4\right)}{\sqrt{14}} \\
\hline
 19 & 3 \sqrt{\frac{2}{7}} \left(-6 \psi _1+5 \psi _2-7 \psi _3+11 \psi _4\right) & -3 \sqrt{\frac{2}{7}} \left(5 \psi _1+7 \psi _2-3 \psi _3+5 \psi
_4\right) & 3 \sqrt{\frac{2}{7}} \left(\psi _1+\psi _2-3 \psi _3+3 \psi
_4\right) & -3 \sqrt{\frac{2}{7}} \left(2 \psi _1+3 \psi _2+\psi _4\right)
\\
\hline
 20 & 3 \sqrt{\frac{2}{7}} \left(17 \psi _1+3 \psi _2+5 \psi _3-5 \psi _4\right) & -3 \sqrt{\frac{2}{7}} \left(5 \psi _1+4 \psi _3-2 \psi _4\right)
& 3 \sqrt{\frac{2}{7}} \left(8 \psi _1+\psi _2+4 \psi _3-4 \psi _4\right) & 3
\sqrt{\frac{2}{7}} \left(13 \psi _1+7 \psi _2+9 \psi _3-\psi _4\right)
\\
\hline
 21 & -9 \sqrt{\frac{2}{7}} \left(6 \psi _1-\psi _2+3 \psi _3-3 \psi _4\right) & \frac{3 \left(-11 \psi _1+2 \psi _2-8 \psi _3+7 \psi _4\right)}{\sqrt{14}}
& \frac{3 \left(-5 \psi _1+11 \psi _2-8 \psi _3+10 \psi _4\right)}{\sqrt{14}} &
\frac{3 \left(58 \psi _1+21 \psi _2+24 \psi _3-5 \psi _4\right)}{\sqrt{14}}
\\
\hline
 22 & 3 \sqrt{\frac{2}{7}} \left(-17 \psi _1+4 \left(\psi _2-3 \psi _3+3 \psi _4\right)\right) & \frac{3 \left(\psi _1-2 \psi _2-3 \psi _4\right)}{\sqrt{14}}
& \frac{3 \left(21 \psi _1-9 \psi _2+16 \psi _3-18 \psi _4\right)}{\sqrt{14}} &
\frac{3 \left(12 \psi _1-7 \psi _2+18 \psi _3-9 \psi _4\right)}{\sqrt{14}}
\\
\hline
 23 & -3 \sqrt{\frac{2}{7}} \left(11 \psi _1+\psi _2+5 \psi _3-\psi _4\right) & -\frac{3 \left(15 \psi _1+6 \psi _2+6 \psi _3-\psi _4\right)}{\sqrt{14}}
& \frac{3 \left(13 \psi _1+\psi _2+8 \psi _3-6 \psi _4\right)}{\sqrt{14}} & -3 \sqrt{\frac{7}{2}} \left(\psi _2+\psi _4\right) \\
\hline
 24 & -3 \sqrt{\frac{2}{7}} \left(9 \psi _1-9 \psi _2+5 \psi _3-9 \psi _4\right) & \frac{3 \left(15 \psi _1+6 \psi _2+6 \psi _3-\psi _4\right)}{\sqrt{14}}
& -\frac{3 \left(13 \psi _1+\psi _2+8 \psi _3-6 \psi _4\right)}{\sqrt{14}} & \frac{3 \left(4 \psi _1-\psi _2-5 \psi _4\right)}{\sqrt{14}} \\
\hline
 25 & 3 \sqrt{\frac{2}{7}} \left(\psi _1+\psi _2-3 \psi _3+3 \psi _4\right) & \frac{3 \left(15 \psi _1+6 \psi _2+6 \psi _3-\psi _4\right)}{\sqrt{14}}
& -\frac{3 \left(13 \psi _1+\psi _2+8 \psi _3-6 \psi _4\right)}{\sqrt{14}} & -\frac{3 \left(4 \psi _1+13 \psi _2+9 \psi _4\right)}{\sqrt{14}} \\
\hline
 26 & 3 \sqrt{\frac{2}{7}} \left(13 \psi _1+3 \psi _2+6 \psi _3-2 \psi _4\right) & -3 \sqrt{\frac{7}{2}} \left(\psi _1-2 \psi _2+2 \psi _3-3 \psi
_4\right) & -\frac{3 \left(19 \psi _1+9 \psi _2+2 \left(\psi _3+\psi
_4\right)\right)}{\sqrt{14}} & 3 \sqrt{\frac{7}{2}} \left(\psi _2+\psi
_4\right)
\\
\hline
 27 & -3 \sqrt{\frac{2}{7}} \left(\psi _1+4 \psi _2-6 \psi _3+2 \psi _4\right) & -3 \sqrt{14} \left(\psi _1+\psi _2+\psi _3\right) & 3 \sqrt{\frac{2}{7}}
\left(6 \psi _1+\psi _2+\psi _3+\psi _4\right) & -3 \sqrt{\frac{2}{7}} \left(5 \psi _1+7 \psi _2-3 \psi _3+5 \psi _4\right) \\
\hline
 28 & -3 \sqrt{\frac{2}{7}} \left(6 \psi _1+3 \psi _2+6 \psi _3-2 \psi _4\right) & 3 \sqrt{\frac{7}{2}} \left(-\psi _1+\psi _4\right) & \frac{3 \left(19
\psi _1-5 \psi _2+16 \psi _3-12 \psi _4\right)}{\sqrt{14}} & -\frac{3 \left(10 \psi _1+21 \psi _2-6 \psi _3+17 \psi _4\right)}{\sqrt{14}} \\
\hline
 29 & 3 \sqrt{\frac{2}{7}} \left(12 \psi _1+11 \psi _2+6 \psi _4\right) & \frac{3 \left(\psi _1+10 \psi _2-12 \psi _3+7 \psi _4\right)}{\sqrt{14}}
& \frac{3 \left(-9 \psi _1+13 \psi _2-2 \psi _3+10 \psi _4\right)}{\sqrt{14}} & \frac{3 \left(2 \psi _1+3 \psi _2+\psi _4\right)}{\sqrt{14}} \\
\hline
 30 & 3 \sqrt{\frac{2}{7}} \left(2 \psi _1-3 \psi _2+6 \psi _3-4 \psi _4\right) & -\frac{3 \left(27 \psi _1+30 \psi _2+17 \psi _4\right)}{\sqrt{14}}
& \frac{3 \left(-17 \psi _1+\psi _2-2 \psi _3+6 \psi _4\right)}{\sqrt{14}} & \frac{3 \left(2 \psi _1+3 \psi _2+\psi _4\right)}{\sqrt{14}} \\
\hline
 31 & -3 \sqrt{\frac{2}{7}} \left(13 \psi _2-6 \psi _3+12 \psi _4\right) & -\frac{3 \left(7 \psi _1+2 \psi _2+12 \psi _3-3 \psi _4\right)}{\sqrt{14}}
& \frac{3 \left(-13 \psi _1+5 \psi _2-14 \psi _3+18 \psi _4\right)}{\sqrt{14}} & \frac{3 \left(2 \psi _1+3 \psi _2+\psi _4\right)}{\sqrt{14}} \\
\hline
 32 & -3 \sqrt{\frac{2}{7}} \left(16 \psi _1+11 \psi _2+6 \psi _3+3 \psi _4\right) & \frac{3 \left(7 \psi _1+6 \psi _2-6 \psi _3+5 \psi _4\right)}{\sqrt{14}}
& \frac{3 \left(-5 \psi _1+3 \psi _2+8 \psi _4\right)}{\sqrt{14}} & \frac{3 \left(12 \psi _1-7 \psi _2+18 \psi _3-9 \psi _4\right)}{\sqrt{14}} \\
\hline
 33 & 3 \sqrt{\frac{2}{7}} \left(14 \psi _1+\psi _2+6 \psi _3-5 \psi _4\right) & -\frac{3 \left(27 \psi _1+6 \psi _2+10 \psi _3-3 \psi _4\right)}{\sqrt{14}}
& -\frac{3 \left(27 \psi _1+7 \psi _2+16 \psi _3-8 \psi _4\right)}{\sqrt{14}} & \frac{36 \psi _1-63 \psi _2+54 \psi _3-69 \psi _4}{\sqrt{14}} \\
\hline
 34 & -3 \sqrt{\frac{2}{7}} \left(2 \psi _1+3 \psi _2+\psi _4\right) & -\frac{3 \left(29 \psi _1+2 \psi _2+10 \psi _3-9 \psi _4\right)}{\sqrt{14}}
& \frac{45 \psi _1-75 \psi _2+48 \psi _3-84 \psi _4}{\sqrt{14}} & \frac{3 \left(32 \psi _1+21 \psi _2+6 \psi _3+11 \psi _4\right)}{\sqrt{14}} \\
\hline
 35 & 6 \sqrt{\frac{2}{7}} \left(5 \psi _1+\psi _2+3 \psi _3\right) & -3 \sqrt{\frac{2}{7}} \left(15 \psi _1+6 \psi _2+6 \psi _3-\psi _4\right) &
-3 \sqrt{\frac{2}{7}} \left(13 \psi _1+\psi _2+8 \psi _3-6 \psi _4\right) & -3 \sqrt{\frac{2}{7}} \left(2 \psi _1+3 \psi _2+\psi _4\right) \\
\hline
\end{array}
}\)
\end{doublespace}
} {\tiny
\begin{doublespace}
\noindent\(\pmb{
\begin{array}{|c|c|c|}
\hline
 0 & 13 & 14 \\
\hline
 1 & -3 \sqrt{14} \left(2 \psi _1+\psi _2\right) & 3 \sqrt{\frac{7}{2}} \left(2 \psi _1-3 \psi _2+2 \psi _3-3 \psi _4\right) \\
\hline
 2 & 3 \sqrt{\frac{2}{7}} \left(9 \psi _1+7 \psi _2+3 \psi _3+2 \psi _4\right) & \frac{3 \left(18 \psi _1-\psi _2+14 \psi _3-5 \psi _4\right)}{\sqrt{14}}
\\
\hline
 3 & -9 \sqrt{\frac{2}{7}} \left(5 \psi _1+4 \psi _3-2 \psi _4\right) & \frac{3 \left(2 \psi _1+3 \psi _2+\psi _4\right)}{\sqrt{14}} \\
\hline
 4 & 3 \sqrt{\frac{2}{7}} \left(13 \psi _1+7 \psi _2+9 \psi _3-\psi _4\right) & 3 \sqrt{\frac{2}{7}} \left(-20 \psi _1+5 \psi _2-14 \psi _3+11 \psi
_4\right) \\
\hline
 5 & 3 \sqrt{\frac{2}{7}} \left(20 \psi _1+7 \psi _2+9 \psi _3-\psi _4\right) & \frac{6 \psi _1-33 \psi _2-39 \psi _4}{\sqrt{14}} \\
\hline
 6 & 3 \sqrt{\frac{2}{7}} \left(-13 \psi _1+3 \psi _2-5 \psi _3+7 \psi _4\right) & \frac{3 \left(16 \psi _1+13 \psi _2+4 \psi _3+7 \psi _4\right)}{\sqrt{14}}
\\
\hline
 7 & -3 \sqrt{\frac{2}{7}} \left(4 \psi _1-9 \psi _2+\psi _3-7 \psi _4\right) & 3 \sqrt{\frac{2}{7}} \left(11 \psi _1+5 \psi _2+\psi _3\right) \\
\hline
 8 & 3 \sqrt{\frac{2}{7}} \left(-11 \psi _1+2 \psi _2-8 \psi _3+7 \psi _4\right) & -\frac{3 \left(20 \psi _1+25 \psi _2-2 \psi _3+21 \psi _4\right)}{\sqrt{14}}
\\
\hline
 9 & -3 \sqrt{\frac{2}{7}} \left(11 \psi _1+5 \psi _2+\psi _3\right) & -\frac{3 \left(34 \psi _1+11 \psi _2+12 \psi _3-7 \psi _4\right)}{\sqrt{14}}
\\
\hline
 10 & 3 \sqrt{\frac{2}{7}} \left(8 \psi _1+\psi _2+4 \psi _3-4 \psi _4\right) & -\frac{3 \left(22 \psi _1+3 \psi _2+16 \psi _3-7 \psi _4\right)}{\sqrt{14}}
\\
\hline
 11 & -3 \sqrt{\frac{2}{7}} \left(13 \psi _1+\psi _2+8 \psi _3-6 \psi _4\right) & \frac{3 \left(4 \psi _1-11 \psi _2+10 \psi _3-11 \psi _4\right)}{\sqrt{14}}
\\
\hline
 12 & 3 \sqrt{\frac{2}{7}} \left(5 \psi _1+4 \psi _3-2 \psi _4\right) & 3 \sqrt{\frac{2}{7}} \left(9 \psi _1+7 \psi _2+3 \psi _3+2 \psi _4\right)
\\
\hline
 13 & 3 \sqrt{\frac{2}{7}} \left(-14 \psi _1+\psi _2-8 \psi _3+9 \psi _4\right) & -\frac{3 \left(8 \psi _1-3 \psi _2+8 \psi _3-5 \psi _4\right)}{\sqrt{14}}
\\
\hline
 14 & -3 \sqrt{\frac{2}{7}} \left(2 \psi _1+3 \psi _2+\psi _4\right) & -\frac{9 \left(12 \psi _1+3 \psi _2+8 \psi _3-3 \psi _4\right)}{\sqrt{14}}
\\
\hline
 15 & 3 \sqrt{\frac{2}{7}} \left(12 \psi _1+3 \psi _2+8 \psi _3-3 \psi _4\right) & \frac{-57 \psi _2+36 \psi _3-51 \psi _4}{\sqrt{14}} \\
\hline
 16 & -3 \sqrt{\frac{2}{7}} \left(2 \psi _1+3 \psi _2+\psi _4\right) & \frac{3 \left(14 \psi _1+3 \psi _2+4 \psi _3-\psi _4\right)}{\sqrt{14}} \\
\hline
 17 & 3 \sqrt{\frac{2}{7}} \left(8 \psi _1+5 \psi _2+4 \psi _4\right) & \frac{3 \left(-28 \psi _1+5 \psi _2-12 \psi _3+17 \psi _4\right)}{\sqrt{14}}
\\
\hline
 18 & 3 \sqrt{\frac{2}{7}} \left(\psi _1+5 \psi _2+4 \psi _4\right) & -\frac{3 \left(14 \psi _1+9 \psi _2-2 \psi _3+11 \psi _4\right)}{\sqrt{14}}
\\
\hline
 19 & -3 \sqrt{\frac{2}{7}} \left(13 \psi _1+16 \psi _2+10 \psi _4\right) & 3 \sqrt{\frac{2}{7}} \left(7 \psi _1-\psi _2+\psi _3-2 \psi _4\right)
\\
\hline
 20 & -3 \sqrt{\frac{2}{7}} \left(6 \psi _1-\psi _2+3 \psi _3-3 \psi _4\right) & 3 \sqrt{\frac{2}{7}} \left(9 \psi _1+7 \psi _2+3 \psi _3+2 \psi
_4\right) \\
\hline
 21 & 3 \sqrt{\frac{2}{7}} \left(\psi _1-2 \psi _2-3 \psi _4\right) & \frac{3 \left(6 \psi _1-15 \psi _2+10 \psi _3-17 \psi _4\right)}{\sqrt{14}}
\\
\hline
 22 & 3 \sqrt{\frac{2}{7}} \left(5 \psi _1+\psi _2+3 \psi _3\right) & \frac{3 \left(-24 \psi _1+\psi _2-16 \psi _3+13 \psi _4\right)}{\sqrt{14}}
\\
\hline
 23 & 3 \sqrt{\frac{2}{7}} \left(7 \psi _1-9 \psi _2+9 \psi _3-11 \psi _4\right) & \frac{3 \left(-6 \psi _1+17 \psi _2-12 \psi _3+21 \psi _4\right)}{\sqrt{14}}
\\
\hline
 24 & 3 \sqrt{\frac{2}{7}} \left(5 \psi _1+7 \psi _2-3 \psi _3+5 \psi _4\right) & \frac{3 \left(42 \psi _1+11 \psi _2+24 \psi _3-13 \psi _4\right)}{\sqrt{14}}
\\
\hline
 25 & 3 \sqrt{\frac{2}{7}} \left(5 \psi _1-5 \psi _2+9 \psi _3-5 \psi _4\right) & \frac{3 \left(6 \psi _1-13 \psi _2+8 \psi _3-13 \psi _4\right)}{\sqrt{14}}
\\
\hline
 26 & -3 \sqrt{\frac{2}{7}} \left(5 \psi _1+4 \psi _3-2 \psi _4\right) & \frac{3 \left(8 \psi _1+9 \psi _2-4 \psi _3+5 \psi _4\right)}{\sqrt{14}}
\\
\hline
 27 & -3 \sqrt{\frac{2}{7}} \left(\psi _1+\psi _2-3 \psi _3+3 \psi _4\right) & -3 \sqrt{\frac{2}{7}} \left(2 \psi _1+3 \psi _2+\psi _4\right) \\
\hline
 28 & 3 \sqrt{\frac{2}{7}} \left(6 \psi _1+\psi _2+\psi _3+\psi _4\right) & -\frac{3 \left(4 \psi _1+3 \psi _2-4 \psi _3+3 \psi _4\right)}{\sqrt{14}}
\\
\hline
 29 & 3 \sqrt{\frac{2}{7}} \left(5 \psi _1-3 \psi _2+7 \psi _3-8 \psi _4\right) & \frac{3 \left(8 \psi _1+17 \psi _2+2 \psi _3+7 \psi _4\right)}{\sqrt{14}}
\\
\hline
 30 & 3 \sqrt{\frac{2}{7}} \left(5 \psi _1+3 \psi _2+\psi _3+4 \psi _4\right) & \frac{3 \left(8 \psi _1+\psi _2-10 \psi _3+3 \psi _4\right)}{\sqrt{14}}
\\
\hline
 31 & 3 \sqrt{\frac{2}{7}} \left(7 \psi _1-\psi _2+\psi _3-2 \psi _4\right) & \frac{3 \left(5 \psi _2+2 \psi _3+3 \psi _4\right)}{\sqrt{14}} \\
\hline
 32 & 3 \sqrt{\frac{2}{7}} \left(4 \psi _1-5 \psi _2+4 \psi _3-6 \psi _4\right) & \frac{3 \left(4 \psi _1+\psi _2-2 \psi _3-\psi _4\right)}{\sqrt{14}}
\\
\hline
 33 & -3 \sqrt{\frac{2}{7}} \left(4 \psi _1-5 \psi _2+4 \psi _3-6 \psi _4\right) & \frac{3 \left(-4 \psi _1-\psi _2+2 \psi _3+\psi _4\right)}{\sqrt{14}}
\\
\hline
 34 & 3 \sqrt{\frac{2}{7}} \left(4 \psi _1-5 \psi _2+4 \psi _3-6 \psi _4\right) & \frac{3 \left(4 \psi _1+\psi _2-2 \psi _3-\psi _4\right)}{\sqrt{14}}
\\
\hline
 35 & 6 \sqrt{\frac{2}{7}} \left(4 \psi _1-5 \psi _2+4 \psi _3-6 \psi _4\right) & -3 \sqrt{\frac{2}{7}} \left(4 \psi _1+\psi _2-2 \psi _3-\psi _4\right)
\\
\hline
\end{array}
}\)
\end{doublespace}
}

\end{document}